\documentclass[aps, prl, floatfix,
  reprint, 
  superscriptaddress, 
  longbibliography, noeprint, nofootinbib, nobibnotes
]{revtex4-2} 
\PassOptionsToPackage{table}{xcolor}
\usepackage{graphicx, tikz, xcolor} 
\usetikzlibrary{arrows.meta,positioning,fit,backgrounds,calc,shapes.geometric}
\usepackage[notrig]{physics} 
\usepackage{amsmath, amssymb}
\usepackage{mathtools} 
\usepackage{csquotes}
\usepackage[pdftex,unicode,pdfusetitle]{hyperref}
\usepackage{natbib, hypernat}
\usepackage[nameinlink,noabbrev]{cleveref}
\usepackage{booktabs}
\hypersetup{
    colorlinks=true,
    citecolor=blue,
    linkcolor=magenta,
    urlcolor=blue,
    breaklinks=true,
    pdfborder={0 0 0}
}
\graphicspath{ {./Figure/} }

\makeatletter
\def\frontmatter@maketitle{%
  \@author@finish
  \title@column\titleblock@produce
  \suppressfloats[t]%
  \let\abstract\@undefined\let\endabstract\@undefined
  \titlepage@sw{%
   \vfil
   \clearpage
  }{}%
  \onecolumn@grid@setup
  \def\set@footnotewidth{\set@footnotewidth@one}%
}%
\makeatother

\newif\ifMain
\newif\ifSupple
\Maintrue
\Suppletrue
\DeclareRobustCommand{\xref}[1]{%
    \ifMain
        \ifSupple
            #1%
        \else
            \begin{NoHyper}#1\end{NoHyper}%
        \fi
    \else
        \begin{NoHyper}#1\end{NoHyper}%
    \fi
}

\makeatletter
\newcommand{\rload}{%
    \begingroup
    \def\newlabel##1##2{%
        \@ifundefined{r@##1}{\global\@namedef{r@##1}{##2}}{}%
    }%
    \InputIfFileExists{refs.aux}{}{%
        \PackageError{refs}{Missing refs.aux}
            {Compile with Maintrue and Suppletrue first.}%
    }%
    \endgroup
}
\ifMain
    \ifSupple
        \newwrite\rout
        \AddToHook{enddocument/afterlastpage}{%
            \immediate\openout\rout=refs.aux\relax
            \let\rlabel\newlabel
            \def\newlabel#1#2{%
                \immediate\write\rout{%
                    \string\newlabel{#1}{\unexpanded{#2}}%
                }%
                \rlabel{#1}{#2}%
            }%
        }
        \AddToHook{enddocument/afteraux}{\immediate\closeout\rout}
    \else
        \AtBeginDocument{\rload}
    \fi
\else
    \AtBeginDocument{\rload}
\fi
\makeatother

\usepackage[resetlabels]{multibib}
\newcites{SM}{ }
\let\origciteSM\citeSM
\RenewDocumentCommand{\citeSM}{O{} O{} m}{%
  \begingroup
  \renewcommand{\citenumfont}[1]{S##1}%
  \origciteSM[#1][#2]{#3}%
  \endgroup
}
\makeatletter
\let\origbibliographySM\bibliographySM
\renewcommand{\bibliographySM}[1]{%
  \begingroup
    \edef\pre@bibdata{\pre@bibdata}%
    \def\bibnumfmt##1{[S##1]}%
    \def\@biblabel##1{[S##1]}%
    \origbibliographySM{#1}%
  \endgroup
}
\makeatother
\usepackage{etoolbox}
\makeatletter
  \robustify\citeSM
\makeatother

\usepackage{amsthm} 
\newtheorem{theorem}{Theorem}
\newtheorem{proposition}{Proposition}
\newtheorem{lemma}{Lemma}

\renewcommand{\qed}{$\hfill\blacksquare$}

\newcommand{\sysSize}{ L }
\newcommand{\Supple}{Supplemental Material~\cite{supplement}}
\newcommand{\Title}{Many-Body Bound States in the Continuum}

\begin{document}
\author{Shoki Sugimoto}
  \email{shoki.sugimoto@ap.t.u-tokyo.ac.jp}
  \affiliation{Department of Applied Physics, the University of Tokyo, 7-3-1 Hongo, Bunkyo-ku, Tokyo 113-0033, Japan}
  \affiliation{Nonequilibrium Quantum Statistical Mechanics RIKEN Hakubi Research Team, RIKEN Pioneering Research Institute (PRI), Wako, Saitama 351-0198, Japan}
\author{Yuto Ashida}
  \affiliation{Department of Physics, the University of Tokyo, 7-3-1 Hongo, Bunkyo-ku, Tokyo 113-0033, Japan}
  \affiliation{Institute for Physics of Intelligence, the University of Tokyo, 7-3-1 Hongo, Bunkyo-ku, Tokyo 113-0033, Japan}
\author{Masahito Ueda}
  \affiliation{Department of Physics, the University of Tokyo, 7-3-1 Hongo, Bunkyo-ku, Tokyo 113-0033, Japan}
  \affiliation{Institute for Physics of Intelligence, the University of Tokyo, 7-3-1 Hongo, Bunkyo-ku, Tokyo 113-0033, Japan}
  \affiliation{Fundamental Quantum Science Program (FQSP), TRIP Headquarters, RIKEN, Wako 351-0198, Japan}
\ifMain
\title{
    \Title
}
\begin{abstract}
    A bound state in the continuum~(BIC) is a spatially localized energy eigenstate embedded in a continuous spectrum of extended eigenstates.
    While diverse types of single-particle BICs have been reported in the literature, understanding how such localization persists in interacting many-body systems remains a fundamental problem.
    Here, we present numerical evidence and a perturbative analysis supporting the existence of many-body BICs in a one-dimensional Bose-Hubbard chain with an attractive impurity potential, a system previously known to host a BIC in the two-particle sector.
    Furthermore, we demonstrate that the Bethe-type construction of the known two-particle BIC breaks down already for three particles even for arbitrary finite superpositions of Bethe waves.
    %
    The resulting BICs violate the eigenstate thermalization hypothesis for several local one- and two-body observables, leading to nonthermal dynamics from experimentally accessible initial states.
\end{abstract}
\maketitle

\paragraph{Introduction.---}  
A bound state in the continuum~(BIC), which is a spatially localized eigenstate in a continuous spectrum of extended eigenstates, was originally proposed by von Neumann and Wigner in 1929~\cite{neumann1929on,note1}.
By modulating the amplitude of the eigenfunction of the free-particle Schrödinger equation and then constructing an artificial potential that supports the modulated eigenfunction, they demonstrated the first example of a single-particle BIC.
As an exact eigenstate, a BIC should neither decay nor couple to the extended eigenstates.
Although the potential they proposed has not been realized so far, a simpler potential using semiconductor epitaxial heterostructures was proposed~\cite{herrick1976construction, stillinger1976potentials}, and a BIC was experimentally observed via bandgap engineering~\cite{capasso1992observation}.
A rich variety of BICs amenable to experimental implementation have been proposed on the basis of symmetry protection~\cite{plotnik2011experimental}, variable separation~\cite{Robnik1986}, parameter tuning known as accidental BICs~\cite{hsu2013observation, hsu2013bloch, corrielli2013observation}, and formation by the interference of two resonances belonging to different channels~\cite{friedrich1985interfering, marinica2008bound, yang2014analytical}.
%
BICs are generic wave phenomena in classical systems, including optics~\cite{hsu2016bound, koshelev2020engineering, joseph2021bound, azzam2021photonic} and acoustics~\cite{tong2020observation, deriy2022bound}, and find applications in nonlinear enhancement~\cite{carletti2018giant, koshelev2019nonlinear, xu2019dynamic, carletti2019high, liu2019high, anthur2020continuous, zograf2022high}, coherent light generation~\cite{kodigala2017lasing, song2018cherenkov, pavlov2018lasing, ha2018directional, wu2020room, azzam2021single, wang2021highly, mohamed2022controlling}, sensors~\cite{yesilkoy2019ultrasensitive, romano2019tuning, tseng2020dielectric, wang2021ultrasensitive, chen2020integrated, altug2022advances}, filters~\cite{foley2014symmetry, doskolovich2019integrated}, and integrated circuits~\cite{bezus2018spatial, yu2019photonic, lin2020chip, bezus2021integrated}.

Beyond single-particle systems, BICs have been studied in interacting lattice models~\cite{zhang2012bound} and waveguide quantum electrodynamics~\cite{ashida2022nonperturbative}.
Recent advances include multiparticle localization and quasi-BIC pumping in interaction-modulated Bose--Hubbard lattices~\cite{Huang_PhysRevLett2024y}, boundary-localized many-body BICs and their circuit simulations~\cite{Sun_CommunPhys2024f}, and the stabilization of three-body resonances into BICs by parameter tuning~\cite{Happ2026-lx}.
A central question is whether many-body BICs can persist beyond the exceptional solvable structure of a known few-particle BIC.
Nonintegrability alone does not answer this question: the two-particle BIC in a one-dimensional Hubbard model with an attractive impurity potential has an exact Bethe wave function despite the nonintegrability of the Hamiltonian~\cite{zhang2012bound}.
Moreover, Liu and Chen have constructed exact many-particle Bethe-type BICs in a different interacting lattice model~\cite{Liu2025-lf}, showing that increasing particle number does not by itself eliminate the Bethe-solvable structure.

In this Letter, we provide numerical evidence and an analytical discussion supporting the existence of BICs in genuinely many-body regimes in the one-dimensional Bose--Hubbard model with an attractive impurity potential. 
We further prove that the Bethe-type construction underlying the known two-particle BIC already breaks down in the three-particle sector, even when arbitrary finite superpositions of Bethe waves are allowed in each free-particle region of configuration space.
This result establishes a precise limitation of the known exact construction and resolves the question of whether the partial integrability identified in Ref.~\cite{zhang2012bound} persists.
Together, these results show that many-body BICs persist in this model beyond the Bethe-solvable structure, rather than being a straightforward continuation of the known two-particle solution.

From a broader perspective, the proposed many-body BICs expand the list of mechanisms that break the eigenstate thermalization hypothesis (ETH)~\cite{von2010proof,deutsch1991quantum,srednicki1994chaos,rigol2008thermalization}, such as integrability~\cite{kinoshita2006quantum, rigol2007relaxation, rigol2009breakdown, rigol2009quantum, cassidy2011generalized, calabrese2011quantum, ilievski2015complete, hamazaki2016generalized, essler2016quench, vidmar2016generalized, fukai2020noncommutative}, many-body localization~\cite{basko2006metal, oganesyan2007localization, vznidarivc2008many, pal2010many, nandkishore2015many, luitz2015many, serbyn2015criterion, choi2016exploring, smith2016many, imbrie2016many}, and quantum many-body scars~\cite{bernien2017probing, shiraishi2017systematic, turner2018weak, turner2018quantum, moudgalya2018exact, bull2019systematic, ho2019periodic, lin2019exact, schecter2019weak, shibata2020onsager, Sanada_PhysRevB2023l}.
For specified observables, the ETH imposes conditions on their matrix elements in the energy eigenbasis;
specifically, it requires their expectation values in \textit{every} energy eigenstate within the relevant energy shell to agree with the corresponding thermal values~\cite{rigol2008thermalization, biroli2010effect, rigol2010quantum, beugeling2014finite, kim2014testing, mondaini2016eigenstate, mondaini2017eigenstate, sugimoto2021test, sugimoto2022eigenstate}.
A crucial observation here is that the many-body BICs exhibit particle distributions distinct from those of the extended eigenstates in the same energy shell, therefore breaking the ETH for observables localized around the impurity.
Since these BICs form only a sparse subset of the spectrum, the resulting ETH violation constitutes a form of weak ergodicity breaking analogous to the phenomenon observed in quantum many-body scars.
The analogy extends to the dynamics: we further demonstrate that simple, experimentally accessible initial states have substantial overlap with the BICs and exhibit nonthermal dynamics for these observables.

\paragraph{Setup.---}
We consider a Bose-Hubbard model on a one-dimensional periodic lattice of $2L+1$ sites with an impurity potential $V$ at the center of the chain:
\begin{multline}
    \hat{H} 
    \coloneqq 
    -t \sum_{x=-\sysSize}^{\sysSize} \qty( \hat{b}_{x+1}^{\dagger} \hat{b}_{x} +\hat{b}_{x}^{\dagger} \hat{b}_{x+1} )
    \\
    + \frac{U}{2} \sum_{x=-\sysSize}^{\sysSize} \hat{n}_{x} (\hat{n}_{x} - 1) + V \hat{n}_{0},
    \label{eq:Hamiltonian}
\end{multline}
where $t$ denotes the hopping amplitude, $U$ the interaction strength, $\hat{b}_{x}$ the annihilation operator of bosons at site $x$ with $\comm*{ \hat{b}_{x} }{ \hat{b}_{y}^{\dagger} } = \delta_{xy}$, and $\hat{n}_{x} \coloneqq \hat{b}_{x}^{\dagger} \hat{b}_{x}$ the particle number operator at site $x$.
This Hamiltonian conserves the total particle number $\hat{N} \coloneqq \sum_{x=-\sysSize}^{\sysSize} \hat{n}_{x}$ and is invariant under reflection $\hat{P}_{\sysSize} \hat{b}_{x} \hat{P}_{\sysSize} \coloneqq \hat{b}_{-x}$.
In Ref.~\cite{zhang2012bound}, the Hamiltonian~\eqref{eq:Hamiltonian} is proved to host a Bethe-type BIC in the $N=2$ sector for a certain region of parameters $(t,U,V)$.
In what follows, we first demonstrate that this Bethe-type mechanism already fails in the $N=3$ sector.
We then present numerical evidence supporting the existence of many-body BICs for $N=4$ and $6$.
Finally, we provide a perturbative argument on the nature of these many-body BICs.

\paragraph{Absence of three-particle Bethe-type bound states.---}
First, we work in the $N=3$ sector of the Hamiltonian~\eqref{eq:Hamiltonian} and consider an infinite chain $L = \infty$.
For an energy eigenstate $\ket*{E}$ satisfying $\hat{H} \ket*{E} = E \ket*{E}$, we denote its wavefunction by $\psi(x_1,x_2,x_3) \coloneqq \matrixel{0}{\hat{b}_{x_1} \hat{b}_{x_2} \hat{b}_{x_3}}{E} / \sqrt{3!}$ \ $(x_1 \leq x_2 \leq x_3)$.
Away from particle contacts and the impurity, the configuration space has four connected chambers,
\begin{equation}
    \mathcal C_m \coloneqq \{x_1<x_2<x_3 \colon x_m < 0 < x_{m+1}\} \subset \mathbb{Z}^{3}
    \label{eq:chambers}
\end{equation}
with $m=0,1,2,3$, $x_{0} = -\infty$ and $x_{4} = +\infty$.
We consider bound energy eigenstates of the finite-superposition Bethe type in these free chambers.
Here, an eigenstate is said to be of the finite-superposition Bethe type if there exist a finite set $\mathcal Z\subset(\mathbb C^{\times})^{3}$ and coefficients
$A_{m,\vb{z}}$ such that, for each $m\in\{0,1,2,3\}$,
\begin{equation}
    \psi(\vb{x})
    = F_m(\vb{x})
    \coloneqq \sum_{\vb{z}\in\mathcal Z} A_{m,\vb{z}} z_1^{x_1}z_2^{x_2}z_3^{x_3},
    \quad
    \forall \vb{x}\in\mathcal C_m.
    \label{eq:finite-bethe}
\end{equation}
We do not assume this Bethe form on particle contacts $x_i=x_j$, on the impurity faces $x_i=0$, or on their intersections.

A subset of the eigenvalue equations at the two-particle contacts and the impurity suffices to establish the following theorem (see \Supple{} for details):
\begin{theorem}
    \label{thm:main}
    Under $tUV\neq0$, there is no nontrivial three-particle eigenstate of the Hamiltonian~\eqref{eq:Hamiltonian} that is normalizable and of finite-superposition Bethe type.
\end{theorem}
Thus, even an arbitrary finite-superposition Bethe-type extension of the \enquote{partial integrability} construction of Ref.~\cite{zhang2012bound} fails already in the three-particle
sector.
This obstruction renders the existence of BICs for $N>2$ highly nontrivial.

\paragraph{Numerical evidence for many-body BICs in four- and six-particle sectors.---}
Below, we provide numerical evidence that BICs nevertheless occur in the $N=4$ and $N=6$ sectors, suggesting that many-body BICs may occur at more general particle numbers.
We employ the exact diagonalization to find many-body BICs of the Hamiltonian~\eqref{eq:Hamiltonian}.
Figure~\ref{fig:022_EnergySpectrum_BoundStates_Main} shows the energy spectrum of $\hat{H}$ in the four-particle sector.
The color of the data points represents the width $\Delta x_{\alpha}$ of the particle distribution in the corresponding eigenstate $\ket*{E_{\alpha}}$ given by
\begin{align}
    \Delta x_{\alpha}
    \coloneqq \sqrt{ \sum_{x=-\sysSize}^{\sysSize} x^2 \rho_{\alpha}(x) - \qty( \sum_{x=-\sysSize}^{\sysSize} x \rho_{\alpha}(x) )^2 },
    \label{eq:WidthOfParticleDistribution}
\end{align}
where $\rho_{\alpha}(x) \coloneqq \expval*{\hat{n}_{x}}{ E_{\alpha} } / N$ is the particle density in the eigenstate $\ket*{ E_{\alpha} }$ with a fixed particle number $\hat{N} \ket*{ E_{\alpha} } = N \ket*{ E_{\alpha} }$.

\begin{figure}[tb]
    \centering
    \includegraphics[width=\linewidth]{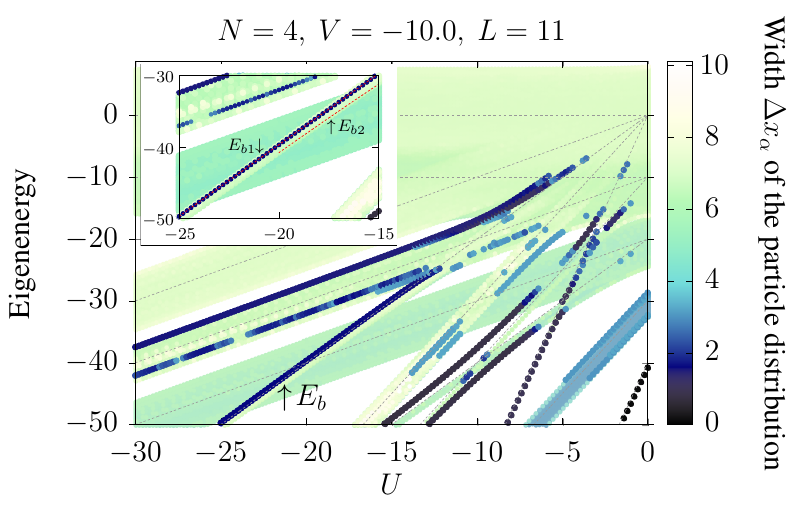}
    \caption{\label{fig:022_EnergySpectrum_BoundStates_Main}
        Energy spectrum of the full Hamiltonian $\hat{H}$~\eqref{eq:Hamiltonian} for $L=11$, $t=1$ and $V=-10$ plotted against $U$.
        The color of data points indicates the width $\Delta x_{\alpha}$ of the particle distribution defined in Eq.~\eqref{eq:WidthOfParticleDistribution}.
        The dashed gray lines in the main panel represent the eigenenergies of $\hat{H}_{0} \coloneqq U\sum_{x=-\sysSize}^{\sysSize} \hat{n}_{x}(\hat{n}_{x}-1)/2 +V\hat{n}_{0}$. 
        The inset shows the spectrum in the region $U\in [-25,-15]$ and $E\in [-50,-30]$, where localized states $\ket*{E_{b}, \pm}$ (dark blue points) cross the continuum of the extended states (light green points).
        The dashed red curves in the inset show the energies ($E_{b1}$ and $E_{b2}$) of the bound states of the effective Hamiltonian $\hat{H}_{(0,2,2)}$ given in Eqs.~\eqref{eq:022_BoundStateEnergy1} and \eqref{eq:022_BoundStateEnergy2}.
        The bound-state energy $E_{b1}$ of $\hat{H}_{(0,2,2)}$ agrees excellently with the energy $E_{b}$ of $\ket*{E_{b}, \pm}$.
    }
\end{figure}

There exist two special eigenstates $\ket*{E_{b}, \pm}$ featuring opposite parities $\pm$ and a narrow width $\Delta x_{b}$ located at the upper envelope of the band with $E_{\alpha} \simeq 2U$.
We observe that these localized eigenstates $\ket*{E_{b}, \pm}$ remain nearly unaffected even when the two bands with $E_{\alpha} \simeq 2U$ and $E_{\alpha} \simeq U + 2V$ intersect at $U \simeq 2V$.
Therefore, these states serve as promising candidates for four-particle BICs.
To determine whether each of $\ket*{E_{b}, \pm}$ is a BIC or a resonance state, we examine the particle distributions $\rho_{b_{\pm}}(x)$ in $\ket*{E_{b}, \pm}$ in Fig.~\ref{fig:ParticleDensity_N4}, finding that they decrease exponentially with increasing $\abs*{x}$ for $2 < \abs*{x} \lesssim 5$ for the system sizes investigated.
The tails of $\rho_{b_{+}}(x)$ with $\abs{x} \gtrsim 5$ decrease with increasing $\sysSize$, which indicates that $\rho(x) \propto e^{ -\order{ \abs*{x} } }$ in the limit $\sysSize\to\infty$.
Therefore, we conclude that $\ket*{E_{b}, +}$ is a four-particle BIC.
On the other hand, the tails of $\rho_{b_{-}}(x)$ with $\abs{x} \gtrsim 5$ do not decrease with increasing $\sysSize$.
Thus, we cannot exclude the possibility that $\ket*{E_{b}, -}$ is not a bound state but a resonance state whose particle density does not vanish for larger $\abs*{x}$.

\begin{figure}[tb]
    \centering
    \includegraphics[width=\linewidth]{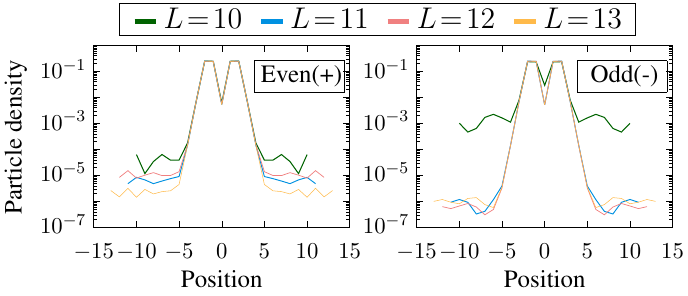}
    \caption{\label{fig:ParticleDensity_N4} 
        Particle distributions of the BIC candidates $\ket*{E_b,\pm}$ at $(t,U,V)=(1,-20,-10)$.
        The decreasing even-parity tails identify $\ket*{E_b,+}$ as a BIC, whereas the odd-parity tails remain inconclusive at the accessible sizes.
    }
\end{figure}

We also investigate the energy spectrum of $\hat{H}$ in the six-particle sector $N=6$, as shown in Fig.~\ref{fig:EnergySpectrum_N6}, which yields significantly more BIC candidates than the four-particle sector.
Specifically, at $(U/t, V/t) = (-15,-20)$, the eigenenergies of the candidate eigenstates for six-particle BICs are labeled with red circles in Fig.~\ref{fig:EnergySpectrum_N6}.
While some candidates emerge at the intersections of two bands, others reside within an individual band away from intersections.
By investigating the particle distribution of these candidate eigenstates, we find numerical evidence for at least seven six-particle BICs as listed in Table~\ref{tab:N6_BoundAndResonanceStates}~(see \Supple{} for details).

\begin{figure}[tb]
    \centering
    \includegraphics[width=\linewidth]{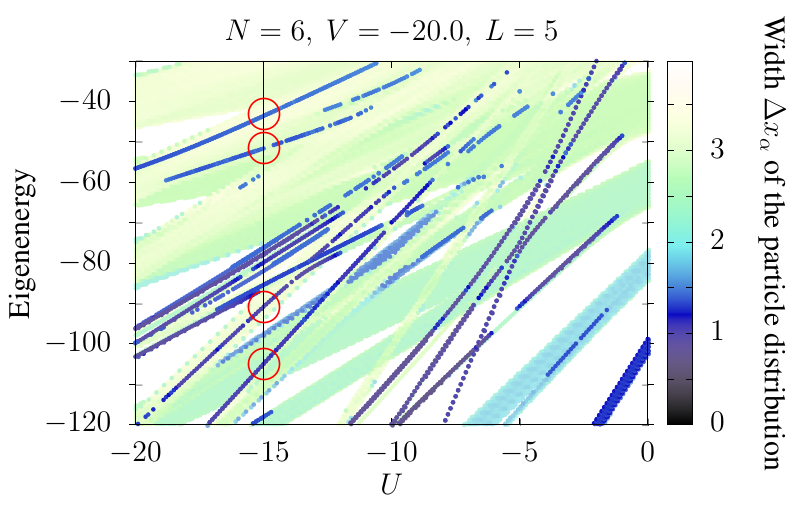}
    \caption{\label{fig:EnergySpectrum_N6}
        Energy spectrum within the six-particle $(N=6)$ sector for $L=5$, $t=1$ and $V=-20$ plotted against $U$.
        At $U=-15$, we find four candidate eigenenergies for BICs, marked by red circles.
    }
\end{figure}
\begin{table}[tb]
    \centering
    \begin{tabular}{|c||c|c|c||c||c|c|c|} \hline
        & {\footnotesize Energy} & {\footnotesize Parity} & & & {\footnotesize Energy} & {\footnotesize Parity} & \\ \hline \hline
         $E_{b1}$ & -106.28 & $\pm$ & BIC 
         & $E_{b5}^{+}$ & -51.661 & $+$ & ?  \\
         $E_{b2}$ & -105.00 & $\pm$ & {\footnotesize Resonance} 
         & $E_{b5}^{-}$ & -51.657 & $-$ & ?  \\
         $E_{b3}$ & -104.74 & $\pm$ & BIC 
         & $E_{b6}$ & -43.623 & $\pm$ & {\footnotesize Resonance} \\
         $E_{b4}^{+}$ & -90.846 & $+$   & BIC
         & $E_{b7}$ & -43.136 & $\pm$ & BIC \\[0.5ex] \hline
    \end{tabular}
    \caption{\label{tab:N6_BoundAndResonanceStates}
        Summary of the candidate eigenstates for six-particle BICs identified along the line $U = -15$ in Fig.~\ref{fig:EnergySpectrum_N6}.
        The unlabeled columns indicate whether each state is a BIC or a resonance state~(see \Supple{} for details).
    }
\end{table}

\paragraph{Perturbative analysis of four-particle BICs.---}
We here give a perturbative analysis supporting our numerical findings presented above.
We decompose the Hamiltonian $\hat{H}$ as $\hat{H} = \hat{H}_{0} +t\hat{T}$,
where $\hat{H}_{0} \coloneqq U\sum_{x=-\sysSize}^{\sysSize} \hat{n}_{x}(\hat{n}_{x}-1)/2 +V\hat{n}_{0}$, and $\hat{T} \coloneqq - \sum_{x=-\sysSize}^{\sysSize} ( \hat{b}_{x+1}^{\dagger} \hat{b}_{x} +\hat{b}_{x}^{\dagger} \hat{b}_{x+1} )$.
The Fock states $\ket*{\vb{n}} \coloneqq \prod_{x=-\sysSize}^{\sysSize} (\hat{b}_{x}^{\dagger})^{n_{x}} \ket*{0}$ with $\vb{n} \coloneqq (n_{-\sysSize}, n_{-\sysSize+1},\dots,n_{\sysSize})$ and $\sum_{x=-\sysSize}^{\sysSize} n_{x} = N$ are eigenstates of $\hat{H}_{0}$.
Since many-body BICs are found in a strongly interacting regime with $|U|, |V|\gg t$, we consider a regime where $\abs*{ E^{(0)}_{\vb{n}} } \gg t$ so that the kinetic term $t\hat{T}$ can be treated perturbatively.
We rewrite$\ket*{\vb{n}}$ as 
\begin{equation}
    \ket*{\vb{n}} \propto \prod_{x=-\sysSize}^{\sysSize} (\hat{b}_{x}^{\dagger})^{n_{x}} \ket*{0} = (\hat{b}_{0}^{\dagger})^{N_{0}} \prod_{\alpha=1}^{M} (\hat{b}_{x_{\alpha}}^{\dagger})^{N_{\alpha}} \ket*{0},
\end{equation}
where $M$ is an integer so that $N_{\alpha} \ (\alpha\geq 1)$ is nonzero.
We change the label of $\ket*{\vb{n}}$ from $\vb{n}$ to $(\vb{N}; \vb{x}) \coloneqq (N_{0},N_{1},\dots, N_{M}; x_{1},\dots, x_{M})$, where $N_{0}$ is the number of particles on the impurity, and $x_{\alpha}$ and $N_{\alpha}$ are the position and the number of bosons of the $\alpha$th cluster.
The eigenenergy of $\hat{H}_{0}$ for the configuration $(\vb{N}; \vb{x})$ is given by
\begin{equation}
    E^{(0)}_{\vb{N}} \coloneqq U\sum_{\alpha=0}^{M} \frac{N_{\alpha}(N_{\alpha}-1)}{2} +VN_{0}\qc
\end{equation}
which is independent of $\vb{x} = (x_{1},\dots,x_{M})$.
Then, we denote the subspace with given $\vb{N}$ by
\begin{equation}
    \mathcal{H}_{\vb{N}} \coloneqq
    \operatorname{span}\!\left\{
        (\hat{b}_{0}^{\dagger})^{N_{0}}
        \prod_{\alpha=1}^{M}
        (\hat{b}_{x_{\alpha}}^{\dagger})^{N_{\alpha}}
        \ket*{0}
        \middle\vert\
        \text{\small$
            \begin{aligned}
                x_{\alpha} &\neq 0 &&\hspace{-1ex} (\forall\alpha) \\[-2pt]
                x_{\alpha} &\neq x_{\beta} &&\hspace{-1ex} (\forall\alpha<\beta)
            \end{aligned}
        $}
    \right\}.
\end{equation}
To be concrete, let us consider an effective Hamiltonian $\hat{H}_{\vb{N}}$ describing the dynamics within the subspace $\mathcal{H}_{\vb{N}}$ with $\vb{N} = (0,2,2)$ and $\vb{N} = (2,1,1)$. 
At their intersection, a four-particle BIC is found numerically, as shown in Fig.~\ref{fig:022_EnergySpectrum_BoundStates_Main}.

\paragraph{Effective Hamiltonian in $\mathcal{H}_{(0,2,2)}$.---}
In the sector with $\vb{N} = (0,2,2)$, the effective Hamiltonian to leading order in $t$ is given by $\hat{H}_{(0,2,2)} = 2U +\hat{H}_{(0,2,2)}^{(2)} +\order{t^4}$ with
\begin{multline}
    \hat{H}_{(0,2,2)}^{(2)} 
    = \frac{8t^2}{U} + \frac{2t^2}{U} \sum_{x(\neq 0)} \hat{A}_{x}^{\dagger} ( \hat{A}_{x-1} + \hat{A}_{x+1} ) \\ 
    -\frac{16t^2}{U} \sum_{x} \hat{n}^{(A)}_{x}\hat{n}^{(A)}_{x+1} +\frac{2t^2 V}{U(U-V)} \left(\hat{n}^{(A)}_{+1} + \hat{n}^{(A)}_{-1} \right),
\end{multline}
where $\hat{A}_{x}$ is an annihilation operator of a two-boson cluster that satisfies
$
   \hat{A} \ket*{0} = 0,\ 
   \hat{A}\hat{A}^{\dagger} \ket*{0} = \ket*{0},\ 
   (\hat{A}^{\dagger})^2 \ket*{0} = 0,\ 
   \hat{A}^{\dagger} \ket*{0} = \frac{1}{\sqrt{2}} (\hat{b}^{\dagger})^2\ket*{0}
$, and we set $\hat{n}^{(A)}_{x} \coloneqq \hat{A}_{x}^{\dagger} \hat{A}_{x}$ and $\hat{A}_{0} = 0$~\cite{note2} (see \Supple{} for the derivation).

Assuming a Bethe-ansatz form for the eigenfunctions, we find three bound states of $\hat{H}_{(0,2,2)}^{(2)}$ near $U = 2V$, where the four-particle BIC is numerically found, which lie \textit{outside} the continuum of $\hat{H}_{(0,2,2)}^{(2)}$~(see Fig.~\ref{fig:022_EnergySpectrum_BoundStates_Main}).
The eigenenergies of the two of them are degenerate and given by
\begin{equation}
    E_{b1} = 2U-\frac{8t^2}{U} +\frac{16t^2}{U} \frac{V^2}{(U-V)(U+7V)}\qc
    \label{eq:022_BoundStateEnergy1}
\end{equation}
and the eigenenergy of the third one is 
\begin{equation}
    E_{b2} = 2U+\frac{8t^2}{U} +\frac{4t^2}{U} \frac{(U-V)^2+V^2}{V(U-V)}.
    \label{eq:022_BoundStateEnergy2}
\end{equation}
The $b1$ and $b2$ eigenstates exist for $-U/3V < 1$ and $0 < U/2V < 1$, respectively.
The eigenenergies $E_{b1}$ and $E_{b2}$ are depicted by red dashed curves in the inset of Fig.~\ref{fig:022_EnergySpectrum_BoundStates_Main}.
The spectrum of $2U+\hat{H}_{(0,2,2)}^{(2)}$ agrees excellently with that of $\hat{H}$ in the region $U\lesssim -12$ for $V=-10$.

The eigenfunctions for the eigenenergy $E_{b1}$ are given by
\begin{align}
    \psi_{\pm}^{(b1)}(x_{1},x_{2})
    &= \pm \psi_{\pm}^{(b1)}(-x_{2},-x_{1})
    \nonumber \\
    &\propto \qty(\frac{U-V}{V})^{x_{1}} \qty(-\frac{V}{U+7V})^{x_{2}}
    \label{Eq:022_BoundState1}
\end{align}
for $0 < x_{1} < x_{2}$, and $\psi_{\pm}^{(b1)}(x_{1},x_{2}) \equiv 0$ for $x_{1} < 0 < x_{2}$.
Here, the subscript $\pm$ denotes the parity of the eigenfunction, i.e., $\psi_{\pm}(-x_{2}, -x_{1}) = \pm \psi_{\pm}(x_{1},x_{2})$.
In fact, we find only the even-parity eigenfunction for the eigenenergy $E_{b2}$, which is given by
\begin{align}
    \psi_{+}^{(b2)}(x_{1},x_{2}) &\propto \qty(\frac{U-V}{V})^{x_{2} - x_{1}},
    \label{Eq:022_BoundState2}
\end{align}
for $x_{1} < 0 < x_{2}$, and $\psi_{+}^{(b2)}(x_{1},x_{2}) \equiv 0$ otherwise.
The Fock states that have the largest overlap with the bound states $\psi_{\pm}^{(b1)}$ are $\frac{1}{2} (\hat{b}_{1}^{\dagger})^{2}(\hat{b}_{2}^{\dagger})^{2} \ket*{0}$ and $ \frac{1}{2} (\hat{b}_{-2}^{\dagger})^{2}(\hat{b}_{-1}^{\dagger})^{2} \ket*{0}$,
while that with $\psi_{+}^{(b2)}$ is $\frac{1}{2} (\hat{b}_{-1}^{\dagger})^{2}(\hat{b}_{1}^{\dagger})^{2} \ket*{0}$.

\paragraph{Effective Hamiltonian in $\mathcal{H}_{(0,2,2)} \oplus \mathcal{H}_{(2,1,1)}$.---}
At $U\simeq 2V$, where the four-particle BIC is numerically found, the subspaces $\vb{N} = (0,2,2)$ and $\vb{N} = (2,1,1)$ become nearly degenerate because $E_{(0,2,2)}^{(0)} = 2U$ and $E_{(2,1,1)}^{(0)} = U+2V$.
In this case, we cannot apply the perturbation theory to these two subspaces separately. 
Instead, we need to consider the union $\mathcal{H}_{(0,2,2)} \oplus \mathcal{H}_{(2,1,1)}$ of these two subspaces.
The effective Hamiltonian in this unified sector is given by $\hat{H}_{(0,2,2)+(2,1,1)} = \hat{H}_{(0,2,2)} + \hat{H}_{(2,1,1)} + \hat{H}_{\mathrm{mix}}$ with $\hat{H}_{(2,1,1)}$ being the effective Hamiltonian in the $\mathcal{H}_{(2,1,1)}$ sector and
\begin{equation}
    \hat{H}_{\mathrm{mix}}
    \coloneqq \frac{4\sqrt{2}t^2}{V} \qty(\hat{A}_{-1}^{\dagger}\hat{A}_{+1}^{\dagger}\hat{B}_{-1}\hat{B}_{+1} +  \hat{B}_{-1}^{\dagger}\hat{B}_{+1}^{\dagger}\hat{A}_{-1}\hat{A}_{+1}).
    \label{Eq:EffectiveHamiltonian_022+211}
\end{equation}
See \Supple{} for the derivation.
Here, $\hat{A}_{x}$ and $\hat{B}_{x}$ are the annihilation operators for the clusters in the subspaces $\vb{N}=(0,2,2)$ and $\vb{N}=(2,1,1)$, respectively, which satisfy $\hat{B} \ket*{0} = 0$, $\hat{B}\hat{B}^{\dagger} \ket*{0} = \ket*{0}$, $(\hat{B}^{\dagger})^2 \ket*{0} = 0$, $\hat{B}^{\dagger} \ket*{0} = \hat{b}^{\dagger}  \ket*{0}$ and $\hat{B} \hat{A}^{\dagger} \ket*{0} = 0$~\cite{note3}.

The inter-sector term $\hat{H}_{\mathrm{mix}}$ is of order $\order{t^2}$ and couples $\frac{1}{2} (\hat{b}_{-1}^{\dagger})^{2}(\hat{b}_{1}^{\dagger})^{2} \ket*{0} \in \mathcal{H}_{(0,2,2)}$ with $\frac{1}{\sqrt{2}} (\hat{b}_{0}^{\dagger})^{2}\, \hat{b}_{-1}^{\dagger}\hat{b}_{1}^{\dagger} \ket*{0} \in \mathcal{H}_{(2,1,1)}$.
Since the leading nontrivial term of $\hat{H}_{(0,2,2)}$ is also of order $\order{t^2}$, the inter-sector term $\hat{H}_{\mathrm{mix}}$ has nonnegligible influence on those eigenstates of $\hat{H}_{(0,2,2)}$ that have a nonnegligible overlap with the state $\frac{1}{2} (\hat{b}_{-1}^{\dagger})^{2}(\hat{b}_{1}^{\dagger})^{2} \ket*{0}$.
The bound state $\psi_{+}^{(b2)}$ is such a state.
Therefore, it mixes up with the states in $\hat{H}_{(2,1,1)}$ when $U\simeq 2V$ and thus disappears.

Meanwhile, the bound states $\psi_{\pm}^{(b1)}$ have no overlap with $\frac{1}{2} (\hat{b}_{-1}^{\dagger})^{2}(\hat{b}_{1}^{\dagger})^{2} \ket*{0}$, and the coupling between $\psi_{\pm}^{(b1)}$ and the states in $\hat{H}_{(2,1,1)}$ is of order $\order{t^3}$.
Therefore, $\psi_{\pm}^{(b1)}$ remain bound eigenstates of the effective Hamiltonian near $U=2V$, subject to a possible correction of order $\order{t}$.
Since the eigenstates of $\hat{H}_{(2,1,1)}$ form a continuum of extended states with almost the same energy as $E_{b1}$, the bound states $\psi_{\pm}^{(b1)}$ can be BICs when $U\simeq 2V$, which is consistent with our numerical finding in Fig.~\ref{fig:022_EnergySpectrum_BoundStates_Main}.

\paragraph{Violation of the ETH and influence of many-body BICs on the dynamics.---}
The particle distribution of a many-body BIC is, by definition, qualitatively different from that of extended eigenstates with comparable energies.
We here quantify the resulting ETH violation for the even-parity four- and six-particle BICs shown in Fig.~\ref{fig:ETH_dynamics} using the following quantity:
\begin{equation}
    \Delta_{\alpha}[\hat{A}] \coloneqq \frac{\abs{  \expval*{\hat{A}}{E_{\alpha}} - \expval*{\hat{A}}^{\mathrm{mc}}(E_{\alpha}) }}{ \| \hat{A} \|_{\infty} },
\end{equation}
for the particle density $\hat n_x$, the nearest-neighbor hopping $\hat b_{x+1}^{\dagger}\hat b_x+\mathrm{H.c.}$, and the onsite pair density $\hat n_x(\hat n_x-1)$.
Here, the microcanonical value is evaluated in the same parity sector, and $\|\!\cdot\!\|_{\infty}$ is the operator norm in the fixed-$N$ sector.
We find that $\Delta_{\alpha}[\hat{A}]$ remains $\order{1}$ in the fixed-$N$ dilute limit for these observables around the impurity ($|x| \lesssim 4$; see Fig.~\ref{fig:ETH_dynamics} and End Matter).
Since the ETH dictates that this quantity must vanish for \textit{every} eigenstate in the relevant energy shell, these BICs clearly violate the ETH for all three observables.

We next examine the dynamics from normalized four- and six-particle initial states proportional to
\begin{equation}
    (\hat{b}_{1}^{\dagger})^{2} (\hat{b}_{2}^{\dagger})^{2} \ket*{0},\
    (\hat{b}_{1}^{\dagger})^{4} (\hat{b}_{2}^{\dagger})^{2} \ket*{0},
    \
    (\hat{b}_{1}^{\dagger})^{2} (\hat{b}_{2}^{\dagger})^{2} (\hat{b}_{3}^{\dagger})^{2} \ket*{0},
    \label{eq:initialStates}
\end{equation}
which have substantial overlaps with the four-particle BIC $\ket*{E_b,+}$ and the six-particle BICs $\ket*{E_{b1},+}$ and $\ket*{E_{b7},+}$.

Following the sector-resolved ETH treatment of Ref.~\cite{hamazaki2016generalized} and accounting for the finite energy width of the initial state, we use the thermal reference value given by
\begin{equation}
    \expectationvalue*{\hat{A}}_{\mathrm{ref}} \coloneqq \sum_{p = \pm 1} \sum_{\alpha} |\braket*{E_{\alpha,p}}{\psi(0)}|^2 \expectationvalue*{\hat{A}}^{\mathrm{mc}}_p(E_{\alpha,p}),
    \label{eq:thermalReference}
\end{equation}
where $\expectationvalue*{\hat{A}}^{\mathrm{mc}}_p(E_{\alpha,p})$ denotes the microcanonical average evaluated in the $p$-parity energy shell centered at energy $E_{\alpha,p}$ with a half-width of $0.2 t$.
If the ETH holds within each parity sector, Eq.~\eqref{eq:thermalReference} gives the predicted value for the long-time average.
The lower panels of Fig.~\ref{fig:ETH_dynamics} show that the local densities do not relax to their thermal values over the time intervals shown; End Matter shows a similar non-thermal behavior for the nearest-neighbor hopping and onsite pair densities.
Moderate variations of the energy-shell half-width slightly shift the thermal reference values but leave our conclusion unaffected.
The persistent oscillations in Fig.~\ref{fig:ETH_dynamics}(b) are attributed to the coherence between the nearby six-particle BICs $\ket*{E_{b1},\pm}$ and $\ket*{E_{b3},\pm}$.
The site-resolved occupation patterns in Eq.~\eqref{eq:initialStates} can be prepared in ultracold atomic gases with single-site control~\cite{bakr2009quantum,wurtz2009experimental}.
Thus, many-body BICs not only violate the ETH for the local observables considered here, but also prevent these observables from thermalizing from simple initial states having substantial overlaps with the BICs.

\begin{figure*}[bth]
\centering
\includegraphics[width=1.05\linewidth]{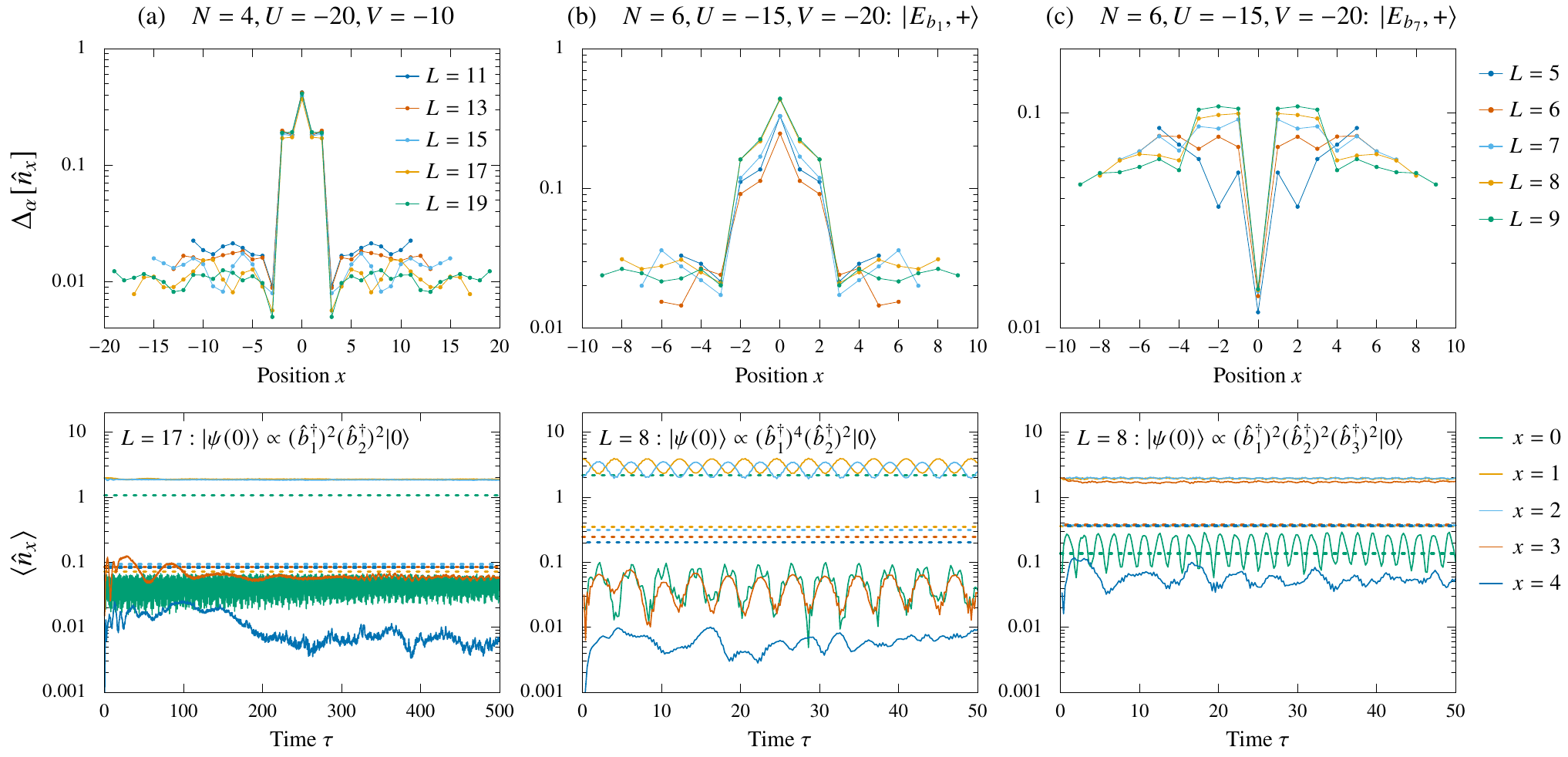}
\caption{
\label{fig:ETH_dynamics}
    Upper panels: System-size dependence of $\Delta_{\alpha}[\hat{n}_x]$ for the even-parity BICs $\ket*{E_b,+}$, $\ket*{E_{b1},+}$, and $\ket*{E_{b7},+}$ in columns~(a)--(c), respectively.
    The microcanonical averages are evaluated with a half-width of $0.2 t$.
    Lower panels: Time evolution of local densities starting from initial states in Eq.~\eqref{eq:initialStates}, with their respective energy expectation values being $2U=-40$, $7U=-105$, and $3U=-45$.
    Dashed lines indicate the corresponding thermal reference values calculated from Eq.~\eqref{eq:thermalReference}.
}
\end{figure*}

\paragraph{Conclusions.---}
We have presented the numerical evidence and perturbative analysis supporting the existence of many-body BICs, i.e., bound eigenstates embedded in the continuum spectrum of extended states.
Specifically, for the Bose-Hubbard chain with an attractive impurity potential,   
we identify a BIC in the four-particle sector that originates from the bound states in the $\vb{N}=(0,2,2)$ sector and does not mix with the extended states of the sector $\vb{N}=(2,1,1)$ at second order in $t$ even when these two sectors become energetically degenerate.
Theorem~\ref{thm:main} shows that the Bethe-type construction of the known two-particle BIC, attributed to ``partial integrability''~\cite{zhang2012bound}, already breaks down at $N=3$, even when allowing arbitrary finite superpositions of Bethe waves in each free-particle region of configuration space.
Together with our numerical results, this shows that many-body BICs persist in this model beyond the Bethe-solvable structure of the known two-particle BIC.
%
By numerically obtaining the energy spectrum and eigenstates, we find at least seven BICs in the six-particle sector as shown in Fig.~\ref{fig:EnergySpectrum_N6} and listed in Table~\ref{tab:N6_BoundAndResonanceStates}.
Beyond the model~\eqref{eq:Hamiltonian} studied in this Letter, 
we expect that many-body BICs can be found in a wide range of many-body quantum systems with local impurity potentials whose strength is comparable to the interaction energy between particles.

Since the particle distributions of BICs and extended states with almost the same eigenenergies markedly disagree, BICs are non-thermalizing states that violate the ETH for observables localized around the impurity.
We have quantified this ETH violation for particle densities, nearest-neighbor hopping, and onsite pair densities around the impurity in the fixed-$N$ dilute limit.
We have also demonstrated that these observables exhibit non-thermal dynamics when starting from the simple initial states in Eq.~\eqref{eq:initialStates}.
Since the one-dimensional Bose-Hubbard model is non-integrable regardless of the presence of the impurity at the center, our results suggest that introducing a local impurity can induce weak ergodicity breaking in a non-integrable system.
It remains an intriguing open problem to rigorously prove the existence of the many-body BICs discovered here in the limit $L\to\infty$ beyond the perturbative regime. 
Exploring the existence of BICs in different quantum lattice models also merits future investigation.

\begin{acknowledgments}
We are very grateful to Synge Todo and Tilman Hartwig for their help in our numerical calculation.
This work was supported by KAKENHI Grant Numbers JP22H01152 from the Japan Society for the Promotion of Science (JSPS).
S.~S. was supported by KAKENHI Grant Number JP22J14935 from the Japan Society for the Promotion of Science (JSPS) and Forefront Physics and Mathematics Program to Drive Transformation (FoPM), a World-leading Innovative Graduate Study (WINGS) Program, the University of Tokyo.
Y.~A. acknowledges support from the Japan Society for the Promotion of Science through Grant No.~JP19K23424 and from JST FOREST Program (Grant Number~JPMJFR222U, Japan).
M. U. acknowledges support by the RIKEN TRIP initiative.
ChatGPT (OpenAI, 5.6 Sol and 6 Astra) was used during manuscript revision to assist with analytical calculations, algebraic checks, and presentation.
The authors formulated the problems and assumptions, developed the proof strategy, and independently verified all arguments and calculations before incorporating them into the manuscript.
\end{acknowledgments}

\bibliography{reference}

\begin{thebibliography}{103}
\bibitem [{not({\natexlab{b}})}]{note2}%
  \BibitemOpen
  \bibinfo {note} {{An example of such an operator $\hat{A}_{x}$ is \begin{align} \hat{A}_{x} = \sqrt{ \frac{ \sqrt{2} \cos\qty( \frac{\pi}{4} \hat{n}_{x} ) \cos\qty( \frac{\pi}{4} (\hat{n}_{x}+1) ) }{ (\hat{n}_{x}+2) (\hat{n}_{x}+1) } }\ \hat{b}_{x}^2, \end{align} where $\hat{n}_{x} \coloneqq \hat{b}_{x}^{\dagger} \hat{b}_{x}$ }}\BibitemShut {NoStop}%
\bibitem [{not({\natexlab{c}})}]{note3}%
  \BibitemOpen
  \bibinfo {note} {{An example of such a pair of operators $\hat{A}_{x}, \hat{B}_{x}$ is \begin{gather} \hat{A}_{x} = \sqrt{ \frac{ \sqrt{2} \cos\qty( \frac{\pi}{4} \hat{n}_{x} ) \cos\qty( \frac{\pi}{4} (\hat{n}_{x}+1) ) }{ (\hat{n}_{x}+2) (\hat{n}_{x}+1) } }\ \hat{b}_{x}^2\qc \\ \hat{B}_{x} = \frac{ 4\cos^2\qty( \frac{2\pi}{3} \hat{n}_{x} ) - 1 }{3 \sqrt{ \hat{n}_{x}+1 }}\ \hat{b}_{x}, \end{gather} where $\hat{n}_{x} \coloneqq \hat{b}_{x}^{\dagger} \hat{b}_{x}$ }}\BibitemShut {NoStop}%
\end{thebibliography}
\begin{figure*}
\section{End Matter}\nopagebreak
    \centering
    \includegraphics[width=\linewidth]{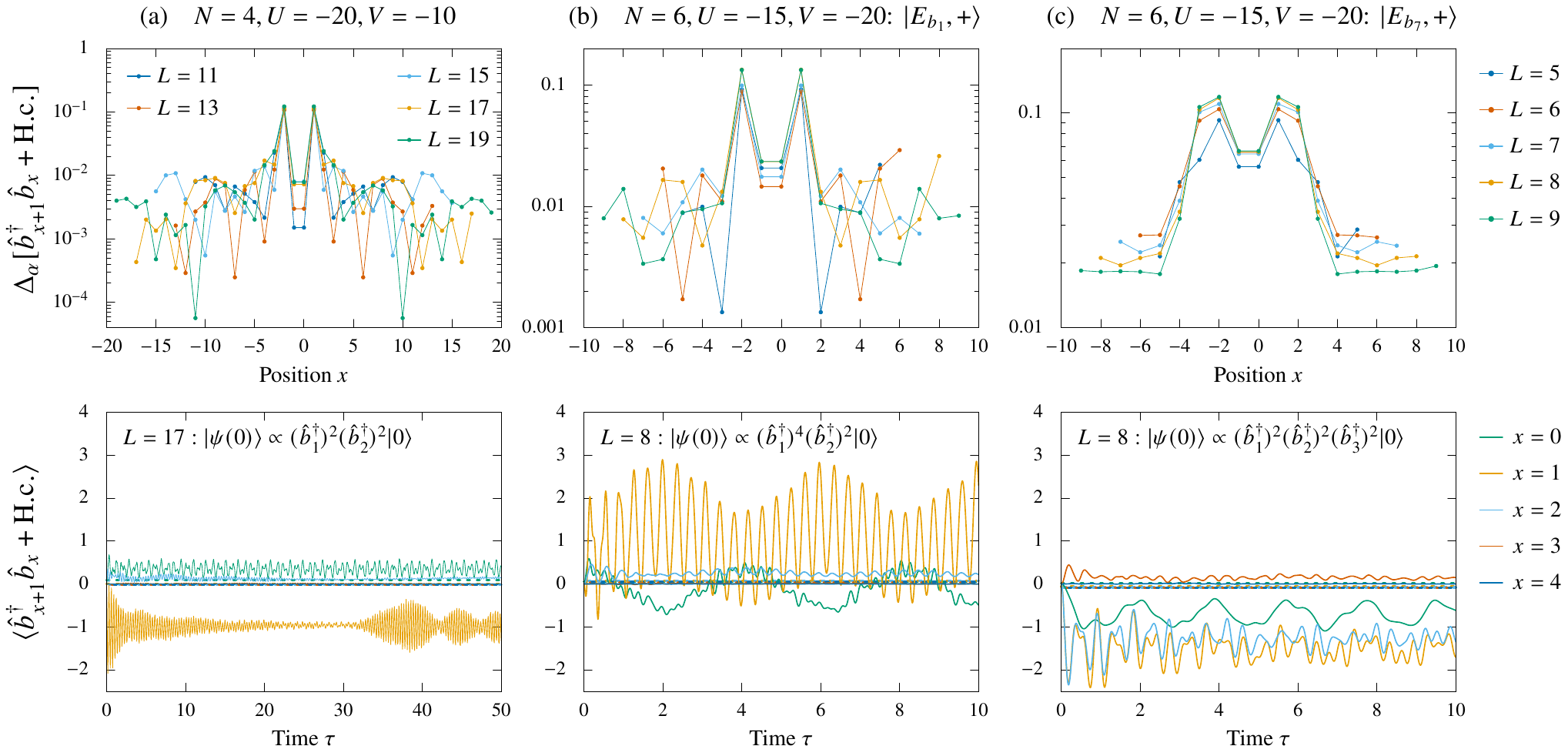}
    \caption{
        ETH violation and dynamics for the nearest-neighbor hopping $\hat{b}_{x+1}^{\dagger}\hat{b}_x+\hat{b}_x^{\dagger}\hat{b}_{x+1}$.
        The upper panels show the system-size dependence of the deviation $\Delta_\alpha$ in the even-parity BICs, and the lower panels display the expectation-value dynamics starting from the corresponding
        normalized Fock initial states.
        All BICs, initial states, parameters, system sizes, and color/line conventions are identical to those in Fig.~\ref{fig:ETH_dynamics}.
    }
    \label{fig:ETH_Dynamics-hop}
    \vspace{1em}
    \includegraphics[width=\linewidth]{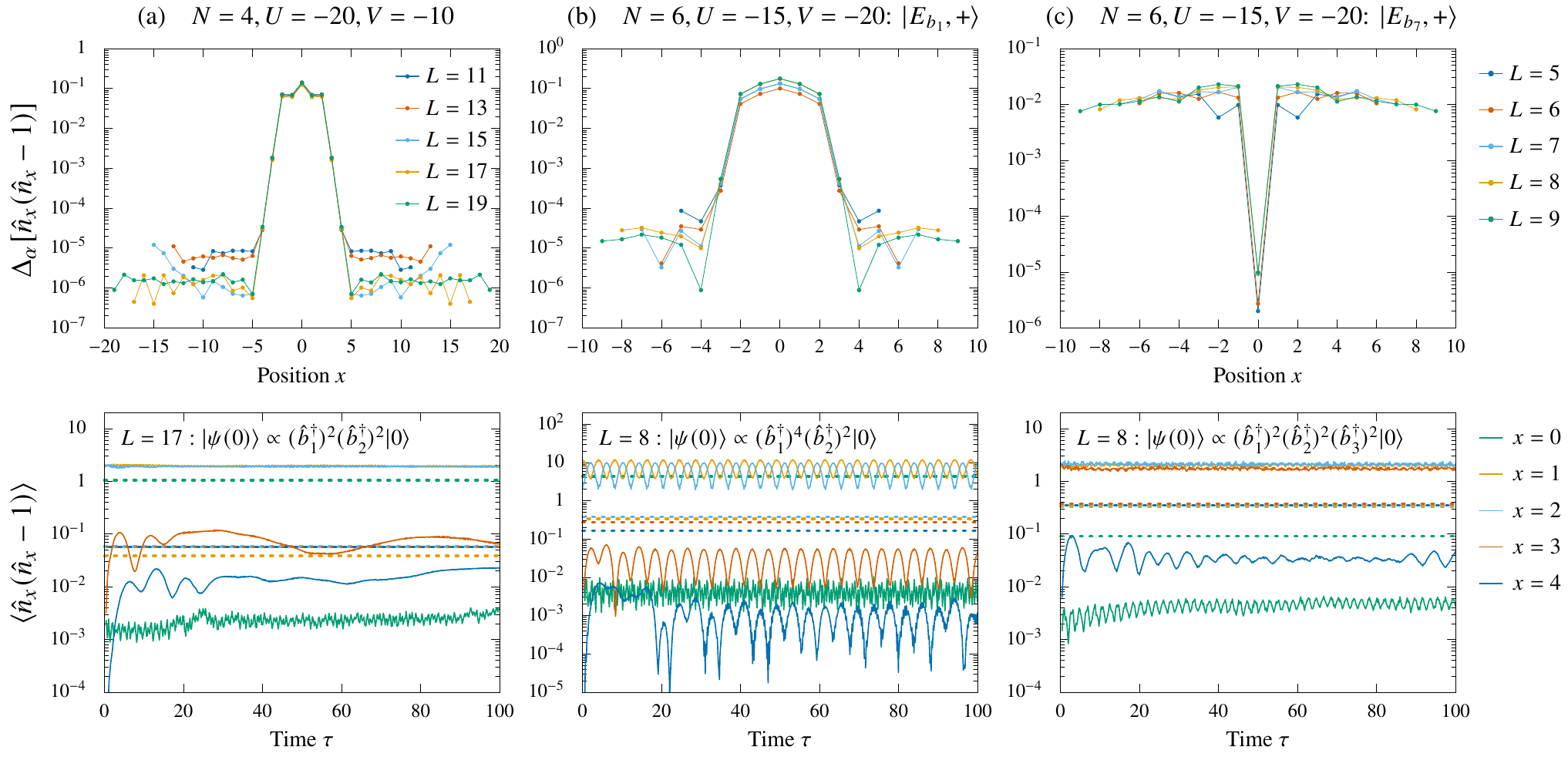}
    \caption{
        ETH violation and dynamics for the onsite pair density $\hat{n}_x(\hat{n}_x-1)$.
        The upper panels show the system-size dependence of the deviation $\Delta_\alpha$ in the even-parity BICs, and the lower panels display the expectation-value dynamics starting from the corresponding
        normalized Fock initial states.
        All BICs, initial states, parameters, system sizes, and color/line conventions are identical to those in Fig.~\ref{fig:ETH_dynamics}.
    }
    \label{fig:ETH_Dynamics-corr}
\end{figure*}
\fi

\ifSupple
\clearpage\clearpage

\setcounter{equation}{0}
\setcounter{figure}{0}
\setcounter{section}{0}
\setcounter{table}{0}
\renewcommand{\theequation}{S\arabic{equation}}
\renewcommand{\thefigure}{S\arabic{figure}}
\renewcommand{\thetable}{S\arabic{table}}
\renewcommand{\theHequation}{\theequation}
\renewcommand{\theHfigure}{\thefigure}
\renewcommand{\theHtable}{\thetable}

\title{
    Supplemental Material: \Title
}
\date{\today}
\maketitle

\section*{I. Proof of Theorem~\xref{\ref{thm:main}}}
Here, we provide a proof of Theorem~\xref{\ref{thm:main}} in the main text.
The boundedness condition implies that the wave function $\psi(\vb{x})$ is square summable.
This square summability is defined over the bosonic three-particle Hilbert space.
Since the multiplicity weights on the ordered configuration space are bounded between $1$ and $3!$, this requirement is equivalent to the ordinary $\ell^2$-summability of the amplitudes in the ordered coordinates.

The structure of the proof is outlined in Fig.~\ref{fig:proof_logic}.
First, we translate square summability into strict decay conditions on the Bethe waves in the four chambers (Lemma~\ref{lem:cones}).
Second, we derive adjacent-particle contact equations and impurity equations from a subset of the exact eigenvalue equations.
Third, we group the exponential terms into orbits under permutations and inversions of their bases and show that the equations do not couple different orbits.
Within each orbit, we combine the contact and impurity equations with the decay conditions to show that all Bethe coefficients vanish.
Finally, the eigenvalue equations and square summability imply that the wave function also vanishes on particle contacts and the impurity, thereby completing the proof.

Since the Hamiltonian is symmetric under reflection, if $\hat{H} \psi(\vb{x}) = E \psi(\vb{x})$, then $\hat{H} \hat{P} \psi(\vb{x}) = E \hat{P} \psi(\vb{x})$.
Here, the action of the reflection operator on the wavefunction is defined by
\begin{equation}
    \hat{P} \psi(x_1,x_2,x_3) = \psi(-x_3,-x_2,-x_1).
\end{equation}
If we have an eigenfunction $\psi(\vb{x})$, we construct an eigenfunction which is also an eigenfunction of the reflection operator by setting
\begin{equation}
    \frac{\hat{I} \pm \hat{P}}{2} \psi(x_1,x_2,x_3).
\end{equation}
Away from particle contacts, this symmetrized eigenfunction is also of a finite-superposition Bethe type.
Therefore, we can assume without loss of generality that the finite-superposition Bethe eigenfunction under consideration satisfies either $\hat{P} \psi(\vb{x}) = +\psi(\vb{x})$ or $\hat{P} \psi(\vb{x}) = -\psi(\vb{x})$.
This implies that the coefficients must satisfy
\begin{equation}
    A_{m,\vb{z}} = \pm A_{3-m,(z_3^{-1}, z_2^{-1}, z_1^{-1})}
    \label{eq_reflection_orig}
\end{equation}
for all $m=0,1,2,3$ and all $\vb{z} \in \mathcal{Z}$.

\begin{figure*}[p]
    \centering
    \begin{tikzpicture}[
      font=\small,
      >={Latex[length=2.5mm]},
      main/.style={draw,rounded corners=1.2mm,align=center,inner sep=2.2mm,text width=10.0cm,minimum height=5mm},
      side/.style={draw,rounded corners=1.2mm,align=center,inner sep=2mm,text width=5.0cm,minimum height=5mm},
      core/.style={draw,rounded corners=1.2mm,align=center,inner sep=2mm,text width=4.0cm,minimum height=16mm},
      decision/.style={draw,diamond,aspect=2.4,align=center,inner sep=1.2mm,text width=3.2cm,minimum height=5mm},
      result/.style={draw,double,rounded corners=1.2mm,align=center,inner sep=2.2mm,text width=5.0cm,minimum height=5mm},
      note/.style={draw,dashed,rounded corners=1mm,align=left,inner sep=2mm,text width=4.6cm,font=\footnotesize},
      flow/.style={->,thick},
      aux/.style={->,thick,dashed}
    ]
    
    \node[draw,rounded corners=1.2mm,align=center,inner sep=2.2mm,text width=12.0cm,minimum height=5mm] (assume) {Assume a nonzero square-summable finite-superposition Bethe-type eigenstate $\psi$.};
    \node[draw,rounded corners=1.2mm,align=center,inner sep=2.2mm,text width=10.0cm,minimum height=5mm,
    below=5mm of assume] (parity) {Parity reduction: $\hat P\psi=\varepsilon\psi$, \; $\varepsilon\in\{+1,-1\}$. \\
    $\Longrightarrow$ It suffices to consider coefficients in $\mathcal{C}_0$ and $\mathcal{C}_1$ chambers.};
    
    \node[draw,rounded corners=1.2mm,align=center,inner sep=2.2mm,text width=6cm,minimum height=5mm,
    side,below right=5mm and -10mm of parity] (cones) {Finite exponential-sum \\ independence and chamberwise strict decay conditions.};
    \node[side,below=5mm of parity] (matching) {Weak exact matching system:\\ $r=2$ contact equations~\eqref{eq_r2-contact-equation} and impurity equations~\eqref{eq:impurity-orbit}.};
    
    \node[main,below=25mm of parity] (orbit) {\Cref{prop:orbit-separation} \\ Orbit reduction: decompose the coefficient set \\ into $W_3=\mathfrak S_3\ltimes(\mathbb Z_2)^3$-orbits and work on one orbit at a time.};
    
    \node[draw,rounded corners=1.2mm,align=center,inner sep=2mm,text width=5.0cm,minimum height=16mm,below=31mm of cones] (table) {\Cref{prop:universal-decay-zeros}, \Cref{tab:universally-forbidden-permutations} \\ Universally vanishing coefficients.};
    \node[core,below=10mm of orbit] (transfer) {\Cref{lem:impurity-vanishing-propagation} \\ Impurity transfer: \\ vanishing in $\mathcal C_{0}$ if and only if vanishing in $\mathcal C_1$.};
    
    \node[decision,below=5mm of transfer] (pzero) {Some $p_i=0$?};
    \node[draw,double,rounded corners=1.2mm,align=center,inner sep=2.2mm,text width=4.0cm,minimum height=14mm,
    left=10mm of pzero] (orbzero1) {\Cref{lem:vanishing-p,lem:impurity-vanishing-propagation} \\ The whole orbit vanishes.};
    \node[draw,rounded corners=1.2mm,align=center,inner sep=2.2mm,text width=4.0cm,minimum height=5mm,
    right=10mm of pzero] (seeds) {\Cref{prop:diagonal-products} \\ $K \coloneqq A^{\varnothing}_{0,\sigma} A^{\varnothing}_{0,\sigma(13)}$ is independent of $\sigma$.};
    
    \node[decision,below left=5mm and 0mm of seeds] (kbranch) {$K\neq0$?};
    \node[side,below left=5mm and 0mm of kbranch] (knz) {$A^{\varnothing}_{0,\sigma} \neq 0$ for all $\sigma \in \mathfrak{S}_3$. \\
    The impurity/contact comparison gives $p_i=p_j$ for some $i\neq j$.};
    \node[side,below right=5mm and 0mm of kbranch] (kz) {$A^{\varnothing}_{0,\sigma} = 0$ for all $\sigma \in \mathfrak{S}_3$.};

    \node[side,below=5mm of kz] (C1z) {All $\mathcal{C}_1$ coefficients vanish.};
    
    \node[result,below=5mm of knz] (contrad) {$p_i=p_j$ implies either $z_i=z_j$ or $z_iz_j=-1$, and both alternatives imply $A^{\varnothing}_{0,(ij)} = 0$, contradiction.};
    \node[result,below=5mm of C1z] (orbzero2) {The whole orbit vanishes \\ by \Cref{lem:impurity-vanishing-propagation}.};
    
    \node[main,below=44mm of kbranch] (openzero) {Since the fixed orbit was arbitrary, all Bethe coefficients vanish.\\
        Hence $F_m\equiv0$ for $m=0,1,2,3$. };
    \node[main,below=5mm of openzero] (support) {\Cref{prop:F0-implies-psi0} \\ If an $\ell^2$ eigenfunction satisfies $\psi(\vb{x}) = 0$ for all $\vb{x} \in \bigcup_{m=0}^{3} \mathcal{C}_m$, \\ then it vanishes everywhere.};
    \node[draw,double,rounded corners=1.2mm,align=center,inner sep=2.2mm,text width=12.0cm,minimum height=5mm,below=5mm of support] (final) {Therefore $\psi\equiv0$, contradicting the initial assumption. This proves \xref{\Cref{thm:main}}.};
    
    \draw[flow] (assume) -- (parity);
    \draw[flow] (assume.east) -| (cones.north);
    \draw[flow] (parity) -- (matching);
    \draw[flow] (cones) -- (matching);
    \draw[flow] (cones) -- (table);
    
    \draw[flow] (matching) -- (orbit);
    \draw[flow] (orbit) -- (table);
    \draw[flow] (orbit) -- (transfer);
    \draw[flow] (table) -- (pzero);
    \draw[flow] (transfer) -- (pzero);
    \draw[flow] (pzero) -- node[above,sloped,font=\footnotesize]{yes} (orbzero1);
    \draw[flow] (pzero) -- node[above,sloped,font=\footnotesize]{no} (seeds);
    \draw[flow] (seeds) -- (kbranch);
    \draw[flow] (kbranch) -- node[above left,font=\footnotesize]{yes} (knz);
    \draw[flow] (kbranch) -- node[above right,font=\footnotesize]{no} (kz);
    \draw[flow] (knz) -- (contrad);
    \draw[flow] (kz) -- (C1z);
    \draw[flow] (C1z) -- (orbzero2);
    \draw[flow] (orbzero1.south) |- (openzero.west);
    \draw[flow] (orbzero2) -- (openzero);
    \draw[flow] (openzero) -- (support);
    \draw[flow] (support) -- (final);
    
    \begin{scope}[on background layer]
    \node[
      draw,
      rounded corners=2mm,
      fill=black!3,
      fit=(table)(transfer)(pzero)(orbzero1)(seeds)(kbranch)(knz)(kz)(contrad)(orbzero2),
      inner sep=4mm,
      label={[font=\bfseries,anchor=north east, xshift=1cm, yshift=-0.5cm]above left:Orbitwise elimination core}
    ] {};
    \end{scope}
    
    \end{tikzpicture}%
    \caption{
        Logical structure of the proof of \xref{\Cref{thm:main}}.
        Following parity reduction and decomposition into $W_3$-orbits, orbitwise elimination is performed using the universal decay zeros, $r=2$ contact equations, impurity transfer, and a case split defined by the constant diagonal product $K$.
        \Cref{prop:F0-implies-psi0} excludes an eigenfunction supported only on configurations with particles at contact and on impurity.
    }
    \label{fig:proof_logic}
\end{figure*}

\section{Finite exponential sums on the four chambers}
\begin{lemma}[One-dimensional finite exponential sums]
    \label{lem:one-dimensional}
    Let $\rho_1,\ldots,\rho_M\in\mathbb C^\times$ be distinct.  If
    \begin{equation}
        f(n)=\sum_{\alpha=1}^M c_\alpha\rho_\alpha^n
    \end{equation}
    belongs to $\ell^2(\mathbb N)$, then $c_\alpha=0$ for every $\alpha$ with $|\rho_\alpha|\geq1$.
\end{lemma}
\begin{proof}
    Let $R=\max\{|\rho_\alpha|:c_\alpha\neq0\}$.
    If $R>1$, divide by $R^n$ and use the linear independence of the finitely many phases with modulus one to obtain a nonzero bounded subsequence, contradicting $f(n)\to0$.
    If $R=1$, the Ces\`aro mean of $|f(n)|^2$ tends to
    \begin{equation}
        \sum_{|\rho_\alpha|=1}|c_\alpha|^2
    \end{equation}
    after grouping equal phases; hence an $\ell^2$ sequence has no nonzero unit-modulus component.
    Equivalently, the same conclusion follows by applying an invertible Vandermonde matrix to consecutive values of $f$.
\end{proof}

\begin{lemma}[Conewise independence and decay]
\label{lem:cones}
    For a monomial $z_1^{x_1}z_2^{x_2}z_3^{x_3}$, square summability in each chamber is equivalent to 
    the condition that the coefficients $A_{m,\vb{z}}$ are zero unless $\vb{z}$ satisfies the following strict inequalities:
    \begin{center}
        \begin{tabular}{c@{\quad}l}
        \toprule
        Chamber & \\
        \midrule
        $\mathcal C_0$ & $|z_1 z_2 z_3| < 1,\ |z_2 z_3| < 1,\ |z_3| < 1$\\
        $\mathcal C_1$ & $|z_1^{-1}| < 1,\ |z_2 z_3| < 1,\ |z_3| < 1$\\
        $\mathcal C_2$ & $|(z_1z_2)^{-1}| < 1,\ |z_1^{-1}| < 1,\ |z_3| < 1$\\
        $\mathcal C_3$ & $|(z_1z_2z_3)^{-1}| < 1,\ |(z_1z_2)^{-1}| < 1,\ |z_1^{-1}| < 1$\\
        \bottomrule
        \end{tabular}
    \end{center}
    Moreover, in each row the map from $(z_1,z_2,z_3)$ to the three displayed bases is injective.
    Consequently, a forbidden monomial cannot be canceled by a finite sum of distinct ordered exponentials.
\end{lemma}

\begin{proof}
    For $\mathcal C_0$, let $\vb{x} = (n,n+r,n+r+s)$ with $n,r,s\geq1$ so that the corresponding product can be expressed as
    \begin{equation}
        z_1^{x_1}z_2^{x_2}z_3^{x_3}
        =
        (z_1z_2z_3)^n(z_2z_3)^r z_3^s.
    \end{equation}
    The other chambers are obtained from the coordinates
    \begin{align}
        \mathcal C_1:&\quad (n,r,s) \coloneqq (-x_1,\ x_2,\ x_3-x_2),\\
        \mathcal C_2:&\quad (n,r,s) \coloneqq (-x_2,\ x_2-x_1,\ x_3),\\
        \mathcal C_3:&\quad (n,r,s) \coloneqq (-x_3,\ x_3-x_2,\ x_2-x_1).
    \end{align}
    Applying \Cref{lem:one-dimensional} successively to these three coordinates proves both necessity and sufficiency.  The injectivity follows immediately by solving for $x_1, x_2$ and $x_3$ in terms of $n, r$, and $s$.
\end{proof}

In every chamber $\mathcal{C}_m$, the eigenvalue equation at the points $\vb{x} \in \mathcal{C}$ with $\vb{x} \pm \vb{e}_j \in \mathcal{C}_m$ for all $j=1,2,3$ gives
\begin{equation}
    E = -t\sum_{j=1}^{3}(z_j+z_j^{-1})
    \label{eq:energy}
\end{equation}
for all $\vb{z} \in \mathcal{Z}$ with $A_{m,\vb{z}} \neq 0$.
Below, we use
\begin{equation}
    p(z) \coloneqq t(z-z^{-1}).
    \label{eq:p-definition}
\end{equation}

\section{A weak exact matching system}
The proof below utilizes the eigenvalue equations only at the points required to establish $F_m \equiv 0$.
Specifically, we employ the contact equations governing the exchange of the second and third particles, as well as the impurity equations connecting the adjacent chambers.
Proving that this weaker system admits only the zero open-chamber coefficient vector suffices to prove \xref{\Cref{thm:main}}.

\subsection{Adjacent-particle contact}
\begin{lemma}[Contact equation]
    \label{lem:contact}
    Suppose two adjacent ordered factors $z_r, z_{r+1}\, (r=1,2)$ satisfy $z_r z_{r+1} \neq -1$.
    The eigenvalue equation on the corresponding two-particle contact face implies
    \begin{multline}
        \bigl[U+p(z_r)-p(z_{r+1})\bigr] A_{m,\vb{z}}
        \\
        +
        \bigl[U-p(z_r)+p(z_{r+1})\bigr] A_{m,\vb{z}^{(r,r+1)}}
        =0.
        \label{eq:contact}
    \end{multline}
    If the exchanged monomial violates the chamber decay inequalities, its coefficient is interpreted as zero due to \Cref{lem:cones}.
\end{lemma}

\begin{proof}
    Consider the points $\vb{x} = (n,\ell,\ell+1)$ with $\ell > n + 1$ and $|n| \geq 2$.
    The Hamiltonian action at this point reads
    \begin{align}
        &(\hat{H} - E) \psi(n,\ell,\ell+1)
        \nonumber \\
        &= \hat{H} \psi(n,\ell,\ell+1) - E F(n,\ell,\ell+1)
        \nonumber \\
        &= -t [ (\psi - F)(n,\ell+1,\ell+1) + (\psi - F)(n,\ell,\ell) ].
    \end{align}
    This implies that there exists a function $C_1(n)$ such that
    \begin{equation}
        (\psi - F)(n,\ell,\ell) = (-1)^{\ell} C_1(n)
        \label{eq:psi-two-body-contact}
    \end{equation}
    for $\ell > n + 1$ and $|n| \geq 2$.

    Next, consider the points $\vb{x} = (n,\ell,\ell)$ with $\ell > n + 1$ and $|n| \geq 2$.
    The Hamiltonian action at this point reads
    \begin{align}
        &(\hat{H} - E) \psi(n,\ell,\ell)
        \nonumber \\
        &= (\hat{H} - U) \psi(n,\ell,\ell) - (E - U)\psi(n,\ell,\ell)
        \nonumber \\
        &= (\hat{H} - U) \psi(n,\ell,\ell) - (E - U) [F(n,\ell,\ell) + (-1)^{\ell} C_1(n)]
        \nonumber \\
        &= U F(n,\ell,\ell) - (E - U) (-1)^{\ell} C_1(n) 
        \nonumber \\
        &\quad -t \left[
        \begin{aligned}
            &(\psi - F)(n-1,\ell,\ell) + (\psi - F)(n+1,\ell,\ell) \\
            &\quad + F(n,\ell-1,\ell) - F(n,\ell,\ell-1) \\
            &\quad + F(n,\ell,\ell+1) - F(n,\ell+1,\ell)
        \end{aligned}
        \right]
        \nonumber \\
        &= -(-1)^{\ell} \{t[C_1(n-1) + C_1(n+1)] + (E - U) C_1(n)\}
        \nonumber \\
        &\quad +\left[
        \begin{aligned}
            &U F(n,\ell,\ell) \\
            &\quad -t \left(
            \begin{aligned}
                &F(n,\ell-1,\ell) - F(n,\ell,\ell-1) \\
                &+ F(n,\ell,\ell+1) - F(n,\ell+1,\ell)    
            \end{aligned}
            \right)
        \end{aligned}
        \right].
    \end{align}
    Therefore, in order for the eigenstate equation to be satisfied for fixed $n$ such that the number of admissible $\ell$ satisfying $\ell > n+1$ without touching the impurity is sufficiently large, the quantity
    \begin{equation}
        S_1(n) \coloneqq (-1)^{\ell} \left[
        \begin{aligned}
            &U F(n,\ell,\ell) \\
            &\ -t \left(
            \begin{aligned}
                &F(n,\ell-1,\ell) - F(n,\ell,\ell-1) \\
                &+ F(n,\ell,\ell+1) - F(n,\ell+1,\ell)    
            \end{aligned}
            \right)
        \end{aligned}
        \right]
    \end{equation}
    must be independent of $\ell$.
    Expanding $F$ in coefficients, we obtain
    \begin{align}
        S_1(n)
        &= \sum_{\vb{z} \in \mathcal{Z}} z_1^n (-z_2 z_3)^{\ell} A_{m,\vb{z}} [U + p(z_2) - p(z_3)].
    \end{align}
    Here, $m$ takes the value $0,1,3$, depending on where the point $\vb{x} = (n,\ell,\ell+1)$ and its neighbors lie.
    This independence from $\ell$ holds if and only if 
    \begin{equation}
        \sum_{\substack{\vb{z} \in \mathcal{Z} \\ z_2 z_3 = - \zeta}} z_1^n A_{m,\vb{z}} [U + p(z_2) - p(z_3)] = 0
    \end{equation}
    for all $\zeta \neq 1$.
    By \Cref{lem:one-dimensional}, we further obtain
    \begin{equation}
        \sum_{\substack{\vb{z} \in \mathcal{Z} \\ z_1 = \eta, z_2 z_3 = - \zeta}} A_{m,\vb{z}} [U + p(z_2) - p(z_3)] = 0
        \label{eq_TwoBodyContact1}
    \end{equation}
    for all $\zeta \neq 1$ and $\eta \neq 0$.

    Here, we argue that the energy condition~\eqref{eq:energy} and the values $\eta$ and $\zeta \ (\neq 1)$ uniquely determine the triplets $\vb{z}$ up to the permutation $z_2 \leftrightarrow z_3$.
    To see this, notice that we can write
    \begin{equation}
        z_2 + z_3 = \frac{ \zeta }{\zeta-1} \left( - \frac{E}{t} - \eta - \frac{1}{\eta} \right).
    \end{equation}
    Since the sum $z_2 + z_3$ and the product $z_2 z_3$ are determined from $E, \eta$ and $\zeta$, the unordered pair $\{z_2, z_3\}$ is uniquely determined from these data as solutions of a quadratic equation.
    Consequently, Eq.~\eqref{eq_TwoBodyContact1} can equivalently be written as
    \begin{multline}
        A_{m,\vb{z}} [U + p(z_2) - p(z_3)] \\
        + A_{m,\vb{z}^{(23)}} [U + p(z_3) - p(z_2)] = 0,
    \end{multline}
    which proves Eq.~\eqref{eq:contact} for $r = 2$.
    The case $r = 1$ can be proved similarly.
\end{proof}

In the ordered bosonic configuration space, the independent adjacent contacts within each chamber are
\begin{equation}
    \begin{array}{c|c}
        \mathcal C_0&(1,2),(2,3)\\
        \mathcal C_1&(2,3)\\
        \mathcal C_2&(1,2)\\
        \mathcal C_3&(1,2),(2,3).
    \end{array}
    \label{eq:contact-pairs}
\end{equation}

Before proceeding, we note that the $r=2$ contact equation can indeed also be used for the exchange of resonant pairs $z_2 z_3 = -1$ under \Cref{lem:cones}.
\begin{lemma}[Restoration of the $r=2$ contact equation for resonant pairs]
    \label{lem:resonant-r2-contact}
    Suppose that $z_2 z_3 = -1$.
    (This implies $p(z_2) = p(z_3)$.)
    Then, under \Cref{lem:cones}, we have 
    \begin{equation}
        A_{m,\vb{z}} = A_{m,\vb{z}^{(23)}} = 0,\quad (m=0,1).
        \label{eq:vanishing-resonant-coeffs}
    \end{equation}
    In particular, the $r=2$ contact equation, which exchanges these two resonant factors, is recovered as
    \begin{align}
        U \bigl[A_{m,\vb{z}} + A_{m,\vb{z}^{(23)}}] = 0,\quad (m=0,1).
        \label{eq:contact_resonant}
    \end{align}
\end{lemma}
\begin{proof}
    Because of the reflection symmetry~\eqref{eq_reflection},
    it suffices to consider only $\mathcal C_0$ and $\mathcal C_1$.
    Since the second and third components of $\vb{z}$ and $\vb{z}^{(23)}$ have product $z_2 z_3 = -1$, \Cref{lem:cones} gives Eq.~\eqref{eq:vanishing-resonant-coeffs}.
\end{proof}

\subsection{Impurity matching}
For \(i\in\{1,2,3\}\), let \(\bar i\) denote the signed
permutation that inverts the \(i\)th component:
\[
    \vb z^{\bar i}
    \coloneqq
    (z_1,\ldots,z_{i-1},z_i^{-1},z_{i+1},\ldots,z_3).
\]
For distinct \(i_1,\ldots,i_r\), the notation
\(\vb z^{\bar i_1\cdots\bar i_r}\) means that all corresponding
components are inverted.

\begin{lemma}[Impurity equations]
    \label{lem:impurity}
    The eigenvalue equations at the impurity boundaries between $\mathcal{C}_i$ and $\mathcal{C}_{i+1}$ impose the following conditions on the coefficients:
    \begin{equation}
        \begin{aligned}
            A_{0,\vb{z}} + A_{0,\vb{z}^{\bar{1}}} &= A_{1,\vb{z}} + A_{1,\vb{z}^{\bar{1}}}, \\
            A_{1,\vb{z}} + A_{1,\vb{z}^{\bar{2}}} &= A_{2,\vb{z}} + A_{2,\vb{z}^{\bar{2}}}, \\
            A_{2,\vb{z}} + A_{2,\vb{z}^{\bar{3}}} &= A_{3,\vb{z}} + A_{3,\vb{z}^{\bar{3}}},
        \end{aligned}
        \label{eq:impurity-continuity}
    \end{equation}
    where $\vb{z}^{\bar{i}}$ denotes the triplet obtained from $\vb{z}$ by replacing $z_i$ with $z_i^{-1}$.
    They also satisfy
    \begin{equation}
        \begin{aligned}
            V (A_{0,\vb{z}} + A_{0,\vb{z}^{\bar{1}}}) &= p(z_1) (A_{0,\vb{z}} - A_{1,\vb{z}}), \\
            V (A_{1,\vb{z}} + A_{1,\vb{z}^{\bar{2}}}) &= p(z_2) (A_{1,\vb{z}} - A_{2,\vb{z}}), \\
            V (A_{2,\vb{z}} + A_{2,\vb{z}^{\bar{3}}}) &= p(z_3) (A_{2,\vb{z}} - A_{3,\vb{z}}).
        \end{aligned}
        \label{eq:impurity-jump}
    \end{equation}
\end{lemma}
\begin{proof}
    For each impurity boundary, the eigenvalue equations at the two adjacent lattice sites equate the value of the wave function on the impurity boundary with the finite-Bethe sums from the two adjoining chambers, thereby yielding \eqref{eq:impurity-continuity}.
    The eigenvalue equation on the impurity boundary, combined with the corresponding one-particle contribution to the energy relation~\eqref{eq:energy}, then leads to Eq.~\eqref{eq:impurity-jump}.

    We give the derivation for the boundary $x_1=0$.
    Consider the points $\vb{x} = (1,x_2,x_3) \ (3 \leq x_2 < x_3-1)$.
    Under the energy condition~\eqref{eq:energy}, the Hamiltonian action at this point reads
    \begin{equation}
        \hat{H} \psi(1,x_2,x_3) = E F_0(1,x_2,x_3) -t (\psi - F_0)(0,x_2,x_3).
    \end{equation}
    By the finite-Bethe assumption~\xref{\eqref{eq:finite-bethe}}, the eigenvalue equation at this point is equivalent to
    \begin{equation}
        \psi(0,x_2,x_3) = F_0(0,x_2,x_3)
        \label{eq_PsiAtImpurity1}
    \end{equation}

    Next, consider the other side $\vb{x} = (-1,x_2,x_3) \ (2 \leq x_2 < x_3-1)$.
    Under the energy condition~\eqref{eq:energy}, the Hamiltonian action at this point reads
    \begin{equation}
        \hat{H} \psi(-1,x_2,x_3) = E F_1(-1,x_2,x_3) -t (\psi - F_1)(0,x_2,x_3).
    \end{equation}
    By the finite-Bethe assumption~\xref{\eqref{eq:finite-bethe}}, the eigenvalue equation at this point is equivalent to
    \begin{equation}
        \psi(0,x_2,x_3) = F_1(0,x_2,x_3)
        \label{eq_PsiAtImpurity2}
    \end{equation}

    Equations~\eqref{eq_PsiAtImpurity1} and \eqref{eq_PsiAtImpurity2} provide two expressions for the same value $\psi(0,x_2,x_3)$ and must therefore coincide.
    By \Cref{lem:one-dimensional}, the corresponding coefficients must match for each distinct pair $(z_2,z_3)$.
    For a fixed pair $(z_2,z_3)$, the possible first components permitted by the energy condition~\eqref{eq:energy} are $z_1$ and $z_1^{-1}$.
    Hence, we must have
    \begin{equation}
        A_{0,\vb{z}} + A_{0,\vb{z}^{\bar{1}}} = A_{1,\vb{z}} + A_{1,\vb{z}^{\bar{1}}},
        \label{eq_C1C2ImpurityMatch}
    \end{equation}
    where $\vb{z}^{\bar{i}}$ denotes the triplet with $z_i$ replaced with $z_i^{-1}$.

    We now impose the eigenvalue equation on the impurity boundary itself.
    Consider the points $\vb{x} = (0,x_2,x_3) \ (2 \leq x_2 < x_3-1)$.
    The Hamiltonian action at these points reads
    \begin{align}
        &(\hat{H} - E) \psi(0,x_2,x_3)
        \nonumber \\
        &= (\hat{H} - V) \psi(0,x_2,x_3) - (E - V) \psi(0,x_2,x_3)
        \nonumber \\
        &= E F_0(0,x_2,x_3) - (E - V) F_0(0,x_2,x_3)
        \nonumber \\
        &\qquad -t (\psi - F_0)(-1,x_2,x_3)
        \nonumber \\
        &= V F_0(0,x_2,x_3) -t (F_1 - F_0)(-1,x_2,x_3)
        \nonumber \\
        &= \sum_{\vb{z} \in \mathcal{Z}} z_2^{x_2} z_3^{x_3} \left[ V A_{0,\vb{z}} -t (A_{1,\vb{z}} - A_{0,\vb{z}}) z_1^{-1} \right].
    \end{align}
    Therefore, the eigenvalue equation at this point is equivalent to
    \begin{align}
        V (A_{0,\vb{z}} + A_{0,\vb{z}^{\bar{1}}}) 
        &= t \left[ \frac{ A_{1,\vb{z}} - A_{0,\vb{z}} }{z_1} + z_1 (A_{1,\vb{z}^{\bar{1}}} - A_{0,\vb{z}^{\bar{1}}}) \right]
        \nonumber \\
        &= t (z_1 - z_1^{-1}) (A_{0,\vb{z}} - A_{1,\vb{z}}),
    \end{align}
    where we have used Eq.~\eqref{eq_C1C2ImpurityMatch}.

    The same argument at the other two impurity boundaries yields the remaining equations in \eqref{eq:impurity-continuity} and \eqref{eq:impurity-jump}.
\end{proof}

\section{Decomposition by orbits of permutations and inversions}
For $i=1,2,3$, let $\vb{z}^{\bar{i}}$ denote the triplet obtained
from $\vb{z}$ by replacing $z_i$ with $z_i^{-1}$.
For $\sigma\in\mathfrak{S}_3$, define
\begin{equation}
    \vb{z}^{\sigma}
    \coloneqq
    \left(z_{\sigma(1)},z_{\sigma(2)},z_{\sigma(3)}\right).
\end{equation}
Let $W_3 \coloneqq \mathfrak S_3\ltimes(\mathbb Z_2)^3$ be the finite group generated by these permutations
and inversions.
By introducing the missing triplets with zero coefficients, we may assume without loss of generality that $\mathcal{Z}$ is a union of $W_3$-orbits.

\begin{proposition}[Separation of different orbits]
    \label{prop:orbit-separation}
    The necessary conditions~\eqref{eq:contact}, \eqref{eq:impurity-continuity}, and \eqref{eq:impurity-jump} do not couple coefficients belonging to different $W_3$-orbits.
    Consequently, the full system decomposes into a direct sum of independent subsystems, one for each $W_3$-orbit.
\end{proposition}
\begin{proof}
    For $i=1,2,3$, the $i$th equations in \eqref{eq:impurity-continuity} and \eqref{eq:impurity-jump} contain only
    \begin{equation}
        A_{i-1,\vb{z}},\quad A_{i-1,\vb{z}^{\bar{i}}},\quad A_{i,\vb{z}},\quad A_{i,\vb{z}^{\bar{i}}}.
    \end{equation}
    Since $\vb{z}^{\bar{i}}$ is obtained from $\vb{z}$ by one of the
    inversions generating $W_3$, all coefficients appearing in
    either equation are associated with the same $W_3$-orbit.

    Let $\{i,j,k\}=\{1,2,3\}$ and consider the contact equations~\eqref{eq:contact}.
    Each equation in \eqref{eq:contact} is obtained by fixing $z_i$ and
    \begin{equation}
        \zeta = -z_j z_k, \quad \zeta\neq1.
    \end{equation}
    The energy condition~\eqref{eq:energy} then gives
    \begin{equation}
        z_j + z_k = \frac{\zeta}{\zeta-1} \left( -\frac{E}{t}-z_i-z_i^{-1} \right).
    \end{equation}
    Together with $z_j z_k = -\zeta$, this uniquely determines the unordered pair $\{z_j,z_k\}$.
    Therefore, any two triplets contributing to the same contact equation differ at most by exchanging $z_j$ and $z_k$, and hence belong to the same $W_3$-orbit.
    Thus, none of the equations \eqref{eq:contact}, \eqref{eq:impurity-continuity}, and \eqref{eq:impurity-jump} couples coefficients belonging to different $W_3$-orbits.
\end{proof}

\subsection{Orbit notation}
Let us fix a representative $\vb{z}$ of a $W_3$-orbit such that
\begin{equation}
    0 < |z_3| \leq |z_2| \leq |z_1| \leq 1.
    \label{eq:ordered-moduli}
\end{equation}
For $I \subseteq [3] \coloneqq\{1,2,3\}$ and $\sigma \in \mathfrak{S}_3$, we introduce the notation
\begin{equation}
    [ (\vb{z}^I)^{\sigma} ]_i \coloneqq z^I_{\sigma(i)} \coloneqq 
    \begin{cases}
        z_{\sigma(i)} & (\sigma(i) \notin I); \\
        z_{\sigma(i)}^{-1} & (\sigma(i) \in I),
    \end{cases}
\end{equation}
and write the Bethe-type function in this $W_3$-orbit as
\begin{equation}
    F_m(\vb{x}) = \sum_{I \subseteq [3]} \sum_{\sigma \in \mathfrak{S}_3} A^I_{m,\sigma} (z^I_{\sigma(1)})^{x_1} (z^I_{\sigma(2)})^{x_2} (z^I_{\sigma(3)})^{x_3},
\end{equation}
where we impose that the coefficients $A^I_{m,\sigma}$ satisfy
\begin{equation}
    A^I_{m,\sigma} = \frac{ A_{m,(z^I_{\sigma(1)},z^I_{\sigma(2)},z^I_{\sigma(3)})} }{\#\{(J,\tau)\colon (\vb{z}^{J})^{\tau} = (\vb{z}^{I})^{\sigma} \}}.
\end{equation}
For example, if $z^I_1 = z^I_2$, then $A^I_{m,\operatorname{id}} = A^I_{m,(12)}$.
Below, we write $p_i \coloneqq p(z_i)$ and $p^I_i \coloneqq p(z^I_i)$.
With these notations, the reflection-symmetric condition~\eqref{eq_reflection_orig} is written as
\begin{equation}
    A^I_{m,\sigma} = \epsilon A^{I^{\mathrm{c}}}_{3-m,\sigma(13)},
    \label{eq_reflection}
\end{equation}
where $\epsilon = \pm 1$ and $I^{\mathrm{c}}$ denotes the complement of $I \subseteq [3]$.

Under the ordering~\eqref{eq:ordered-moduli}, certain coefficients can be eliminated, as specified by the following proposition.
\begin{proposition}[Universal zeros imposed by square summability]
    \label{prop:universal-decay-zeros}
    The ordering~\eqref{eq:ordered-moduli}, combined with \Cref{lem:cones}, forces the coefficients $A^I_{m,\sigma}$ listed in \Cref{tab:universally-forbidden-permutations} to vanish, regardless of which equalities hold in Eq.~\eqref{eq:ordered-moduli}.
\end{proposition}
\begin{proof}
    To have nonzero $A^I_{m,\sigma} \ (m=0,1)$, we must have $|z^I_{\sigma(3)}| < 1$ from \Cref{lem:cones}.
    However, this condition is violated when $\sigma(3)\in I$, because in this case, we have $|z^I_{\sigma(3)}| = |z_{\sigma(3)}|^{-1} \geq1$.
    Hence, \Cref{lem:cones} forces
    \begin{equation}
        A^I_{m,\sigma} = 0 
        \quad (m=0,1,\ \forall\sigma\in\mathfrak S_3)
        \label{eq:C0-third-inverted}
    \end{equation}
    for $I=\{\sigma(3)\},\ \{\sigma(1),\sigma(3)\},\ \{\sigma(2),\sigma(3)\},\ [3]$.

    The second necessary condition to have nonzero $A^I_{m,\sigma} \ (m=0,1)$ is $|z^I_{\sigma(2)} z^I_{\sigma(3)}| < 1$.
    Under $\sigma(3) \notin I$, this condition is violated when $\sigma(2)\in I$ and $\sigma(2) > \sigma(3)$.
    The permutations corresponding to this case are $(123),\ (13)$, and $(23)$, and we therefore have
    \begin{equation}
        A^I_{m,\sigma} = 0\quad (m=0,1,\ \sigma= (123),\ (13),\ (23),\ I\ni \sigma(2)).
        \label{eq:sigma3-less-sigma2}
    \end{equation}

    The third necessary condition to have nonzero $A^I_{m,\sigma}$ is $|z^I_{\sigma(1)} z^I_{\sigma(2)} z^I_{\sigma(3)}| < 1$ for $m=0$ and $|z^I_{\sigma(1)}| > 1$ for $m=1$.
    The condition for $m=1$ is violated when $\sigma(1) \notin I$.
    Under $\sigma(3) \notin I$, the condition for $m=0$ is violated, in addition to Eq.~\eqref{eq:sigma3-less-sigma2} above, when $I = \{\sigma(1), \sigma(2)\}$ and $\sigma = (132)$.
    Indeed, we have 
    \begin{equation}
        \frac{ |z_2| }{ |z_3 z_1| } \geq \frac{ |z_2| }{ |z_2 z_1| } = \frac{1}{|z_1|} \geq 1.
    \end{equation}
    This completes the proof.
\end{proof}

\begin{table}[t]
    \caption{
    For each $\sigma$-dependent inversion set $I$ shown in the first column, the table lists the permutations $\sigma\in\mathfrak S_3$ for which $A^I_{m,\sigma}=0$ is imposed by \Cref{lem:cones}.
    Due to the reflection symmetry~\eqref{eq_reflection} it is sufficient to consider $\mathcal C_0$ and $\mathcal C_1$.
    }
    \label{tab:universally-forbidden-permutations}
    \centering
    \begingroup
    \setlength{\tabcolsep}{7pt}
    \renewcommand{\arraystretch}{1.35}
    
    \begin{tabular}{|c||c|c|}
        \hline
        $I$
        & $\mathcal C_0$ $(m=0)$
        & $\mathcal C_1$ $(m=1)$
        \\
        \hline\hline
    
        $\varnothing$
        & $\varnothing$
        & $\mathfrak S_3$
        \\
        \hline\hline
    
        $\{\sigma(1)\}$
        & $\varnothing$
        & $\varnothing$
        \\
        \hline
    
        $\{\sigma(2)\}$
        & $(123),(13),(23)$
        & $\mathfrak S_3$
        \\
        \hline
    
        $\{\sigma(3)\}$
        & $\mathfrak S_3$
        & $\mathfrak S_3$
        \\
        \hline\hline
    
        $\{\sigma(1),\sigma(2)\}$
        & $(123),(13),(132),(23)$
        & $(123),(13),(23)$
        \\
        \hline
    
        $\{\sigma(1),\sigma(3)\}$
        & $\mathfrak S_3$
        & $\mathfrak S_3$
        \\
        \hline
    
        $\{\sigma(2),\sigma(3)\}$
        & $\mathfrak S_3$
        & $\mathfrak S_3$
        \\
        \hline\hline
    
        $\{1,2,3\}$
        & $\mathfrak S_3$
        & $\mathfrak S_3$
        \\
        \hline
    \end{tabular}
    
    \endgroup
\end{table}

In the following, we employ the $r=2$ contact equations and the impurity equations as necessary (but not sufficient) conditions for $\psi$ to be a finite-superposition Bethe-type eigenstate of the Hamiltonian.
Here, we note that the $r=2$ contact equations are available for both $m=0,1$, irrespective of whether $z^I_{\sigma(2)} z^I_{\sigma(3)} = -1$ holds by virtue of \Cref{lem:resonant-r2-contact}.

Within the orbit notation, the $r=2$ contact equations can be written as
\begin{equation}
    \boxed{
    \begin{multlined}
        [U + p^I_{\sigma(2)} - p^I_{\sigma(3)}] A^{I}_{m,\sigma} \\
        \qquad\qquad + [U - p^I_{\sigma(2)} + p^I_{\sigma(3)}] A^{I}_{m,\sigma(23)} = 0
    \end{multlined}
    }
    \label{eq_r2-contact-equation}
\end{equation}
for $m=0,1$, $I \subseteq [3]$, and $\sigma \in \mathfrak{S}_3$.
For convenience, we put the Cayley tree of the $\mathfrak{S}_3$ group as Fig.~\eqref{fig:S3-Cayley-tree}, from which we can read off which coefficients are connected by the $r=2$ contact equation.
The impurity equations can be written as
\begin{align}
    \boxed{
    \begin{aligned}
        A^I_{r-1,\sigma} + A^{I\triangle\{\sigma(r)\}}_{r-1,\sigma}
        &= A^I_{r,\sigma} + A^{I\triangle\{\sigma(r)\}}_{r,\sigma}, \\
        V [A^I_{r-1,\sigma} + A^{I\triangle\{\sigma(r)\}}_{r-1,\sigma}] &= p^I_{\sigma(r)} [A^I_{r-1,\sigma} - A^I_{r,\sigma}]
    \end{aligned}}
    \label{eq:impurity-orbit}
\end{align}
for $r= 1,2,3$, $I \subseteq [3]$, and $\sigma \in \mathfrak{S}_3$.
Here, $S \triangle T$ denotes the symmetric difference of the sets $S$ and $T$ defined by
\begin{equation}
    S \triangle T \coloneqq (S \cup T) \setminus (S \cap T).
\end{equation}

\begin{figure}
    \centering
\begin{tikzpicture}[
    scale=1,
    every node/.style={
        circle,
        draw,
        minimum size=9mm,
        inner sep=1pt,
        font=\small
    },
    s-edge/.style={
        draw=blue,
        very thick
    },
    t-edge/.style={
        draw=red,
        very thick
    }
]
    \node (e)   at (0,  2) {$e$};
    \node (s)   at (1.8,  1) {$(12)$};
    \node (st)  at (1.8, -1) {$(123)$};
    \node (sts) at (0, -2) {$(13)$};
    \node (ts)  at (-1.8,-1) {$(132)$};
    \node (t)   at (-1.8, 1) {$(23)$};
    
    \draw[s-edge] (e)   -- node[draw=none, fill=none, midway, right] {$(12)$} (s);
    \draw[s-edge] (st)  -- node[draw=none, fill=none, midway, right] {$(12)$} (sts);
    \draw[s-edge] (ts)  -- node[draw=none, fill=none, midway, left]  {$(12)$} (t);
    
    \draw[t-edge] (e)   -- node[draw=none, fill=none, midway, left]  {$(23)$} (t);
    \draw[t-edge] (s)   -- node[draw=none, fill=none, midway, right] {$(23)$} (st);
    \draw[t-edge] (sts) -- node[draw=none, fill=none, midway, left]  {$(23)$} (ts);
\end{tikzpicture}
    \caption{Cayley tree of the permutation group $\mathfrak{S}_3$. 
    We employ only the $r=2$ contact equations~\eqref{eq_r2-contact-equation}, which correspond to the red edges permuting the second and the third components.}
    \label{fig:S3-Cayley-tree}
\end{figure}

Since the system of contact equations and impurity equations are invariant under the reflection~\eqref{eq_reflection}, it is sufficient to consider coefficients in $\mathcal{C}_0$ and $\mathcal{C}_1$, and we focus on them in the following.
In particular, the impurity equations with $r = 2$ give self-consistent equations in $\mathcal{C}_1$:
\begin{equation}
    \begin{aligned}
        A^I_{1,\sigma} + A^{I\triangle\{\sigma(2)\}}_{1,\sigma}
        &= \epsilon \Big[ A^{I^c}_{1,\sigma(13)} + A^{I^c\triangle\{\sigma(2)\}}_{1,\sigma(13)} \Big], \\
        V [A^I_{1,\sigma} + A^{I\triangle\{\sigma(2)\}}_{1,\sigma}] &= p^I_{\sigma(2)} [A^I_{1,\sigma} - \epsilon A^{I^c}_{1,\sigma(13)}].
    \end{aligned}
    \label{eq:impurity-orbit-C1}
\end{equation}

With the impurity equations, we can prove the following two lemmas.
\begin{lemma}
    \label{lem:vanishing-p}
    If $p_i = 0$ for some $i$, then we have
    \begin{equation}
        A^I_{\sigma^{-1}(i) - 1,\sigma} = A^I_{\sigma^{-1}(i), \sigma} = 0
        \label{eq:self-inverse-two-chambers-zero}
    \end{equation}
    for all $I \subseteq [3]$ and $\sigma \in \mathfrak{S}_3$.
\end{lemma}
\begin{proof}
    Note that $p_i = 0$ implies $z_i = z_i^{-1}$ and hence $A^I_{m,\sigma} = A^{I\triangle\{i\}}_{m,\sigma}$ for arbitrary $I \subseteq [3]$.
    Then, the impurity equation~\eqref{eq:impurity-orbit} with $r = \sigma^{-1}(i)$ reads
    \begin{equation}
        A^I_{\sigma^{-1}(i) - 1,\sigma} = A^{I\triangle\{i\}}_{\sigma^{-1}(i),\sigma},\quad
        2V A^I_{\sigma^{-1}(i) - 1,\sigma} = 0.
    \end{equation}
    Since $V \neq 0$, this gives
    \begin{equation}
        A^I_{\sigma^{-1}(i) - 1,\sigma} = A^{I\triangle\{i\}}_{\sigma^{-1}(i),\sigma} = 0.
    \end{equation}
    Since $I$ is arbitrary, this proves Eq.~\eqref{eq:self-inverse-two-chambers-zero}.
\end{proof}
\begin{lemma}[Vanishing chamber propagation]
    \label{lem:impurity-vanishing-propagation}
    Let $m \in \{0,1\}$ and $\sigma \in \mathfrak{S}_3$ be fixed.
    Then the impurity equation~\eqref{eq:impurity-orbit} implies that
    \begin{align}
        \forall I,\ A^I_{0,\sigma} = 0 &\quad\iff\quad \forall I,\ A^I_{1,\sigma} = 0. \label{eq:impurity-vanishing-propagation}
    \end{align}
\end{lemma}
\begin{proof}
    Solving the impurity equation~\eqref{eq:impurity-orbit} with $r=1$ for $A^I_{1,\sigma}$ and $A^{I\triangle\{\sigma(1)\}}_{1,\sigma}$, we obtain
    \begin{multline}
        p^I_{\sigma(1)}
        \begin{pmatrix}
            A^I_{1,\sigma} \\ A^{I\triangle\{\sigma(1)\}}_{1,\sigma}
        \end{pmatrix}
        \\
        = 
        \begin{pmatrix}
            p^I_{\sigma(1)}-V & -V \\
            V & p^I_{\sigma(1)}+V
        \end{pmatrix}
        \begin{pmatrix}
            A^I_{0,\sigma} \\ A^{I\triangle\{\sigma(1)\}}_{0,\sigma}
        \end{pmatrix}
    \end{multline}
    for any $I \subseteq [3]$.
    Here, the determinant of the coefficient matrix is $(p^I_{\sigma(1)})^2$.
    If $p_{\sigma(1)} = 0$, then \Cref{lem:vanishing-p} with $i = \sigma(1)$ gives Eq.~\eqref{eq:impurity-vanishing-propagation}.
    If $p_{\sigma(1)} \neq 0$, then the equation is invertible, and therefore gives Eq.~\eqref{eq:impurity-vanishing-propagation}.
\end{proof}
Combining \Cref{lem:vanishing-p} and \Cref{lem:impurity-vanishing-propagation}, we conclude that 
\begin{equation}
    \boxed{p_i = 0 \Longrightarrow A^{I}_{m,\sigma} = 0 \quad \text{for all } I, m, \sigma.}
\end{equation}
In the following, we therefore assume 
\begin{equation}
    p_1\neq0,\quad p_2\neq0,\quad p_3\neq0.
    \label{eq:core-zero-nonzero-p}
\end{equation}

\subsection{Impurity equations}
\begin{lemma}[Impurity equations with \Cref{tab:universally-forbidden-permutations}]
    Solving the impurity equations with $r=1$, we obtain for all $\sigma \in \mathfrak{S}_3$:
    \begin{equation}
        \begin{aligned}
            A^{\sigma(1)}_{1,\sigma} &= \frac{p_{\sigma(1)}}{V}  A^{\varnothing}_{0,\sigma}, \\
            A^{\sigma(1)}_{0,\sigma} &= \left( \frac{p_{\sigma(1)}}{V} -1\right) A^{\varnothing}_{0,\sigma}.
        \end{aligned}
        \label{eq:impurityEq_r1_I0}
    \end{equation}
    and
    \begin{equation}
        \begin{aligned}
            A^{\sigma(1) \sigma(2)}_{1,\sigma} &= \frac{p_{\sigma(1)}}{V} A^{\sigma(2)}_{0,\sigma}, \\
            A^{\sigma(1) \sigma(2)}_{0,\sigma} &= \left(\frac{p_{\sigma(1)}}{V} - 1\right) A^{\sigma(2)}_{0,\sigma}.
        \end{aligned}
        \label{eq:impurityEq_r1_Isj}
    \end{equation}
\end{lemma}
\begin{proof}[Proof of Eq.~\eqref{eq:impurityEq_r1_I0}]
    The impurity equation~\eqref{eq:impurity-orbit} with $r = 1$ and $I = \varnothing$ reads
    \begin{equation}
        \begin{aligned}
            A^{\varnothing}_{0,\sigma} + A^{\sigma(1)}_{0,\sigma} &= A^{\sigma(1)}_{1,\sigma}, \\
            V [A^{\varnothing}_{0,\sigma} + A^{\sigma(1)}_{0,\sigma}] &= p_{\sigma(1)} A^{\varnothing}_{0,\sigma}.
        \end{aligned}
    \end{equation}
    Here, we have $A^{\varnothing}_{1,\sigma} = 0$ for all $\sigma \in \mathfrak{S}_{3}$ from \Cref{tab:universally-forbidden-permutations}.
    Solving the impurity equation with this input, we obtain Eq.~\eqref{eq:impurityEq_r1_I0}.
\end{proof}
\begin{proof}[Proof of Eq.~\eqref{eq:impurityEq_r1_Isj}]
    The impurity equations~\eqref{eq:impurity-orbit} with $(r,I) = (1,\{\sigma(j)\}) \ (j=2,3)$ read
    \begin{align}
        A^{\sigma(j)}_{0,\sigma} + A^{\sigma(1) \sigma(j)}_{0,\sigma} &= A^{\sigma(j)}_{1,\sigma} + A^{\sigma(1) \sigma(j)}_{1,\sigma}, \\
        V [A^{\sigma(j)}_{0,\sigma} + A^{\sigma(1) \sigma(j)}_{0,\sigma}] &= p_{\sigma(1)} A^{\sigma(j)}_{0,\sigma}.
    \end{align}
    Here, from \Cref{tab:universally-forbidden-permutations}, we have $A^{i}_{1,\sigma} = 0$ for all $\sigma \in \mathfrak{S}_{3}$ such that $i \neq \sigma(1)$.
    In other words, we have $A^{\sigma(j)}_{1,\sigma} = 0$ for all $j=2,3$ and all $\sigma \in \mathfrak{S}_{3}$.
    Solving the impurity equation with this input, we obtain Eq.~\eqref{eq:impurityEq_r1_Isj}.
\end{proof}

\begin{proposition}[Constant diagonal product]
    \label{prop:diagonal-products}
    There exists a constant $K$ such that
    \begin{equation}
        A^{\varnothing}_{0,\sigma} A^{\varnothing}_{0,\sigma(13)} = K\quad (\forall \sigma \in \mathfrak{S}_3). 
        \label{eq:diagonal-products}
    \end{equation}
\end{proposition}
\begin{proof}
    First, suppose that $z_1 z_2 \neq -1$.
    Then, all of the contact equations in $\mathcal{C}_0$ are available because of \Cref{lem:resonant-r2-contact}:
    \begin{equation}
        \begin{multlined}
            [U + p_{\sigma(r)} - p_{\sigma(r+1)}] A^{\varnothing}_{0,\sigma} \\
            + [U - p_{\sigma(r)} + p_{\sigma(r+1)}] A^{\varnothing}_{0,\sigma(r,r+1)} = 0,
        \end{multlined}
        \label{eq_contactEq_m0_I0}
    \end{equation}
    where $r \in \{1,2\}$.
    Replacing $\sigma$ with $\sigma(13)$, using
    \begin{gather}
        (r,r+1) = (13)(\bar{r},\bar{r}+1)(13),\\ \sigma(13)(r) = \sigma(\bar{r}+1),\quad \sigma(13)(r+1) = \sigma(\bar{r}),
    \end{gather}
    with $\bar{r} \coloneqq 3-r$, and rewriting $\bar{r} \in \{1,2\}$ as $r$ again, we obtain
    \begin{equation}
        \begin{multlined}
            [U + p_{\sigma(r+1)} - p_{\sigma(r)}] A^{\varnothing}_{0,\sigma(13)} \\
            + [U - p_{\sigma(r+1)} + p_{\sigma(r)}] A^{\varnothing}_{0,\sigma(r,r+1)(13)} = 0.
        \end{multlined}
        \label{eq_contactEq_m0_I0-s13}
    \end{equation}
    
    Multiplying Eq.~\eqref{eq_contactEq_m0_I0} by $A^{\varnothing}_{0,\sigma(13)}$ and Eq.~\eqref{eq_contactEq_m0_I0-s13} by $A^{\varnothing}_{0,\sigma(r,r+1)}$ and subtracting the resulting two equations, we obtain
    \begin{multline}
        (U + p_{\sigma(r)} - p_{\sigma(r+1)}) \\
        \times \left( A^{\varnothing}_{0,\sigma} A^{\varnothing}_{0,\sigma(13)} - A^{\varnothing}_{0,\sigma(r,r+1)} A^{\varnothing}_{0,\sigma(r,r+1)(13)} \right) = 0.
    \end{multline}
    Similarly, multiplying Eq.~\eqref{eq_contactEq_m0_I0} by $A^{\varnothing}_{0,\sigma(r,r+1)(13)}$ and Eq.~\eqref{eq_contactEq_m0_I0-s13} by $A^{\varnothing}_{0,\sigma}$ and subtracting the resulting two equations, we obtain
    \begin{multline}
        (U - p_{\sigma(r)} + p_{\sigma(r+1)}) \\
        \times \left( A^{\varnothing}_{0,\sigma} A^{\varnothing}_{0,\sigma(13)} - A^{\varnothing}_{0,\sigma (r,r+1)} A^{\varnothing}_{0,\sigma(r,r+1)(13)} \right) = 0.
    \end{multline}
    Adding these two equations and using $U\neq0$, we obtain
    \begin{equation}
        A^{\varnothing}_{0,\sigma} A^{\varnothing}_{0,\sigma(13)} = A^{\varnothing}_{0,\sigma (r,r+1)} A^{\varnothing}_{0,\sigma(r,r+1)(13)}.
    \end{equation}
    Since $(12)$ and $(23)$ generate all elements of $\mathfrak{S}_3$, this implies Eq.~\eqref{eq:diagonal-products}.

    If $z_1 z_2 = -1$, we have $A^{\varnothing}_{0,(13)} = A^{\varnothing}_{0,(123)} = 0$ from \Cref{lem:resonant-r2-contact}.
    Then, we have
    \begin{equation}
        A^{\varnothing}_{0,\operatorname{id}} A^{\varnothing}_{0,(13)} = A^{\varnothing}_{0,(23)} A^{\varnothing}_{0,(123)} = 0.
    \end{equation}
    Using the $r=2$ contact equation with $\sigma = (132)$, we obtain
    \begin{equation}
        (U + p_1 - p_2) A^{\varnothing}_{0,(132)} = U A^{\varnothing}_{0,(132)} = 0,
    \end{equation}
    where we use $p_1 = p_2$ from $z_1 z_2 = -1$.
    Thus, we have $A^{\varnothing}_{0,(132)} = 0$ and $A^{\varnothing}_{0,(12)} A^{\varnothing}_{0,(132)} = 0$.
    This proves \Cref{prop:diagonal-products}.
\end{proof}

\section{Case: $K \neq 0$}
In this case, we have
\begin{equation}
    A^{\varnothing}_{0,\sigma} \neq 0 \quad (\forall \sigma \in \mathfrak{S}_3).
    \label{eq_all_A0_nonvanishing}
\end{equation}
Consider the self-consistent impurity equations in $\mathcal{C}_1$, which is given in Eq.~\eqref{eq:impurity-orbit-C1}.
Since $A^{23}_{1,(123)} = A^{13}_{1,(23)} = 0$ from \Cref{tab:universally-forbidden-permutations}, we set $(I,\sigma) = (\{2\}, (123))$ (equivalently, $(\{1\}, (23))$), obtaining
\begin{equation}
    A^2_{1,(123)} = \epsilon A^{1}_{1,(23)},\quad V A^2_{1,(123)} = p_3 A^2_{1,(123)}.
    \label{eq:impurity-orbit-C1-I2-s123}
\end{equation}
Since $A^{2}_{1,(123)} = (p_2/V) A^{\varnothing}_{0,(123)} \neq 0$ from Eq.~\eqref{eq:impurityEq_r1_I0} , the second equation in Eq.~\eqref{eq:impurity-orbit-C1-I2-s123} implies
\begin{equation}
    p_3 = V.
\end{equation}

Here, we use the contact equations with $(m,r) = (0,2)$, which are valid from \Cref{lem:resonant-r2-contact} irrespective of whether $z_2^I z_3^I = -1$ or not.
For $I = \{1\}$, we have $A^1_{0,(123)} = 0$ from \Cref{tab:universally-forbidden-permutations}.
Thus, setting $(I,\sigma) = (\{1\}, (123))$ in Eq.~\eqref{eq:contact}, we obtain 
\begin{align}
    (U - V - p_1) A^1_{0,(12)} &= 0.
    \label{eq_contact_m0_I1_s123}
\end{align}
For $I = \{2\}$, we have $A^2_{0,(23)} = 0$ from \Cref{tab:universally-forbidden-permutations}.
Thus, setting $(I,\sigma) = (\{2\}, (23))$ in Eq.~\eqref{eq:contact}, we obtain 
\begin{equation}
    (U - p_2 - V) A^2_{0,\operatorname{id}} = 0.
    \label{eq_contact_m0_I2_s23}
\end{equation}
On the other hand, Eq.~\eqref{eq:impurityEq_r1_Isj} with $j=2$ gives
\begin{align}
    A^{12}_{1,(12)} = \frac{p_2}{V} A^1_{0,(12)},\quad A^{12}_{1,\operatorname{id}} = \frac{p_1}{V} A^2_{0,\operatorname{id}}.
    \label{eq:impurityEq_r1_Is2}
\end{align}
The $r=2$ impurity equations which these coefficients enter are
\begin{gather}
    \begin{aligned}
    A^{12}_{1,(12)} + A^{2}_{1,(12)} &= \epsilon [A^{3}_{1,(132)} + A^{13}_{1,(132)} ], \\
    V [A^{12}_{1,(12)} + A^{2}_{1,(12)}] &= -p_1 [A^{12}_{1,(12)} - \epsilon A^{3}_{1,(132)}],
    \end{aligned}
    \label{eq:impurity-m1_I12_s12}
    \\
    \begin{aligned}
    A^{12}_{1,\operatorname{id}} + A^{1}_{1,\operatorname{id}} &= \epsilon [A^{3}_{1,(13)} + A^{23}_{1,(13)}], \\
    V [A^{12}_{1,\operatorname{id}} + A^{1}_{1,\operatorname{id}}] &= -p_2 [A^{12}_{1,\operatorname{id}} - \epsilon A^{3}_{1,(13)}].
    \label{eq:impurity-m1_I12_sid}
    \end{aligned}
\end{gather}
Substituting $A^{23}_{1,(13)} = 0$ from \Cref{tab:universally-forbidden-permutations} into the impurity equation~\eqref{eq:impurity-m1_I12_sid}, we obtain
\begin{equation}
    A^{12}_{1,\operatorname{id}} = \epsilon \frac{p_2-V}{p_2} A^{3}_{1,(13)} = \epsilon \frac{p_2-V}{p_2} A^{\varnothing}_{0,(13)}.
\end{equation}
Multiplying by $(U-p_2-V)$ and using Eqs.~\eqref{eq:impurityEq_r1_Is2} and \eqref{eq_contact_m0_I2_s23}, we obtain
\begin{equation}
    (U-V-p_2)(p_2-V) = 0,
    \label{eq_Kneq0-A}
\end{equation}
where we use $A^{\varnothing}_{0,(13)} \neq 0$.

In addition, the $r=2$ contact equation involving $A^{13}_{1,(132)}$ gives
\begin{equation}
    (U-p_1-p_2) A^{13}_{1,(132)} = 0,
\end{equation}
where we use $A^{13}_{1,(13)} = 0$ from \Cref{tab:universally-forbidden-permutations}.
Thus, multiplying the impurity equation~\eqref{eq:impurity-m1_I12_s12} by $(U-p_1-p_2)$ and solving the resulting equation, we obtain
\begin{multline}
    (U-p_1-p_2) A^{12}_{1,(12)} \\ = (U-p_1-p_2) (p_1-V) \frac{p_2}{V^2} A^{\varnothing}_{0,(12)}.
\end{multline}
Multiplying by $(U-V-p_1)$ and using Eqs.~\eqref{eq:impurityEq_r1_Is2} and \eqref{eq_contact_m0_I1_s123}, we obtain
\begin{equation}
    (U-p_1-p_2)(U-V-p_1)(p_1-V) = 0,
    \label{eq_Kneq0-B}
\end{equation}
where we use $A^{\varnothing}_{0,(12)} \neq 0$.

Equations~\eqref{eq_Kneq0-A} and \eqref{eq_Kneq0-B} imply that $p_1$ and $p_2$ must be either $V$ or $U - V$.
Therefore, the three variables $p_1$, $p_2$, and $p_3$ take only two possible values $V$ or $U-V$.
By the pigeonhole principle, there must exist a coinciding pair, which we denote $p_i = p_j$.
Since
\begin{equation}
    p_i - p_j = t (z_i - z_j) \left( 1 + \frac{1}{z_i z_j} \right),
\end{equation}
the equality $p_i = p_j$ implies either
\begin{equation}
    z_i = z_j \quad\text{or}\quad z_i z_j = -1.
\end{equation}

In the former case, we have $A^{\varnothing}_{0,\tau(12)} = A^{\varnothing}_{0,\tau(123)}$ by definition of $A^I_{0,\sigma}$, where 
\begin{equation}
        \tau\coloneqq
        \begin{pmatrix}
            1&2&3\\
            i&k&j
        \end{pmatrix}.
    \end{equation}
Then, the $r=2$ contact equation connecting these identical coefficients gives
\begin{equation}
    0 = U (A^{\varnothing}_{0,\tau(12)} + A^{\varnothing}_{0,\tau(123)}) = 2 U A^{\varnothing}_{0,\tau(12)}.
\end{equation}
In the latter case, the product of the second and third components of $A^{\varnothing}_{0,\tau(12)}$ and $A^{\varnothing}_{0,\tau(123)}$ are $-1$.
Thus, we must have $A^{\varnothing}_{0,\tau(12)} = A^{\varnothing}_{0,\tau(123)} = 0$ by \Cref{lem:cones}.
Therefore, in both cases we have $A^{\varnothing}_{0,\tau(12)} = 0$, which contradicts Eq.~\eqref{eq_all_A0_nonvanishing} and the nontrivial solution with $K \neq 0$ does not exist.
\qed

\section{Case: $K = 0$}
In this case, we show that, for every $\sigma\in\mathfrak S_3$, all coefficients in $\mathcal{C}_0$ vanish.
\Cref{lem:impurity-vanishing-propagation} then implies that all the coefficients $A^I_{m,\sigma}$ vanish, excluding nontrivial Bethe solution.

We start with the following lemma.
\begin{lemma}
    \label{lem:All-A00-vanish}
    Suppose that $A^{\varnothing}_{0,\sigma} A^{\varnothing}_{0,\sigma(13)} = 0$ for all $\sigma \in \mathfrak{S}_3$, and $p_i \neq 0$ for all $i = 1,2,3$.
    Then, we have
    \begin{equation}
        A^{\varnothing}_{0,\sigma} = 0 \quad (\sigma\in\mathfrak S_3).
        \label{eq:all-C0-outer-zero}
    \end{equation}
\end{lemma}
\begin{proof}
    We consider the self-consistent impurity equation~\eqref{eq:impurity-orbit-C1} involving the coefficients $A^{\sigma(1)}_{1,\sigma}$, which reads
    \begin{equation}
        \begin{aligned}
            A^{\sigma(1)}_{1,\sigma} + A^{\sigma(1) \sigma(2)}_{1,\sigma} &= \epsilon [A^{\sigma(2)\sigma(3)}_{1,\sigma(13)} + A^{\sigma(3)}_{1,\sigma(13)}], \\
            V [A^{\sigma(1)}_{1,\sigma} + A^{\sigma(1) \sigma(2)}_{1,\sigma}] &= p_{\sigma(2)} [ A^{\sigma(1)}_{1,\sigma} - \epsilon A^{\sigma(2)\sigma(3)}_{1,\sigma(13)} ].
        \end{aligned}
        \label{eq:self-consistent-s1}
    \end{equation}
    Solving these equations for $A^{\sigma(1) \sigma(2)}_{1,\sigma}$, we obtain
    \begin{equation}
        (p_{\sigma(2)} + V) A^{\sigma(1) \sigma(2)}_{1,\sigma} = p_{\sigma(2)} \epsilon A^{\sigma(3)}_{1,\sigma(13)} - V A^{\sigma(1)}_{1,\sigma}.
    \end{equation}
    Substituting Eq.~\eqref{eq:impurityEq_r1_I0}, we obtain
    \begin{equation}
        (p_{\sigma(2)} + V) A^{\sigma(1) \sigma(2)}_{1,\sigma} = \frac{p_{\sigma(2)} p_{\sigma(3)}}{V} \epsilon A^{\varnothing}_{0,\sigma(13)} - p_{\sigma(1)} A^{\varnothing}_{0,\sigma}.
        \label{eq:impurityEq_r1_Is1_s1s2}
    \end{equation}
    Here, we have $A^{\sigma(1) \sigma(2)}_{1,\sigma} = 0$ for $\sigma = (123), (13), (23)$.
    Substituting these values, we obtain
    \begin{equation}
        \frac{p_{\sigma(2)} p_{\sigma(3)}}{V} \epsilon A^{\varnothing}_{0,\sigma(13)} = p_{\sigma(1)} A^{\varnothing}_{0,\sigma} \quad(\sigma = (123),(13),(23)).
    \end{equation}
    Solving this equation with $A^{\varnothing}_{0,\sigma} A^{\varnothing}_{0,\sigma(13)} = 0$ gives
    \begin{equation}
        A^{\varnothing}_{0,(123)} = A^{\varnothing}_{0,(23)} = A^{\varnothing}_{0,(13)} = A^{\varnothing}_{0,\operatorname{id}} = 0.
    \end{equation}

    For the remaining coefficients $A^{\varnothing}_{0,\sigma}$ $(\sigma = (12), (132))$, we use the $r=2$ contact equation involving these coefficients.
    Since $(12)(23) = (123)$ and $(132)(23) = (13)$, these equations read
    \begin{equation}
        (U+p_{\sigma(2)}-p_{\sigma(3)}) A^{\varnothing}_{0,\sigma} = 0.
        \label{eq:contact-m0-r2-s12-s132}
    \end{equation}
    We also have $A^{\sigma(1)\sigma(2)}_{1,\sigma(23)} = 0$ for $\sigma = (12),(132)$.
    Thus, the $r=2$ contact equations involving $A^{\sigma(1)\sigma(2)}_{1,\sigma}$ reads
    \begin{multline}
        0 = (U-p_{\sigma(2)}-p_{\sigma(3)}) A^{\sigma(1)\sigma(2)}_{1,\sigma} \\
        = \frac{p_{\sigma(1)}}{V} (U-p_{\sigma(2)}-p_{\sigma(3)}) A^{\sigma(2)}_{0,\sigma}.
    \end{multline}
    Thus, multiplying Eq.~\eqref{eq:impurityEq_r1_Is1_s1s2} by $(U-p_{\sigma(2)}-p_{\sigma(3)})$, we obtain
    \begin{align}
        (U-p_{\sigma(2)}-p_{\sigma(3)}) \left(\frac{p_{\sigma(2)} p_{\sigma(3)}}{V} \epsilon A^{\varnothing}_{0,\sigma(13)} - p_{\sigma(1)} A^{\varnothing}_{0,\sigma}\right) = 0.
    \end{align}
    Multiplying both sides by $A^{\varnothing}_{0,\sigma}$ and using $A^{\varnothing}_{0,\sigma} A^{\varnothing}_{0,\sigma(13)} = 0$, we obtain
    \begin{equation}
        (U-p_{\sigma(2)}-p_{\sigma(3)}) \left(A^{\varnothing}_{0,\sigma}\right)^2 = 0.
        \label{eq:temp}
    \end{equation}
    Multiplying Eq.~\eqref{eq:contact-m0-r2-s12-s132} by $A^{\varnothing}_{0,\sigma}$, and subtracting Eq.~\eqref{eq:temp}, we obtain
    \begin{equation}
        2 p_{\sigma(2)} \left(A^{\varnothing}_{0,\sigma}\right)^2 = 0\quad (\sigma = (12),(132)).
    \end{equation}
    Since $p_{\sigma(2)} \neq 0$, this implies $A^{\varnothing}_{0,(12)} = A^{\varnothing}_{0,(132)} = 0$, completing the proof.
\end{proof}

Combining \Cref{lem:All-A00-vanish} with Eq.~\eqref{eq:impurityEq_r1_I0}, we obtain
\begin{equation}
    A^{\sigma(1)}_{1,\sigma} = 0\quad (\forall \sigma \in \mathfrak{S}_3).
\end{equation}
The other coefficients in $\mathcal{C}_1$ that are not forced to vanish in \Cref{tab:universally-forbidden-permutations} are
\begin{equation}
    A^{\sigma(1) \sigma(2)}_{1,\sigma}\quad (\sigma = \operatorname{id}, (12), (132)).
\end{equation}
Substituting the result of \Cref{lem:All-A00-vanish} into the self-consistent impurity equation~\eqref{eq:self-consistent-s1}, we obtain
\begin{align}
    A^{\sigma(1)\sigma(2)}_{1,\sigma} &= \epsilon A^{\sigma(2)\sigma(3)}_{1,\sigma(13)},\\
    (p_{\sigma(2)} + V) A^{\sigma(1)\sigma(2)}_{1,\sigma} &= 0.
\end{align}
Since $A^{23}_{1,(13)} = 0$ from \Cref{tab:universally-forbidden-permutations}, substituting $\sigma = \operatorname{id}$ gives
\begin{equation}
    A^{12}_{1,\operatorname{id}} = 0.
\end{equation}
On the other hand, substituting $\sigma = (12)$ gives
\begin{equation}
    A^{12}_{1,(12)} = \epsilon A^{13}_{1,(132)},\quad (p_1+V) A^{12}_{1,(12)} = 0.
    \label{eq:p1V_A1_I12_s12}
\end{equation}
Since $A^{13}_{0,(132)} = 0$ from \Cref{tab:universally-forbidden-permutations}, Eq.~\eqref{eq:impurityEq_r1_Isj} gives
\begin{equation}
    A^{13}_{1,(132)} = A^{1}_{0,(132)},\quad (p_3-V) A^{13}_{1,(132)} = 0.
    \label{eq:p3V_A1_I12_s132}
\end{equation}
Combining Eqs.~\eqref{eq:p1V_A1_I12_s12} and \eqref{eq:p3V_A1_I12_s132}, we obtain
\begin{equation}
    (p_1+p_3)A^{12}_{1,(12)} = 0.
    \label{eq:p1p3_A1_I12_s12}
\end{equation}
Since $A^{12}_{1,(123)} = 0$ from \Cref{tab:universally-forbidden-permutations}, the $r=2$ contact equation involving $A^{12}_{1,(12)}$ gives
\begin{equation}
    (U-p_1-p_3) A^{12}_{1,(12)} = U A^{12}_{1,(12)} = 0,
\end{equation}
where we use Eq.~\eqref{eq:p1p3_A1_I12_s12}.
Since $U \neq 0$, this gives 
\begin{equation}
    A^{12}_{1,(12)} = A^{13}_{1,(132)} = 0.
\end{equation}
At this point, we have shown that all the coefficients in $\mathcal{C}_1$ vanish.
Then, \Cref{lem:impurity-vanishing-propagation} implies that all coefficients in $\mathcal{C}_0$ also vanish, which demonstrates that $F\equiv 0$ whenever $K=0$.
\qed

Having completed the proof of $F\equiv 0$, we conclude the proof of \xref{\Cref{thm:main}} by showing that $F \equiv 0$ implies $\psi \equiv 0$.
In other words, there can exist no nonzero eigenstates supported only at configurations with multiple occupancies or particles on the impurity.
\begin{proposition}[$F_m\equiv0$ implies $\psi\equiv0$]
    \label{prop:F0-implies-psi0}
    Suppose that the eigenstate $\psi$ is square summable:
    \begin{equation}
        \sum_{x_1\leq x_2\leq x_3}
        |\psi(x_1,x_2,x_3)|^2<\infty.
    \end{equation}
    If $F_m\equiv0$ for all $m\in\{0,1,2,3\}$, then
    \begin{equation}
        \psi(x_1,x_2,x_3)=0 \quad (x_1\leq x_2\leq x_3).
    \end{equation}
\end{proposition}
\begin{proof}
    Below, we assume $x_1\leq x_2 \leq x_3$ and prove that 
    \begin{equation}
        \psi(x_1,x_2,x_3) = 0\quad \text{for sufficiently large $x_3$}.
        \label{eq:psi-finite-support}
    \end{equation}
    Due to the reflection symmetry $\psi(x_1,x_2,x_3) = \epsilon \psi(-x_3,-x_2,-x_1)$, Eq.~\eqref{eq:psi-finite-support} implies that the support of $\psi$ is finite: there exists $L < \infty$ such that $\psi(x_1,x_2,x_3)$ can be nonzero only if $-L < x_1 \leq x_2 \leq x_3 < L$.
    
    Given Eq.~\eqref{eq:psi-finite-support}, \Cref{prop:F0-implies-psi0} can be proved as follows. 
    If $\psi\not\equiv0$, choose a point $\vb{x}^{\ast}$ in the support of $\psi$ whose third coordinate is maximal.
    Then, the maximality of $\vb{x}^{\ast}$ implies that $\psi(\vb{x}^{\ast} + \vb{e}_3) = 0$, and that $\psi$ vanishes at the neighbors of $\vb{x}^{\ast} + \vb{e}_3$ except for $\vb{x}^{\ast}$.
    Therefore, the eigenvalue equation at $\vb{x}^{\ast} + \vb{e}_3$ gives $t \psi(\vb{x}^{\ast}) = 0$.
    Since $t \neq 0$, this contradicts $\psi(\vb{x}^{\ast}) \neq 0$, and we must have $\psi \equiv 0$.

    Below, we prove Eq.~\eqref{eq:psi-finite-support}.
    Throughout the remainder of the proof, $\ell$ is assumed to be sufficiently large so that any previously established identity remains valid when $\ell$ is replaced by $\ell\pm1$.
    
    Since $F_m\equiv0$ in every open chamber,
    \begin{equation}
        \psi(x_1,x_2,x_3)=0
        \quad
        (x_1<x_2<x_3,\ x_1x_2x_3\neq0).
        \label{eq:open-chamber-zero}
    \end{equation}
    For $|n|\geq2$ and $\ell\geq\max\{2,n+2\}$, the eigenvalue equation at $(n,\ell,\ell+1)$ gives
    \begin{align}
        0 &= (\hat H-E)\psi(n,\ell,\ell+1)
        \nonumber \\
        &= -t\left[ \psi(n,\ell,\ell) + \psi(n,\ell+1,\ell+1) \right].
        \label{eq:right-contact-alternation}
    \end{align}
    Thus, $|\psi(n,\ell,\ell)|$ is independent of $\ell$ on this half-line.
    Square summability implies $\lim_{\ell \to +\infty} \psi(n,\ell,\ell) = 0$, and therefore
    \begin{equation}
        \psi(n,\ell,\ell)=0
        \quad
        \left(
            |n|\geq2,\ 
            \ell\geq\max\{2,n+2\}
        \right).
        \label{eq:right-contact-away-zero}
    \end{equation}
    
    For all sufficiently large positive $\ell$, the eigenvalue
    equations at the following four zero amplitudes reduce
    successively to
    \begin{align}
        0
        &=
        (\hat H-E)\psi(-2,\ell,\ell)
        =
        -t\psi(-1,\ell,\ell),
        \label{eq:right-contact-minus-one}\\
        0
        &=
        (\hat H-E)\psi(-1,\ell,\ell)
        =
        -t\psi(0,\ell,\ell),
        \label{eq:right-contact-zero}\\
        0
        &=
        (\hat H-E)\psi(2,\ell,\ell)
        =
        -t\psi(1,\ell,\ell),
        \label{eq:right-contact-one}\\
        0
        &=
        (\hat H-E)\psi(\ell-2,\ell,\ell)
        =
        -t\psi(\ell-1,\ell,\ell).
        \label{eq:right-contact-nearest}
    \end{align}
    Every omitted term vanishes by Eq.~\eqref{eq:open-chamber-zero} or \eqref{eq:right-contact-away-zero}.
    Since $t\neq0$,
    \begin{equation}
        \psi(n,\ell,\ell)=0
        \quad (\forall n < \ell)
        \label{eq:all-right-contacts-zero}
    \end{equation}
    for all sufficiently large $\ell$.
    
    Next, the eigenvalue equations at strictly ordered points around the impurity
    give
    \begin{align}
        0
        &=
        (\hat H-E)\psi(-1,n,\ell)
        =
        \begin{multlined}[t]
            -t\psi(0,n,\ell) \\
            \qquad (2\leq n\leq\ell-1),    
        \end{multlined}
        \label{eq:first-impurity-away}\\
        0
        &=
        (\hat H-E)\psi(n,1,\ell)
        =
        -t\psi(n,0,\ell)
        \quad (n\leq-2).
        \label{eq:second-impurity-away}
    \end{align}
    Here again all omitted terms vanish by Eqs.~\eqref{eq:open-chamber-zero} and \eqref{eq:all-right-contacts-zero}.

    Next, we consider the points of the form $(n,n,\ell)$.
    For $n\leq-3$, the eigenvalue equation at
    $(n,n+1,\ell)$ gives
    \begin{equation}
        \psi(n,n,\ell) + \psi(n+1,n+1,\ell) =0\quad (n\leq-3).
        \label{eq:left-contact-negative-alternation}
    \end{equation}
    Square summability as $n\to-\infty$ therefore gives
    \begin{equation}
        \psi(n,n,\ell)=0
        \quad
        (n\leq-2).
        \label{eq:left-contact-negative-zero}
    \end{equation}
    Using Eqs.~\eqref{eq:second-impurity-away} and \eqref{eq:left-contact-negative-zero}, the following equations reduce successively to
    \begin{align}
        0
        &=
        (\hat H-E)\psi(-2,-1,\ell)
        =
        -t\psi(-1,-1,\ell),
        \label{eq:left-contact-minus-one}\\
        0
        &=
        (\hat H-E)\psi(-1,-1,\ell)
        =
        -2t\psi(-1,0,\ell),
        \label{eq:second-impurity-minus-one}\\
        0
        &= (\hat H-E)\psi(-1,1,\ell) 
        = -t \psi(0,1,\ell),
        \label{eq:first-impurity-one}\\
        0
        &=
        (\hat H-E)\psi(0,1,\ell)
        \nonumber\\
        &=
        -t\left[
            \psi(0,0,\ell)
            +
            \psi(1,1,\ell)
          \right].
        \label{eq:zero-one-relation}
    \end{align}
    Consequently, we obtain
    \begin{gather}
        \psi(-1,-1,\ell) = \psi(-1,0,\ell) = \psi(0,1,\ell) = 0, 
        \label{eq:minus-one-one-zero} \\
        \text{and}\quad \psi(1,1,\ell)=-\psi(0,0,\ell).
        \label{eq:zero-one-contact-relation}
    \end{gather}
    Equations~\eqref{eq:open-chamber-zero}, \eqref{eq:all-right-contacts-zero}--\eqref{eq:second-impurity-away}, and \eqref{eq:minus-one-one-zero} now give
    \begin{equation}
        \psi(x_1,x_2,\ell)=0
        \quad
        (\forall x_1<x_2\leq\ell)
    \end{equation}
    for all sufficiently large $\ell$.
    
    It remains to prove that, for sufficiently large $\ell$, $\psi(n,n,\ell) = 0$ for $n \geq 0$ and $\psi(\ell,\ell,\ell) = 0$.
    
    The eigenvalue equations at $(0,0,\ell)$ and
    $(1,1,\ell)$ now read
    \begin{align}
        &\begin{multlined}[b]
            (E-U-2V)\psi(0,0,\ell) \\
            \qquad = -t\left[ \psi(0,0,\ell-1) + \psi(0,0,\ell+1) \right],
        \end{multlined}
        \label{eq:double-impurity-equation}\\
        &\begin{multlined}[b]
            (E-U)\psi(1,1,\ell) \\
            \qquad = -t\left[ \psi(1,1,\ell-1) + \psi(1,1,\ell+1) \right].
        \end{multlined}
        \label{eq:one-one-equation}
    \end{align}
    Using Eq.~\eqref{eq:zero-one-contact-relation} with $\ell-1$, $\ell$, and $\ell+1$ in
    Eq.~\eqref{eq:one-one-equation}, we obtain
    \begin{equation}
        (E-U)\psi(0,0,\ell)
        =
        -t\left[
            \psi(0,0,\ell-1)
            +
            \psi(0,0,\ell+1)
          \right].
    \end{equation}
    Subtracting Eq.~\eqref{eq:double-impurity-equation} then gives $2V\psi(0,0,\ell)=0$.
    Since $V\neq0$, we obtain
    \begin{equation}
        \psi(0,0,\ell) = \psi(1,1,\ell) =0.
        \label{eq:zero-one-contacts-zero}
    \end{equation}
    For $1\leq n\leq\ell-2$, the eigenvalue equation at
    $(n,n+1,\ell)$ gives
    \begin{equation}
        \psi(n,n,\ell)
        +
        \psi(n+1,n+1,\ell)
        =0.
        \label{eq:left-contact-positive-alternation}
    \end{equation}
    Equation~\eqref{eq:zero-one-contacts-zero} and repeated
    use of Eq.~\eqref{eq:left-contact-positive-alternation} yield
    \begin{equation}
        \psi(n,n,\ell)=0
        \quad
        (1\leq n<\ell).
        \label{eq:left-contact-positive-zero}
    \end{equation}
    Together with Eqs.~\eqref{eq:left-contact-negative-zero} and \eqref{eq:zero-one-contacts-zero}, this proves
    \begin{equation}
        \psi(n,n,\ell)=0
        \quad
        (n<\ell).
        \label{eq:all-left-contacts-zero}
    \end{equation}
    
    Finally, the eigenvalue equation at $(\ell-1,\ell,\ell)$ gives
    \begin{align}
        0
        &= (\hat H-E)\psi(\ell-1,\ell,\ell)
        \nonumber\\
        &= -t\left[ \psi(\ell,\ell,\ell) + 2\psi(\ell-1,\ell-1,\ell) \right]
        \nonumber\\
        &= -t\psi(\ell,\ell,\ell),
    \end{align}
    where we use Eqs.~\eqref{eq:all-right-contacts-zero} and \eqref{eq:all-left-contacts-zero}.
    Thus, we have proved
    \begin{equation}
        \psi(x_1,x_2,x_3)=0
    \end{equation}
    whenever $x_3$ is sufficiently large.
\end{proof}

\Cref{prop:F0-implies-psi0} completes the proof of \xref{\Cref{thm:main}}.

\clearpage
\makeatletter
   	\c@secnumdepth=4
    \def\@pointsize{10}
	\expandafter\@process@pointsize\expandafter{\@pointsize@default}%
	\appdef\setup@hook{\normalsize}%
	\setup@hook
\makeatother
\setcounter{section}{1}
\onecolumngrid
\section{Numerical results for the six-particle sector}
In this section, we numerically analyze the particle distribution of the candidate eigenstates for BICs found in Fig.~\xref{\ref{fig:EnergySpectrum_N6}} in the main text.
For this purpose, we first examine the width $\Delta x_{\alpha}$ of the particle distribution $\rho_{\alpha}(x) \coloneqq \expval*{ \hat{n}_{x} }{ E_{\alpha} }/ N$ defined in Eq.~\xref{\eqref{eq:WidthOfParticleDistribution}} in the main text by varying the system size $L$ in Fig.~\ref{fig:WidthOfParticleDistribution_N6}.
We find seven eigenenergies inside the continuum of extended states for which the width of the particle distribution does not vary with increasing the system size.
These eigenenergies are highlighted with red vertical lines in Fig.~\ref{fig:WidthOfParticleDistribution_N6}.

\begin{figure}[htb]
    \centering
    \includegraphics[width=\linewidth]{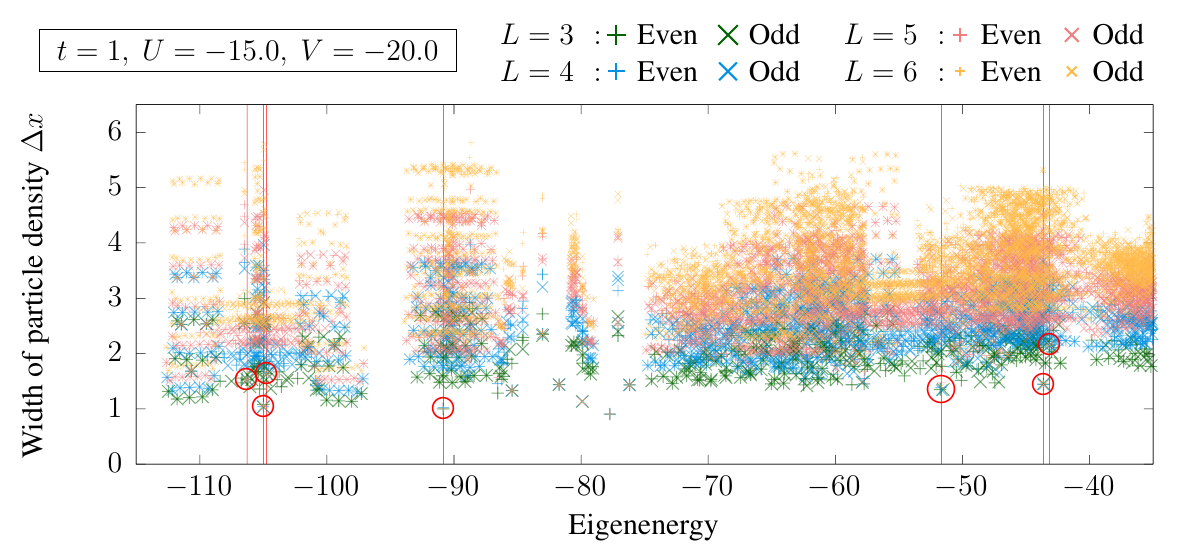}
    \caption{\label{fig:WidthOfParticleDistribution_N6}
        Width $\Delta x_{\alpha}$ of the particle distribution for eigenstates of $\hat{H}$.
        The data are shown for the six-particle sector with eigenenergies satisfying $-115\leq E_{\alpha} \leq -35$, while the parameters of the Hamiltonian are set to be $(t,U,V) = (1,-15,-20)$.
        Seven eigenenergies are highlighted with red vertical lines in the region $-115\leq E_{\alpha} \leq -35$ such that the width $\Delta x_{\alpha}$ of an eigenstate belonging to these eigenenergies (marked with red circles) does not increase when increasing the system size.
        Therefore, these eigenstates are candidates for BICs.
        To determine whether these eigenstates are bound states or resonance states, we examine the system-size dependence of the particle distribution in these eigenstates in Figs.~\ref{fig:ParticleDensity_N6_Eb-150} and \ref{fig:ParticleDensity_N6_Eb-55}.
    }
\end{figure}

To distinguish bound states from resonance states, we examine the particle density of the candidate eigenstates shown in Figs.~\ref{fig:ParticleDensity_N6_Eb-150} and~\ref{fig:ParticleDensity_N6_Eb-55}.
For bound states, the particle density vanishes as both the system size $L$ and the distance from the impurity site $r$ approach infinity.
For candidate eigenstates with eigenenergies
\begin{align}
    E_{b1} &= -106.2761\qc & E_{b3} &= -104.7417\qc \\
    E^{+}_{b4} &= -90.84575\qc & E_{b7} &= -43.13639\qc
\end{align}
we find that the tails of the particle distribution exponentially decrease with increasing the distance $r$ from the impurity site and do not increase with increasing the system size.
Thus, we conclude that these states are BICs.

On the other hand, for eigenstates with eigenenergies
\begin{align}
    E_{b2} &= -105.0034\qc & E_{b6} &= -43.62337\qc
\end{align}
the tails of the particle distribution increase with increasing $L$.
Therefore, we conclude that these states are not BICs but resonant states.

Finally, for the remaining eigenstates with eigenenergies
\begin{align}
    E_{b5}^{+} &= -51.66085\qc & E_{b5}^{-} &= -51.65726\qc 
\end{align}
where the superscripts $+$ and $-$ denote the parity of the state, the tails of the particle distribution show non-monotonic dependences on $L$.
Therefore, we cannot decide whether these two eigenstates are BICs or resonances with the currently available system size.
These conclusions are summarized in Table~\xref{\ref{tab:N6_BoundAndResonanceStates}} in the main text.

\begin{figure}[htb]
    \vspace{-1.5cm}
    \centering
    \includegraphics[width=0.85\linewidth]{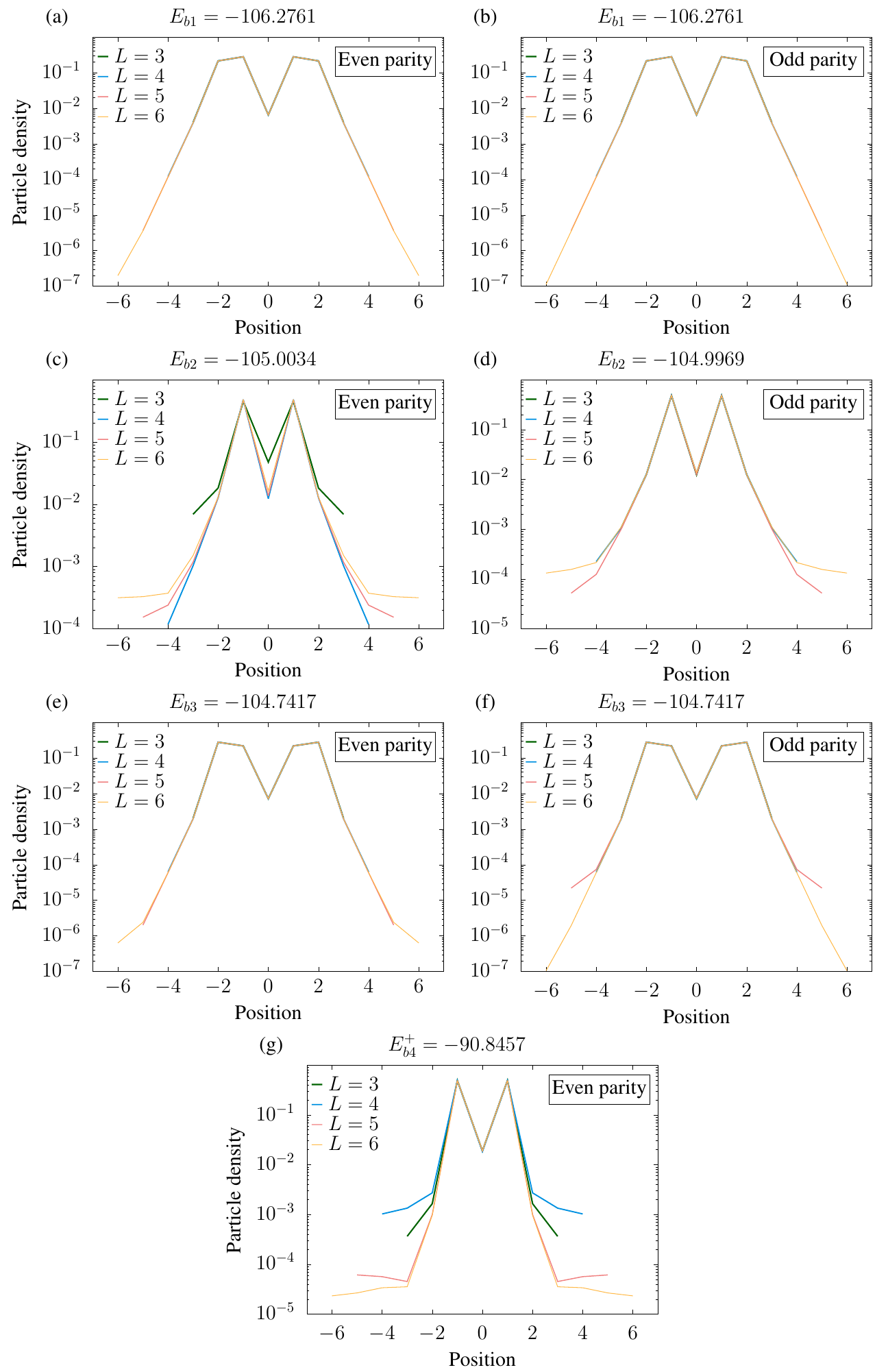}
    \caption{\label{fig:ParticleDensity_N6_Eb-150}
        Particle density distributions of eigenstates for candidate BICs in the six-particle sector, corresponding to eigenenergies (a)(b) $E_{b1}$, (c)(d) $E_{b2}$, (e)(f) $E_{b3}$, and (g) $E_{b4}$.
        The parameters of the Hamiltonian are set to be $(t,U,V) = (1,-15,-20)$.
        (a)(b)(e)(f)(g) The tails of the distributions do not increase with increasing $L$ for $b1,b3$, and $b4$, indicating that these states are bound states.
        (c)(d) The tails for $b2$ increase with increasing $L$, indicating that these states are resonant states.
    }
\end{figure}
\begin{figure}[htb]
    \centering
    \includegraphics[width=0.95\linewidth]{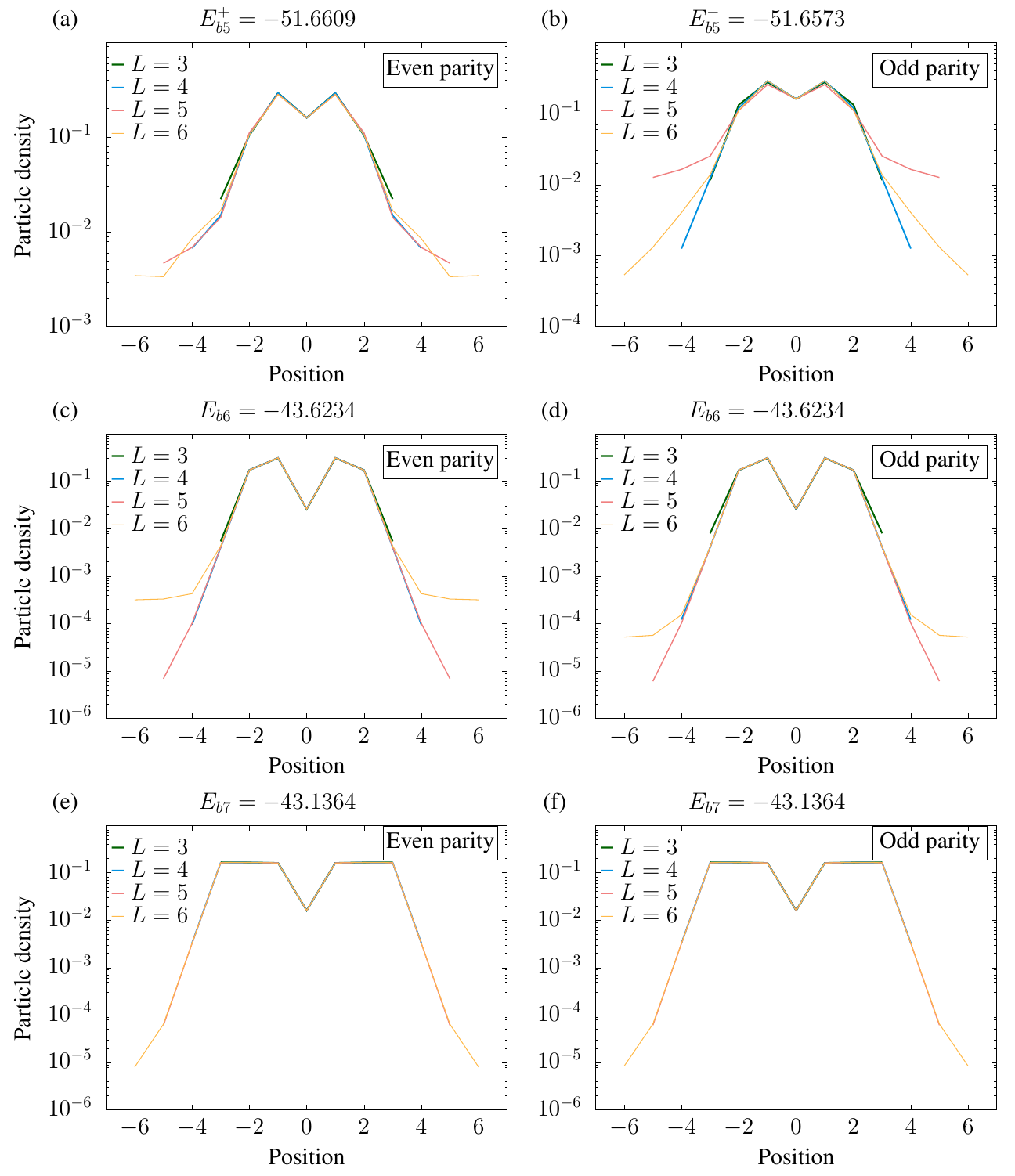}
    \caption{\label{fig:ParticleDensity_N6_Eb-55}
        Particle density distributions of eigenstates for candidate BICs in the six-particle sector, corresponding to eigenenergies (a)(b) $E_{b5}$, (c)(d) $E_{b6}$, and (e)(f) $E_{b7}$.
        The parameters of the Hamiltonian are set to be $(t,U,V) = (1,-15,-20)$.
        (a)(b) The tails for $b5$ show non-monotonic behavior with respect to $L$.
        Although the tails seem not to increase with increasing $L$, we cannot decide whether the states belonging to eigenenergy $E_{b5}$ are bound or resonant states with the currently available system size.
        (c)(d) The tails for $b6$ increase with increasing $L$, indicating that these states are resonant states.
        (e)(f) The tails of the distributions do not increase with increasing $L$ for $b7$, indicating that these states are bound states.
    }
\end{figure}

\clearpage
\section{Derivation of the effective Hamiltonian in the $\vb{N} = (0,2,2)$ sector}
In this section, we derive the effective Hamiltonian in the $\vb{N} = (0,2,2)$ sector defined by
\begin{equation}
    \mathcal{H}_{(0,2,2)} \coloneqq \operatorname{span}\Bqty{ \ket*{x_{1},x_{2}} \coloneqq \frac{1}{2}(\hat{b}_{x_{1}}^{\dagger})^2 (\hat{b}_{x_{2}}^{\dagger})^2 \ket*{0} \mid x_{1} < x_{2},\ x_{1}\neq 0,\ x_{2}\neq 0 },
\end{equation}
whose unperturbed energy is $2U$.
We denote the projector onto $\mathcal{H}_{(0,2,2)}$ by $\hat{\Pi}$ and define $\hat{Q} \coloneqq 1 - \hat{\Pi}$.

For the first-order perturbation, we have $\hat{\Pi}\hat{T}\hat{\Pi} = 0$.
Following Ref.~\citeSM{mila2011strong}, the second-order perturbation $\hat{H}_{(0,2,2)}^{(2)}$ is calculated as
\begin{align}
    &\quad 
    \hat{H}_{(0,2,2)}^{(2)} \ket*{x_{1},x_{2}} \nonumber \\
    &= t^2 \hat{\Pi}\hat{T}\hat{Q}\frac{1}{2U-\hat{Q}\hat{H}_{0}\hat{Q}}\hat{Q}\hat{T}\hat{\Pi} \ket*{x_{1},x_{2}} \nonumber \\
    &= -t^2 \hat{\Pi}\hat{T}\frac{\hat{Q}}{2U-\hat{Q}\hat{H}_{0}\hat{Q}} 
    \bqty{ \hat{b}_{x_{1}-1}^{\dagger}\hat{b}_{x_{1}}^{\dagger} (\hat{b}_{x_{2}}^{\dagger})^2 
    +(\hat{b}_{x_{1}}^{\dagger})^2 \hat{b}_{x_{2}}^{\dagger}\hat{b}_{x_{2}+1}^{\dagger} +\hat{b}_{x_{1}}^{\dagger}\hat{b}_{x_{1}+1}^{\dagger} (\hat{b}_{x_{2}}^{\dagger})^2 +(\hat{b}_{x_{1}}^{\dagger})^2 \hat{b}_{x_{2}-1}^{\dagger}\hat{b}_{x_{2}}^{\dagger} } \ket*{0}.
\end{align}
When $x_{1}+1 < x_{2}$, we obtain
\begin{align}
    \hat{H}_{(0,2,2)}^{(2)} \ket*{x_{1},x_{2}}
    &= -t^2 \hat{\Pi}\hat{T} \bqty{ \frac{ \hat{b}_{x_{1}-1}^{\dagger}\hat{b}_{x_{1}}^{\dagger} (\hat{b}_{x_{2}}^{\dagger})^2 }{U-\delta_{x_{1},1} V}
    +\frac{ (\hat{b}_{x_{1}}^{\dagger})^2  \hat{b}_{x_{2}}^{\dagger}\hat{b}_{x_{2}+1}^{\dagger} }{U-\delta_{x_{2},-1} V} +\frac{ \hat{b}_{x_{1}}^{\dagger}\hat{b}_{x_{1}+1}^{\dagger}  (\hat{b}_{x_{2}}^{\dagger})^2 }{U-\delta_{x_{1},-1} V}
    +\frac{ (\hat{b}_{x_{1}}^{\dagger})^2 \hat{b}_{x_{2}-1}^{\dagger}\hat{b}_{x_{2}}^{\dagger} }{U-\delta_{x_{2},1} V} } \ket*{0} 
    \nonumber \\
    &= t^2 \hat{\Pi}\left[ 
    \begin{multlined}
        \frac{ (\hat{b}_{x_{1}-1}^{\dagger})^2 (\hat{b}_{x_{2}}^{\dagger})^2 +(\hat{b}_{x_{1}}^{\dagger})^2 (\hat{b}_{x_{2}}^{\dagger})^2 }{U-\delta_{x_{1},1} V}
        +\frac{ (\hat{b}_{x_{1}}^{\dagger})^2 (\hat{b}_{x_{2}+1}^{\dagger})^2 +(\hat{b}_{x_{1}}^{\dagger})^2 (\hat{b}_{x_{2}}^{\dagger})^2 }{U-\delta_{x_{2},-1} V}
        \\
        +\frac{ (\hat{b}_{x_{1}+1}^{\dagger})^2 (\hat{b}_{x_{2}}^{\dagger})^2 +(\hat{b}_{x_{1}}^{\dagger})^2 (\hat{b}_{x_{2}}^{\dagger})^2 }{U-\delta_{x_{1},-1} V}
        +\frac{ (\hat{b}_{x_{1}}^{\dagger})^2 (\hat{b}_{x_{2}-1}^{\dagger})^2 +(\hat{b}_{x_{1}}^{\dagger})^2 (\hat{b}_{x_{2}}^{\dagger})^2 }{U-\delta_{x_{2},1} V}
    \end{multlined}
    \right] \ket*{0}
    \nonumber \\
    &= 2t^2 \hat{\Pi}\bqty{ 
        \frac{ \ket*{x_{1}-1,x_{2}} }{U-\delta_{x_{1},1} V}
        +\frac{ \ket*{x_{1},x_{2}+1} }{U-\delta_{x_{2},-1} V}
        +\frac{ \ket*{x_{1}+1,x_{2}} }{U-\delta_{x_{1},-1} V}
        +\frac{ \ket*{x_{1},x_{2}-1} }{U-\delta_{x_{2},1} V} } \nonumber \\
    &\qquad +2t^2\bqty{ \frac{2}{U} + \frac{1}{U-\delta_{x_{1}^2,1} V } +\frac{1}{U-\delta_{x_{2}^2,1} V } } \ket*{x_{1},x_{2}} \nonumber \\
    &= \frac{2t^2}{U} \hat{\Pi}\bqty\Big{ 
        \ket*{x_{1}-1,x_{2}}
        +\ket*{x_{1},x_{2}+1}
        +\ket*{x_{1}+1,x_{2}}
        +\ket*{x_{1},x_{2}-1} } \nonumber \\
    &\qquad +2t^2\bqty{ \frac{4}{U} + \frac{V}{U(U-V)} \left(\delta_{x_{1}^2,1} + \delta_{x_{2}^2,1}\right) } \ket*{x_{1},x_{2}}.
    \label{eq_EffectiveHamiltonian022_MatElem1}
\end{align}
Here, we use the identity $(U-\delta_{x_{j}^2,1} V)^{-1} = U^{-1} + \delta_{x_{j}^2,1} V/U(U-V)$.
On the other hand, when $x_{1}+1 = x_{2}$, we have
\begin{align}
    \hat{H}_{(0,2,2)}^{(2)} \ket*{x_{1},x_{2}}
    &= -t^2 \hat{\Pi}\hat{T} \bqty{ \frac{ \hat{b}_{x_{1}-1}^{\dagger}\hat{b}_{x_{1}}^{\dagger} (\hat{b}_{x_{2}}^{\dagger})^2 }{U-\delta_{x_{1},1} V}
    +\frac{ (\hat{b}_{x_{1}}^{\dagger})^2  \hat{b}_{x_{2}}^{\dagger}\hat{b}_{x_{2}+1}^{\dagger} }{U-\delta_{x_{2},-1} V} +\frac{ \hat{b}_{x_{1}}^{\dagger} (\hat{b}_{x_{2}}^{\dagger})^3 +(\hat{b}_{x_{1}}^{\dagger})^3 \hat{b}_{x_{2}}^{\dagger} }{-U} } \ket*{0} \nonumber \\
    &= t^2 \hat{\Pi}\bqty{ \frac{ (\hat{b}_{x_{1}-1}^{\dagger})^2 (\hat{b}_{x_{2}}^{\dagger})^2 + (\hat{b}_{x_{1}}^{\dagger})^2 (\hat{b}_{x_{2}}^{\dagger})^2 }{U-\delta_{x_{1},1} V}
    +\frac{ (\hat{b}_{x_{1}}^{\dagger})^2  (\hat{b}_{x_{2}+1}^{\dagger})^2 + (\hat{b}_{x_{1}}^{\dagger})^2 (\hat{b}_{x_{2}}^{\dagger})^2 }{U-\delta_{x_{2},-1} V} +\frac{ 6(\hat{b}_{x_{1}}^{\dagger})^2 (\hat{b}_{x_{2}}^{\dagger})^2 }{-U} } \ket*{0} \nonumber \\
    &= \frac{2t^2}{U} \hat{\Pi}\bqty\Big{ 
        \ket*{x_{1}-1,x_{2}}
        +\ket*{x_{1},x_{2}+1}
        +\ket*{x_{1}+1,x_{2}}
        +\ket*{x_{1},x_{2}-1} } \nonumber \\
    &\qquad +2t^2\bqty{ -\frac{4}{U} + \frac{V}{U(U-V)} \left(\delta_{x_{1}^2,1} + \delta_{x_{2}^2,1}\right) } \ket*{x_{1},x_{2}}.
    \label{eq_EffectiveHamiltonian022_MatElem2}
\end{align}
By rewriting Eqs.~\eqref{eq_EffectiveHamiltonian022_MatElem1} and \eqref{eq_EffectiveHamiltonian022_MatElem2} in the operator form, we obtain
\begin{align}
    \hat{H}_{(0,2,2)}^{(2)}
    = \frac{2t^2}{U} \sum_{x(\neq 0)} \hat{A}_{x}^{\dagger} ( \hat{A}_{x-1} + \hat{A}_{x+1} ) -\frac{16t^2}{U} \sum_{x} \hat{n}^{(A)}_{x}\hat{n}^{(A)}_{x+1} +\frac{2t^2 V}{U(U-V)} \left(\hat{n}^{(A)}_{+1} + \hat{n}^{(A)}_{-1} \right) +\frac{8t^2}{U},
\end{align}
where $\hat{A}_{x}$ is an annihilation operator of hardcore bosons such that $\ket*{x_{1},x_{2}} = \hat{A}_{x_{1}}^{\dagger} \hat{A}_{x_{2}}^{\dagger} \ket*{0}$.
We also set $\hat{A}_{0} = 0$ and $\hat{n}^{(A)}_x \coloneqq \hat{A}_x^\dag\hat{A}_x$.
This model is a variant of the hardcore Bose-Hubbard model with a repulsive nearest-neighbor interaction.
This perturbation breaks down when $U^2 \ll t^2$ or $U \simeq V$.

\section{Derivation of the effective Hamiltonian in the $\vb{N} = (2,1,1)$ sector}
In this section, we derive the effective Hamiltonian in the $\vb{N} = (2,1,1)$ sector defined by
\begin{equation}
    \mathcal{H}_{(2,1,1)} \coloneqq \operatorname{span}\Bqty{ \ket*{y_{1},y_{2}} \coloneqq \frac{1}{\sqrt{2}} (\hat{b}_{0}^{\dagger})^2 \hat{b}_{y_{1}}^{\dagger} \hat{b}_{y_{2}}^{\dagger} \ket*{0} \mid y_{1} < y_{2},\ y_{1}\neq 0,\ y_{2}\neq 0 },
\end{equation}
whose unperturbed energy is $U+2V$.
Let $\hat{\Pi}$ be the projector onto $\mathcal{H}_{(2,1,1)}$ and set $\hat{Q}\coloneqq 1-\hat{\Pi}$.

For the first-order perturbation, we have
\begin{equation}
    \hat{H}_{(2,1,1)}^{(1)} \ket*{y_{1},y_{2}} 
    = t\hat{\Pi}\hat{T}\hat{\Pi} \ket*{y_{1},y_{2}} 
    = -t\hat{\Pi} \bqty\Big{ \ket*{y_{1}-1,y_{2}} + \ket*{y_{1},y_{2}+1} + \ket*{y_{1}+1,y_{2}} + \ket*{y_{1},y_{2}-1} }.
\end{equation}
Therefore, we obtain
\begin{equation}
    \hat{H}_{(2,1,1)}^{(1)} = -t \sum_{y(\neq 0)} \hat{B}_{y}^{\dagger} ( \hat{B}_{y-1} + \hat{B}_{y+1} ),
\end{equation}
where $\hat{B}_{y}\ (y\neq 0)$ is the annihilation operator of the \textit{cluster} consisting of a single particle and satisfies the commutation and anti-commutation relations of hard-core bosons.
We set $\hat{B}_{0} = 0$.

We then calculate the second-order term given by
\begin{equation}
    \hat{H}_{(2,1,1)}^{(2)} \ket*{y_{1},y_{2}} = t^2 \hat{\Pi}\hat{T}\hat{Q}\frac{1}{U+2V-\hat{Q}\hat{H}_{0}\hat{Q}}\hat{Q}\hat{T}\hat{\Pi} \ket*{y_{1},y_{2}}.
\end{equation}
The first application of the hopping operator $\hat{T}$ followed by the projection $\hat{Q}$ gives
\begin{align}
    \hat{Q} \hat{T} \ket*{y_1,y_2}
    &= \frac{1}{\sqrt{2}} \hat{Q} \hat{T} (\hat{b}_0^\dag)^2 \hat{b}_{y_1}^\dag \hat{b}_{y_2}^\dag \ket*{0}
    \nonumber \\
    &= \frac{-1}{\sqrt{2}} \hat{Q} \left[ 2 \hat{b}_0^\dag (\hat{b}_{1}^\dag + \hat{b}_{-1}^\dag) \hat{b}_{y_1}^\dag \hat{b}_{y_2}^\dag 
    + (\hat{b}_0^\dag)^2 (\hat{b}_{y_1+1}^\dag + \hat{b}_{y_1-1}^\dag) \hat{b}_{y_2}^\dag
    + (\hat{b}_0^\dag)^2 \hat{b}_{y_1}^\dag (\hat{b}_{y_2+1}^\dag + \hat{b}_{y_2-1}^\dag) \right]
    \ket*{0}
    \nonumber \\
    &= \frac{-1}{\sqrt{2}} \left[ 2 \hat{b}_0^\dag (\hat{b}_{1}^\dag + \hat{b}_{-1}^\dag) \hat{b}_{y_1}^\dag \hat{b}_{y_2}^\dag 
    + (\hat{b}_0^\dag)^3 (\delta_{y_1^2,1} \hat{b}_{y_2}^\dag + \hat{b}_{y_1}^\dag \delta_{y_2^2,1})
    + (\hat{b}_0^\dag)^2 \delta_{y_2-y_1,1} \left((\hat{b}_{y_1}^\dag)^2 + (\hat{b}_{y_2}^\dag)^2\right) 
    \right] \ket*{0}.
\end{align}
The successive application of $\hat{Q} (U+2V - \hat{Q} \hat{H}_0 \hat{Q})^{-1}$ gives
\begin{align}
    \hat{Q} \frac{1}{U+2V - \hat{Q} \hat{H}_0 \hat{Q}} \hat{Q} \hat{T} \ket*{y_1,y_2}
    &= -\frac{1}{\sqrt{2}} \left[
    \begin{multlined}
        \frac{ 2 \hat{b}_0^\dag \hat{b}_1^\dag \hat{b}_{y_1}^\dag \hat{b}_{y_2}^\dag }{U+V - U (\delta_{y_1,1} + \delta_{y_2,1})}
        + \frac{ 2 \hat{b}_0^\dag \hat{b}_{-1}^\dag \hat{b}_{y_1}^\dag \hat{b}_{y_2}^\dag }{U+V - U (\delta_{y_1,-1} + \delta_{y_2,-1})}
        \\
        + \frac{(\hat{b}_0^\dag)^3 (\delta_{y_1^2,1} \hat{b}_{y_2}^\dag + \hat{b}_{y_1}^\dag \delta_{y_2^2,1})}{-2U-V}
        + \frac{(\hat{b}_0^\dag)^2 \delta_{y_2-y_1,1} \left((\hat{b}_{y_1}^\dag)^2 + (\hat{b}_{y_2}^\dag)^2\right)}{-U}
    \end{multlined}
    \right] \ket*{0}
\end{align}
Here, since $\delta_{y_1,s} + \delta_{y_2,s} \ (s= \pm 1)$ takes either $+1$ or $0$, we have
\begin{multline}
    \frac{ 2 \hat{b}_0^\dag \hat{b}_1^\dag \hat{b}_{y_1}^\dag \hat{b}_{y_2}^\dag }{U+V - U (\delta_{y_1,1} + \delta_{y_2,1})}
    + \frac{ 2 \hat{b}_0^\dag \hat{b}_{-1}^\dag \hat{b}_{y_1}^\dag \hat{b}_{y_2}^\dag }{U+V - U (\delta_{y_1,-1} + \delta_{y_2,-1})}
    \\
    \begin{aligned}[b]
        &= \frac{2 \hat{b}_0^\dag (\hat{b}_1^\dag+\hat{b}_{-1}^\dag) \hat{b}_{y_1}^\dag \hat{b}_{y_2}^\dag}{U+V} + \frac{U}{V(U+V)} 2 \hat{b}_0^\dag \hat{b}_{y_1}^\dag \hat{b}_{y_2}^\dag \left[\hat{b}_1^\dag (\delta_{y_1,1}+\delta_{y_2,1}) + \hat{b}_{-1}^\dag (\delta_{y_1,-1}+\delta_{y_2,-1}) \right]
        \\
        &= \frac{2 \hat{b}_0^\dag (\hat{b}_1^\dag+\hat{b}_{-1}^\dag) \hat{b}_{y_1}^\dag \hat{b}_{y_2}^\dag}{U+V} + \frac{U}{V(U+V)} 2 \hat{b}_0^\dag \hat{b}_{y_1}^\dag \hat{b}_{y_2}^\dag (\delta_{y_1^2,1} \hat{b}_{y_1}^\dag + \delta_{y_2^2,1} \hat{b}_{y_2}^\dag).
        \\
        &= \frac{2 \hat{b}_0^\dag (\hat{b}_1^\dag+\hat{b}_{-1}^\dag) \hat{b}_{y_1}^\dag \hat{b}_{y_2}^\dag}{U+V} + \frac{U}{V(U+V)} 2 \hat{b}_0^\dag \left[\delta_{y_1^2,1} (\hat{b}_{y_1}^\dag)^2 \hat{b}_{y_2}^\dag + \delta_{y_2^2,1} \hat{b}_{y_1}^\dag (\hat{b}_{y_2}^\dag)^2 \right].
    \end{aligned}
\end{multline}
Therefore, the second-order term is calculated as
\begin{align}
    \hat{H}_{(2,1,1)}^{(2)} \ket*{y_{1},y_{2}} 
    &= -\frac{t^2}{\sqrt{2}} \hat{\Pi}\hat{T}
    \left[
    \begin{multlined}
        \frac{2 \hat{b}_0^\dag (\hat{b}_1^\dag+\hat{b}_{-1}^\dag) \hat{b}_{y_1}^\dag \hat{b}_{y_2}^\dag}{U+V} 
        + \frac{U}{V(U+V)} 2 \hat{b}_0^\dag \left(\delta_{y_1^2,1} (\hat{b}_{y_1}^\dag)^2 \hat{b}_{y_2}^\dag + \delta_{y_2^2,1} \hat{b}_{y_1}^\dag (\hat{b}_{y_2}^\dag)^2 \right)
        \\
        + \frac{(\hat{b}_0^\dag)^3 (\delta_{y_1^2,1} \hat{b}_{y_2}^\dag + \hat{b}_{y_1}^\dag \delta_{y_2^2,1})}{-2U-V}
        + \frac{(\hat{b}_0^\dag)^2 \delta_{y_2-y_1,1} \left((\hat{b}_{y_1}^\dag)^2 + (\hat{b}_{y_1+1}^\dag)^2\right)}{-U}
    \end{multlined}
    \right] \ket*{0}
    \nonumber \\
    &= \frac{t^2}{\sqrt{2}} \hat{\Pi} (\hat{b}_0^\dag)^2
    \left[
    \begin{multlined}
        \frac{4\hat{b}_{y_1}^\dag \hat{b}_{y_2}^\dag}{U+V} 
        + \frac{2 (\hat{b}_1^\dag+\hat{b}_{-1}^\dag) (\delta_{y_1^2,1} \hat{b}_{y_2}^\dag + \delta_{y_2^2,1} \hat{b}_{y_1}^\dag)}{U+V} 
        \\
        + \frac{4U}{V(U+V)} (\delta_{y_1^2,1} + \delta_{y_2^2,1}) \hat{b}_{y_1}^\dag \hat{b}_{y_2}^\dag
        + \frac{3 (\hat{b}_1^\dag+\hat{b}_{-1}^\dag) (\delta_{y_1^2,1} \hat{b}_{y_2}^\dag + \hat{b}_{y_1}^\dag \delta_{y_2^2,1})}{-2U-V}
        \\
        + \frac{2 \delta_{y_2-y_1,1} \left(\hat{b}_{y_1-1}^\dag \hat{b}_{y_1}^\dag + 2 \hat{b}_{y_1}^\dag \hat{b}_{y_1+1}^\dag + \hat{b}_{y_1+1}^\dag \hat{b}_{y_1+2}^\dag\right)}{-U}
    \end{multlined}
    \right] \ket*{0}
    \nonumber \\
    &= \frac{t^2}{\sqrt{2}} \hat{\Pi} (\hat{b}_0^\dag)^2
    \left[
    \begin{multlined}
        \left(
            \frac{4}{U+V}
            + \frac{8U^2+5UV-V^2}{V(U+V)(2U+V)}
              (\delta_{y_1^2,1}+\delta_{y_2^2,1})
            - \frac{4}{U}\delta_{y_2-y_1,1}
        \right)
        \hat{b}_{y_1}^\dag \hat{b}_{y_2}^\dag
        \\
        + \frac{U-V}{(U+V)(2U+V)}
        \left(
            \delta_{y_1^2,1}
            \hat{b}_{-y_1}^\dag \hat{b}_{y_2}^\dag
            + \delta_{y_2^2,1}
            \hat{b}_{y_1}^\dag \hat{b}_{-y_2}^\dag
        \right)
        \\
        - \frac{2}{U}\delta_{y_2-y_1,1}
        \left(
            \hat{b}_{y_1-1}^\dag \hat{b}_{y_1}^\dag
            + \hat{b}_{y_2}^\dag \hat{b}_{y_2+1}^\dag
        \right)
    \end{multlined}
    \right] \ket*{0}
    \nonumber \\
    &= \begin{multlined}[t]
        t^2 \left( \frac{4}{U+V} + \frac{8U^2+5UV-V^2}{V(U+V)(2U+V)} (\delta_{y_1^2,1}+\delta_{y_2^2,1}) - \frac{4}{U}\delta_{y_2-y_1,1} \right) \ket*{y_1,y_2} 
        \\
        + \frac{t^2(U-V)}{(U+V)(2U+V)} \left( \delta_{y_1^2,1}\ket*{-y_1,y_2} + \delta_{y_2^2,1}\ket*{y_1,-y_2} \right)
        \\
        - \frac{2t^2}{U}\delta_{y_2-y_1,1} \Big( \ket*{y_1-1,y_1} + \ket*{y_2,y_2+1} \Big).
    \end{multlined}
    \label{eq_EffectoveHamiltonian211_Elements}
\end{align}
Here, we extend the notation by setting $\ket*{a,b}=\ket*{b,a}$ and $\ket*{0,b}=\ket*{a,0}=\ket*{a,a}=0$.

The operator form of the Hamiltonian is then given by
\begin{align}
    \hat{H}_{(2,1,1)}^{(2)}
    &= 
    \begin{multlined}[t]
        \frac{4t^2}{U+V} 
        + \frac{t^2(8U^2+5UV-V^2)}{V(U+V)(2U+V)} \left(\hat{n}_{-1}^{(B)}+\hat{n}_1^{(B)}\right)
        -\frac{4t^2}{U} \sum_x \hat{n}_x^{(B)}\hat{n}_{x+1}^{(B)}
        \\
        + \frac{t^2(U-V)}{(U+V)(2U+V)} \left( \hat{B}_{-1}^\dag\hat{B}_1 + \hat{B}_1^\dag\hat{B}_{-1} \right)
        - \frac{2t^2}{U} \sum_x \left( \hat{B}_{x-1}^\dag\hat{n}_x^{(B)}\hat{B}_{x+1} + \hat{B}_{x+1}^\dag\hat{n}_x^{(B)}\hat{B}_{x-1} \right),
    \end{multlined}
    \label{Eq:211_EffectiveHamiltonian_2nd}
\end{align}
where $\hat{n}^{(B)}_{y} \coloneqq  \hat{B}_{y}^{\dagger} \hat{B}_{y}$.

\clearpage
\section{Derivation of the effective Hamiltonian \\ in the union of the $\vb{N} = (0,2,2)$ and $\vb{N} = (2,1,1)$ sectors}
We consider the case where $2U \simeq U+2V$, and therefore the subspaces $\mathcal{H}_{(0,2,2)}$ and $\mathcal{H}_{(2,1,1)}$ are mixed by the term $\hat{T}$.
Let $\hat{\Pi}_{1}$ be the projector onto $\mathcal{H}_{(0,2,2)}$ and $\hat{\Pi}_{2}$ be that onto $\mathcal{H}_{(2,1,1)}$, and set $\hat{\Pi} \coloneqq \hat{\Pi}_{1}+\hat{\Pi}_{2}$ and $\hat{Q} \coloneqq 1 - \hat{\Pi}$.
We denote a basis state of $\mathcal{H}_{(0,2,2)}$ by $\ket*{x_{1}, x_{2}}$ and that of $\mathcal{H}_{(2,1,1)}$ by $\ket*{y_{1}, y_{2}}$.

The zeroth-order Hamiltonian is given by
\begin{equation}
    \hat{H}_{\mathrm{eff}}^{(0)} = 2U \hat{\Pi}_{1} + (U+2V) \hat{\Pi}_{2}.
\end{equation}
Since $\hat{\Pi}_{1}\hat{T}\hat{\Pi}_{2} = 0$, the first-order perturbation gives
\begin{equation}
    \hat{H}_{\mathrm{eff}}^{(1)} = t\hat{\Pi}_{2}\hat{T}\hat{\Pi}_{2} = -t \sum_{x(\neq 0)} \hat{B}_{x}^{\dagger} ( \hat{B}_{x-1} + \hat{B}_{x+1} ),
\end{equation}
where $\hat{B}_{x}$ is defined as an annihilation operator of hardcore bosons such that $\ket*{y_{1},y_{2}} \coloneqq \hat{B}_{y_{1}}^{\dagger}\hat{B}_{y_{2}}^{\dagger} \ket*{0} \in \mathcal{H}_{(2,1,1)}$.

We divide the second-order perturbation $\hat{H}_{\mathrm{eff}}^{(2)}$ as follows:
\begin{align}
    \hat{H}_{\mathrm{eff}}^{(2)}
    = \hat{\Pi}_{1}\hat{H}_{\mathrm{eff}}^{(2)}\hat{\Pi}_{1} + \hat{\Pi}_{2}\hat{H}_{\mathrm{eff}}^{(2)}\hat{\Pi}_{2} +(\hat{\Pi}_{1}\hat{H}_{\mathrm{eff}}^{(2)}\hat{\Pi}_{2}+\hat{\Pi}_{2}\hat{H}_{\mathrm{eff}}^{(2)}\hat{\Pi}_{1})
    \quad\eqqcolon \hat{H}_{(0,2,2)}^{(2)} + \hat{H}_{(2,1,1)}^{(2)}+ \hat{H}_{\mathrm{mix}}.
    \label{eq_IntersectionEffectiveHamiltonian_Division}
\end{align}
The last term $\hat{H}_{\mathrm{mix}} \coloneqq \hat{\Pi}_{1}\hat{H}_{\mathrm{eff}}^{(2)}\hat{\Pi}_{2}+\hat{\Pi}_{2}\hat{H}_{\mathrm{eff}}^{(2)}\hat{\Pi}_{1}$ in Eq.~\eqref{eq_IntersectionEffectiveHamiltonian_Division} mixes the sectors $\mathcal{H}_{(0,2,2)}$ and $\mathcal{H}_{(2,1,1)}$.
Since the impurity site is unoccupied in $\mathcal{H}_{(0,2,2)}$, the hopping operator $\hat{T}$ must act on $(\hat{b}_0^\dag)^2$ twice to carry configurations from $\mathcal{H}_{(2,1,1)}$ to $\mathcal{H}_{(0,2,2)}$.
Therefore, we have
\begin{align}
    \hat{\Pi}_{1}\hat{H}_{\mathrm{eff}}^{(2)}\hat{\Pi}_{2} \ket*{y_{1},y_{2}} 
    &= t^2 \hat{\Pi}_{1} \hat{T} \hat{Q} \frac{1}{U+2V-\hat{Q}\hat{H}_0\hat{Q}} \hat{Q} \hat{T}\, \frac{1}{\sqrt{2}}(\hat{b}_{0}^{\dagger})^2 \hat{b}_{y_{1}}^{\dagger} \hat{b}_{y_{2}}^{\dagger} \ket*{0} 
    \nonumber \\
    &= t^2 \hat{\Pi}_{1} \hat{T} \hat{Q} \frac{1}{U+2V-\hat{Q}\hat{H}_0\hat{Q}}\, (-\sqrt{2}) \hat{b}_{0}^{\dagger} (\hat{b}_{-1}^{\dagger}+\hat{b}_{1}^{\dagger})  \hat{b}_{y_{1}}^{\dagger} \hat{b}_{y_{2}}^{\dagger} \ket*{0}
    \nonumber \\
    &= -\sqrt{2} t^2 \hat{\Pi}_{1} \hat{T} \hat{b}_{0}^{\dagger} \qty(\frac{\hat{b}_{-1}^{\dagger}}{U+V - (\delta_{y_1,-1}+\delta_{y_2,-1})U} + \frac{\hat{b}_{1}^{\dagger}}{U+V - (\delta_{y_1,1}+\delta_{y_2,1})U})  \hat{b}_{y_{1}}^{\dagger} \hat{b}_{y_{2}}^{\dagger} \ket*{0}
    \nonumber \\
    &= -\sqrt{2} t^2 \hat{\Pi}_{1} \hat{T} \hat{b}_{0}^{\dagger} \qty(
    \frac{\hat{b}_{1}^{\dagger}+\hat{b}_{-1}^{\dagger}}{U+V} + \frac{U}{V(U+V)} \left( \hat{b}_{-1}^\dag (\delta_{y_1,-1}+\delta_{y_2,-1}) + \hat{b}_1^\dag (\delta_{y_1,1}+\delta_{y_2,1}) \right) )
    \hat{b}_{y_{1}}^{\dagger} \hat{b}_{y_{2}}^{\dagger} \ket*{0}
    \nonumber \\
    &= -\sqrt{2} t^2 \hat{\Pi}_{1} \hat{T} \hat{b}_{0}^{\dagger} 
    \qty( \frac{(\hat{b}_{1}^{\dagger}+\hat{b}_{-1}^{\dagger}) \hat{b}_{y_{1}}^{\dagger} \hat{b}_{y_{2}}^{\dagger}}{U+V} + \frac{U}{V(U+V)} \left( \delta_{y_1^2,1} (\hat{b}_{y_1}^\dag)^2 \hat{b}_{y_2}^\dag + \delta_{y_2^2,1} \hat{b}_{y_1}^\dag (\hat{b}_{y_2}^\dag)^2 \right) )
    \ket*{0}
    \nonumber \\
    &= \sqrt{2} t^2 \hat{\Pi}_{1} (\hat{b}_{-1}^{\dagger} + \hat{b}_{1}^{\dagger}) 
    \qty( \frac{(\hat{b}_{1}^{\dagger}+\hat{b}_{-1}^{\dagger}) \hat{b}_{y_{1}}^{\dagger} \hat{b}_{y_{2}}^{\dagger}}{U+V} + \frac{U}{V(U+V)} \left( \delta_{y_1^2,1} (\hat{b}_{y_1}^\dag)^2 \hat{b}_{y_2}^\dag + \delta_{y_2^2,1} \hat{b}_{y_1}^\dag (\hat{b}_{y_2}^\dag)^2 \right) )
    \ket*{0}
    \nonumber \\
    &= \sqrt{2} t^2 
    \qty( \frac{ 2 \delta_{y_1,-1} \delta_{y_2,1} (\hat{b}_{-1}^{\dagger})^2 (\hat{b}_{1}^{\dagger})^2 }{U+V} + \frac{U}{V(U+V)} \, 2 \delta_{y_1,-1} \delta_{y_2,1} (\hat{b}_{-1}^\dag)^2 (\hat{b}_1^\dag)^2 )
    \ket*{0}
    \nonumber \\
    &= \frac{2\sqrt{2} t^2}{V} \delta_{y_1,-1} \delta_{y_2,1} (\hat{b}_{-1}^{\dagger})^2 (\hat{b}_{1}^{\dagger})^2 
    \ket*{0}
    \nonumber \\
    &= \frac{4\sqrt{2} t^2}{V} \delta_{y_1,-1} \delta_{y_2,1} \ket*{x_1=-1,x_2=+1}.
\end{align}
Therefore, we obtain
\begin{equation}
    \hat{\Pi}_{1}\hat{H}_{\mathrm{eff}}^{(2)}\hat{\Pi}_{2} = \frac{4\sqrt{2}t^2}{V} \hat{A}_{-1}^{\dagger}\hat{A}_{+1}^{\dagger}\hat{B}_{-1}\hat{B}_{+1}.
\end{equation}
Thus, the second-order effective Hamiltonian within the sector $\mathcal{H}_{(0,2,2)} \oplus \mathcal{H}_{(2,1,1)}$ is given by
\begin{align}
    \hat{H}_{\mathrm{eff}}^{(2)}
    &= \hat{H}_{(0,2,2)}^{(2)} + \hat{H}_{(2,1,1)}^{(2)} +\frac{4\sqrt{2}t^2}{V} \qty(\hat{A}_{-1}^{\dagger}\hat{A}_{+1}^{\dagger}\hat{B}_{-1}\hat{B}_{+1} +  \hat{B}_{-1}^{\dagger}\hat{B}_{+1}^{\dagger}\hat{A}_{-1}\hat{A}_{+1}).
\end{align}
Here, the operators $\hat{A}$ and $\hat{B}$ satisfy $\hat{B}^{\dagger}\hat{A}^{\dagger} = 0$, $\hat{B}\hat{A}^{\dagger} \ket*{0} = 0$, and $\hat{A}\hat{B}^{\dagger} \ket*{0} = 0$.

\clearpage
\section{Derivation of the bound eigenstates of the effective Hamiltonian \\ in the $\vb{N} = (0,2,2)$ sector}
In this section, we solve the eigenvalue equation
\begin{equation}
    \hat{H}_{(0,2,2)}^{(2)}\ket*{E^{(2)}}=E^{(2)}\ket*{E^{(2)}},
    \qquad \hat{P}\ket*{E^{(2)}}=\epsilon\ket*{E^{(2)}},
    \qquad \epsilon\in\{+1,-1\},
\end{equation}
where $\hat{P}\ket*{x_1,x_2}=\ket*{-x_2,-x_1}$ denotes the reflection.
In terms of the wave function $\psi(x_1, x_2) \coloneqq \matrixel*{0}{\hat{A}_{x_1} \hat{A}_{x_2}}{E^{(2)}}$, the eigenvalue equation reads
\begin{align}
    \begin{multlined}[b]
        E^{(2)} \psi(x_1,x_2) = \frac{2t^2}{U} \bigl[\psi(x_1-1,x_2)+\psi(x_1+1,x_2) +\psi(x_1,x_2-1)+\psi(x_1,x_2+1)\bigr]
        \\
        +\left[ \frac{8t^2}{U} (1 - 2 \delta_{x_1+1,x_2}) + \frac{2t^2 V}{U (U-V)} (\delta_{x_1^2,1}+\delta_{x_2^2,1}) \right]
        \psi(x_1,x_2)
    \end{multlined}
    ,\quad (x_1<x_2,\ x_1x_2 \neq 0).
    \label{Eq:022_EVequation}
\end{align}
Here, $x_0=-\infty$ and $x_3=+\infty$.
We seek for the bound eigenstates that have the Bethe form in the free configurations:
\begin{gather}
    \psi(x_1,x_2) = F_m(x_1,x_2),\quad 
    \begin{aligned}[t]
        \forall \vb{x} \in \mathcal{C}_m &\coloneqq \{(x_1,x_2) \in \mathbb{Z}^2 \colon x_1 < x_2,\ x_m < 0 < x_{m+1}\},\ (m=0,1,2), \\
        F_m(x_1,x_2) &\coloneqq \sum_{\vb{z} \in \mathcal{Z}} A_{m,\vb{z}} z_1^{x_1} z_2^{x_2} \quad (z_j \in \mathbb{C}\setminus\{0\},\ A_{\vb{z}} \in \mathbb{C}).
    \end{aligned}
    \label{eq_022_BetheForm}
\end{gather}
Here, $\mathcal{Z} \subset (\mathbb{C} \setminus \{0\})^2$ is a finite set closed under any signed permutation $W_2 \coloneqq \mathfrak{S}_2 \ltimes (\mathbb Z_2)^2$ defined by
\begin{equation}
    \sigma \in \mathfrak{S}_2,\quad I \subseteq \{1,2\},\quad (\sigma, I)\colon (z_1, z_2) \mapsto (z^I_{\sigma(1)}, z^I_{\sigma(2)}),
    \quad\text{where}\quad
    z^I_j \coloneqq 
    \begin{cases}
            z_j & (j \notin I); \\ z_j^{-1} & (j \in I).
    \end{cases}
    \label{eq_022_signedPerm}
\end{equation}
The independence of the multivariate exponentials and the reflection symmetry give
\begin{equation}
    A_{m, (z_1,z_2)} = \epsilon A_{2-m, (z_2^{-1}, z_1^{-1})}\quad (m=0,1,2).
\end{equation}
Therefore, below, we focus on coefficients with $m=0$ and $m=1$.

\subsection{Eigenvalue equations in terms of Bethe functions}
At free configurations $\vb{x} \in \mathcal{C}_m$ with $x_1+1 < x_2$ and $x_j^2 \neq 1 \ (j=1,2)$, the eigenvalue equations are satisfied if and only if
\begin{equation}
    E^{(2)} = \frac{2t^2}{U} (z_1 + z_1^{-1} + z_2 + z_2^{-1}) + \frac{8t^2}{U}\quad\text{whenever $A_{m,\vb{z}} \neq 0$.}
    \label{eq_022_freeEVres}
\end{equation}

The points in $\mathcal{C}_0$ not covered by Eq.~\eqref{eq_022_freeEVres} are those with $x_2 = x_1+1$ or $x_1=1$.
Since $\psi(x_1,x_2) = 0$ for points outside $\mathcal{H}_{(0,2,2)}$, 
the eigenvalue equations at these points reduce to the following system of equations:
\begin{equation}
    \begin{aligned}
        - \frac{2t^2}{U} \left[ F_0(x_1+1,x_1+1) + F_0(x_1,x_1) \right] - \frac{16t^2}{U} F_0(x_1,x_1+1) &= 0,\quad (\forall x_1 > 1), \\
        -\frac{2t^2}{U} F_0(0,x_2) + \frac{2t^2 V}{U(U-V)} F_0(1,x_2)& = 0,\quad (\forall x_2 > 2), \\
        -\frac{2t^2}{U} \left[ F_0(0,2) + F_0(2,2) + F_0(1,1) \right] + \left[ -\frac{16t^2}{U} + \frac{2t^2 V}{U(U-V)} \right] F_0(1,2) &= 0.
    \end{aligned}
    \label{eq_002_EVequation_C0}
\end{equation}

The points in $\mathcal{C}_1$ not covered by Eq.~\eqref{eq_022_freeEVres} are those with $x_1=-1$ or $x_2=1$.
The eigenvalue equations at these points are given by
\begin{equation}
    \begin{aligned}
        -\frac{2t^2}{U} F_1(0,x_2) + \frac{2t^2 V}{U(U-V)} F_1(-1,x_2) &= 0,\quad (x_2 > 1), \\
        -\frac{2t^2}{U} F_1(x_1,0) + \frac{2t^2 V}{U(U-V)} F_1(x_1,1) &= 0,\quad (x_1 < -1), \\
        -\frac{2t^2}{U} [F_1(0,1) + F_1(-1,0)] + \frac{4t^2 V}{U(U-V)} F_1(-1,1) &= 0.
    \end{aligned}
    \label{eq_002_EVequation_C1}
\end{equation}

\subsection{Splitting of eigenstate equations into $W_2$ orbits}
Substituting the definition of $F_m$ in Eq.~\eqref{eq_022_BetheForm} and comparing the coefficients of independent exponentials, these equations are satisfied if and only if the coefficients $\{A_{m,\vb{z}}\}$ satisfy, for any $\eta \in \mathbb{C}$,
\begin{equation}
    \sum_{\substack{\vb{z} \in \mathcal{Z} \\ \eta = z_1 z_2}} A_{0,\vb{z}} (\eta + 1 + 8 z_2) = 0,
    \quad 
    \sum_{\substack{\vb{z} \in \mathcal{Z} \\ \eta = z_2}} A_{0,\vb{z}} \left(1 - \frac{V}{U-V} z_1 \right) =0, 
    \quad
    \sum_{\substack{\vb{z} \in \mathcal{Z} \\ \eta = z_2}} A_{1,\vb{z}} \left(1 - \frac{V}{U-V} z_1^{-1} \right) =0.
    \label{eq_022_coeffEVEqs}
\end{equation}
Due to the identity
\begin{equation}
     (z_1+z_1^{-1} + z_2 + z_2^{-1}) = (z_1+z_2) \left(1+\frac{1}{z_1 z_2}\right) = \frac{U}{2t^2}E^{(2)} - 4,
\end{equation}
fixing either $z_1z_2\neq-1$ or $z_2$ determines the pair $(z_1,z_2)$ up to permutation or inversion of $z_1$, respectively; the bound state conditions require $A_{0,\vb z}=0$ when $z_1z_2=-1$.
This implies that the system~\eqref{eq_022_coeffEVEqs} splits into the systems each involving only a single orbit of $\vb{z}$ generated by signed permutations~\eqref{eq_022_signedPerm}.
Therefore, in the following, we focus on each orbit separately.
We choose a representative with $\vb z=(z_1,z_2)$ with $0<|z_2|\leq|z_1|\leq1$ and write $F_m$ for this orbit as
\begin{equation}
    F_m(x_1,x_2) = \sum_{I\subseteq[2]}\sum_{\sigma\in S_2} A^I_{m,\sigma}(z_{\sigma(1)}^I)^{x_1}(z_{\sigma(2)}^I)^{x_2},
    \quad
    A^I_{m,\sigma}=\epsilon A^{I^c}_{2-m,\sigma(12)}.
    \label{eq:022_unified_orbit}
\end{equation}

\subsection{Nontrivial solutions in $\mathcal{C}_1$}
The bound states in $\mathcal{C}_1$ must satisfy
\begin{equation}
    A^I_{1,\sigma} \neq 0 \quad\Longrightarrow\quad |z^I_{\sigma(1)}| > 1,\quad |z^I_{\sigma(2)}| < 1.
\end{equation}
Due to the former condition, $A^I_{1,\sigma}$ and $A^{I\triangle \{\sigma(1)\}}_{1,\sigma}$ cannot be simultaneously nonzero.
Since $0 < |z_2| \leq |z_1| \leq 1$, we have $A^I_{1,\sigma} = 0$ unless $I = \{\sigma(1)\}$.
With the parity condition $A^{\{\sigma(1)\}}_{1,\sigma} = \epsilon A^{\{\sigma(2)\}}_{1,\sigma(12)}$,  Eq.~\eqref{eq_002_EVequation_C1} is equivalent to
\begin{equation}
    A^{\{\sigma(1)\}}_{1,\sigma} = 0 \quad\text{or}\quad z_1 = z_2 = \frac{U-V}{V}.
\end{equation}
For a nonzero solution in $\mathcal C_1$, we therefore have
\begin{equation}
    F_1(x_1,x_2) = (1+\epsilon) A^{\{1\}}_{1,\operatorname{id}} \left(\frac{U-V}{V}\right)^{x_2-x_1}.
\end{equation}
This represents a valid bound state when $\epsilon = +1$ and $|(U-V)/V| < 1$, i.e., when $0 < U/V < 2 \ (U\neq V)$.
This solution corresponds to Eq.~\xref{\eqref{Eq:022_BoundState2}} in the main text.
The corresponding energy gives Eq.~\xref{\eqref{eq:022_BoundStateEnergy2}} in the main text:
\begin{equation}
    E^{(2)} = \frac{8t^2}{U} + \frac{2t^2}{U} (z_1 + z_1^{-1} + z_2 + z_2^{-1}) = \frac{8t^2}{U} + \frac{4t^2}{U} \frac{(U-V)^2 + V^2}{V(U-V)}.
\end{equation}

\subsection{Nontrivial solutions in $\mathcal{C}_0$}
The bound states in $\mathcal{C}_0$ must satisfy
\begin{equation}
    A^I_{0,\sigma} \neq 0 \quad\Longrightarrow\quad |z^I_{\sigma(2)}| < 1,\quad |z^I_1 z^I_2| < 1.
\end{equation}
This requires $A^I_{0,\sigma} = 0$ unless $A^{\varnothing}_{0,\operatorname{id}}$, $A^{\varnothing}_{0,(12)}$, or $A^{\{1\}}_{0,\operatorname{id}}$.
In particular, we have $A^{\{1\}}_{0,(12)} = A^{\{2\}}_{0,(12)} = 0$.
Thus, Eq.~\eqref{eq_002_EVequation_C0} is equivalent to
\begin{alignat}{3}
    A^{\varnothing}_{0,\operatorname{id}} (z_1 z_2 + 1 + 8 z_2) + A^{\varnothing}_{0,(12)} (z_1 z_2 + 1 + 8 z_1) = 0,&\quad &\quad
    A^{\{1\}}_{0,\operatorname{id}} \left(\frac{z_2}{z_1} + 1 + 8 z_2 \right) &= 0,
    \label{eq_002_EVequation_C0_coeff1}
    \\
    A^{\varnothing}_{0,\operatorname{id}} \left( 1 - \frac{V}{U-V} z_1 \right) + A^{\{1\}}_{0,\operatorname{id}} \left( 1 - \frac{V}{U-V} \frac{1}{z_1} \right) = 0,&\quad &\quad
    A^{\varnothing}_{0,(12)} \left( 1 - \frac{V}{U-V} z_2 \right) &= 0.
    \label{eq_002_EVequation_C0_coeff2}
\end{alignat}

\subsubsection{Case: $A^{\varnothing}_{0,\operatorname{id}} (z_1 z_2 + 1 + 8 z_2) = 0$}
In this case, Eq.~\eqref{eq_002_EVequation_C0_coeff1} gives
\begin{equation}
    A^{\varnothing}_{0,\operatorname{id}} z_1 = A^{\varnothing}_{0,\operatorname{id}} \left(-\frac{1+8z_2}{z_2}\right),
    \quad
    A^{\varnothing}_{0,(12)} z_1 = A^{\varnothing}_{0,(12)} \left(-\frac{1}{z_2+8}\right),
    \quad
    A^{\{1\}}_{0,\operatorname{id}} z_1^{-1} = A^{\{1\}}_{0,\operatorname{id}} \left(-\frac{1+8z_2}{z_2}\right).
    \label{eq_002_EVequation_C0_coeff3}
\end{equation}
Therefore, the Bethe function becomes
\begin{equation}
    F_0(x_1,x_2)
    = \left( A^{\varnothing}_{0,\operatorname{id}} + A^{\{1\}}_{0,\operatorname{id}} \right)
    \left(-\frac{1+8z_2}{z_2}\right)^{x_1} z_2^{x_2} + A^{\varnothing}_{0,(12)} z_2^{x_1} \left(-\frac{1}{z_2+8}\right)^{x_2}.
\end{equation}
On the other hand, substituting Eq.~\eqref{eq_002_EVequation_C0_coeff3} into Eq.~\eqref{eq_002_EVequation_C0_coeff2}, we obtain
\begin{equation}
    \left( A^{\varnothing}_{0,\operatorname{id}} + A^{\{1\}}_{0,\operatorname{id}} \right)
    \left(-\frac{1+8z_2}{z_2}\right) = \left( A^{\varnothing}_{0,\operatorname{id}} + A^{\{1\}}_{0,\operatorname{id}} \right)
    \left(\frac{U-V}{V}\right),
    \quad\text{and}\quad
    A^{\varnothing}_{0,(12)} z_2 = A^{\varnothing}_{0,(12)} \left(\frac{U-V}{V}\right).
\end{equation}
Using these results to eliminate $z_2$, we obtain the solution corresponding to Eq.~\xref{\eqref{Eq:022_BoundState1}} in the main text:
\begin{equation}
    F_0(x_1,x_2)
    = \left( A^{\varnothing}_{0,\operatorname{id}} + A^{\varnothing}_{0,(12)} + A^{\{1\}}_{0,\operatorname{id}} \right)
    \left(\frac{U-V}{V}\right)^{x_1}
    \left(-\frac{V}{U+7V}\right)^{x_2},\quad (U \neq -7V),
\end{equation}
It is straightforward to verify that this solution solves Eq.~\eqref{Eq:022_EVequation} with energy given in Eq.~\xref{\eqref{eq:022_BoundStateEnergy1}} in the main text:
\begin{equation}
    E^{(2)}
    = \frac{8t^2}{U} +\frac{2t^2}{U}(z_1+z_1^{-1}+z_2+z_2^{-1})
    = -\frac{8t^2}{U} + \frac{16t^2}{U}\frac{V^2}{(U-V)(U+7V)}.
\end{equation} 
This solution represents a bound state when $|V/(U+7V)|<1$ and $|(U-V)/(U+7V)|<1$, which are equivalent to $U/V>-3$.

\subsubsection{Case: $A^{\varnothing}_{0,\operatorname{id}} (z_1 z_2 + 1 + 8 z_2) \neq 0$}
In this case, Eq.~\eqref{eq_002_EVequation_C0_coeff1} implies $A^{\varnothing}_{0,(12)} (z_1 z_2 + 1 + 8 z_1) \neq 0$, and Eq.~\eqref{eq_002_EVequation_C0_coeff2} then implies $z_2 = (U-V)/V$.
Multiplying Eq.~\eqref{eq_002_EVequation_C0_coeff2} by $z_1^{-1} z_2 + 1 + 8 z_2$ and using Eq.~\eqref{eq_002_EVequation_C0_coeff1}, we obtain $(z_1^{-1} z_2 + 1 + 8 z_2) (z_1 - z_2) = 0$.
This implies
\begin{equation}
    (z_1^{-1} z_2 + 1 + 8 z_2) F_0(x_1,x_2) = (z_1^{-1} z_2 + 1 + 8 z_2) (A^{\varnothing}_{0,\operatorname{id}} + A^{\varnothing}_{0,(12)}) z_1^{x_1+x_2} = 0,
\end{equation}
where we use Eq.~\eqref{eq_002_EVequation_C0_coeff1} multiplied by $z_1^{-1} z_2 + 1 + 8 z_2$ in deriving the second equality.
Since $F_0 \not\equiv 0$, we must have
\begin{equation}
    z_1 = - \frac{z_2}{1+8z_2} = - \frac{U-V}{8U-7V},
    \qquad z_2 = \frac{U-V}{V}.
\end{equation}
Then, Eqs.~\eqref{eq_002_EVequation_C0_coeff1} and \eqref{eq_002_EVequation_C0_coeff2} determine $A^{\varnothing}_{0,(12)}$ and $A^{\{1\}}_{0,\operatorname{id}}$, and the resulting solution is
\begin{equation}
    F_0(x_1,x_2) = A^{\varnothing}_{0,\operatorname{id}} \left[ 
    \begin{multlined}
        \left( -\frac{U-V}{8U-7V} \right)^{x_1} \left(\frac{U-V}{V}\right)^{x_2} + \frac{(7U-6V)(9U-8V)}{U(U-2V)} \left(\frac{U-V}{V}\right)^{x_1} \left( -\frac{U-V}{8U-7V} \right)^{x_2}
        \\
        - \frac{2(4U-3V)}{8U-7V} \frac{(U-V)^2}{U^2+6UV-6V^2} \left( -\frac{8U-7V}{U-V} \right)^{x_1} \left(\frac{U-V}{V}\right)^{x_2}
    \end{multlined}
     \right].
\end{equation}
This solution represents a valid bound state when
\begin{equation}
    \frac{U}{V} \in \left(\frac{3}{4}, \frac{6}{7}\right) \cup \left(\frac{8}{9}, 1\right).
\end{equation}
This is outside the focus $(U \simeq 2V)$ of the main text.
The corresponding energy is given by
\begin{equation}
    E^{(2)} 
    = \frac{8t^2}{U} + \frac{2t^2}{U} (z_1 + z_1^{-1} + z_2 + z_2^{-1}) 
    = -\frac{8t^2}{U} + \frac{16 t^2}{U} \frac{(U-V)^2}{V(8U-7V)}.
\end{equation}

\bibliographystyleSM{apsrev4-2}
\bibliographySM{supplement}
\fi 

\end{document}